\documentclass[11pt]{article}
\usepackage{jcappub}
\usepackage{bm}
\usepackage{color}
\usepackage{graphicx}
\usepackage{multirow}

\newcommand{\qm}{\left (q \over m_N \right )}
\renewcommand{\qm}{\q}
\newcommand{\qmsq}{\qmsq}
\renewcommand{\qmsq}{\q^2}
\newcommand{\del}{\nabla}

\newcommand{\N}{N} % symbol for nucleon

\newcommand{\be}{\begin{equation}}
\newcommand{\ee}{\end{equation}}
\newcommand{\een}{\end{subequations}}
\newcommand{\ben}{\begin{subequations}}
\newcommand{\beq}{\begin{eqalignno}}
\newcommand{\eeq}{\end{eqalignno}}

\newcommand{\lsim}{\mathrel{\mathop{\kern 0pt \rlap
      {\raise.2ex\hbox{$<$}}}\lower.9ex\hbox{\kern-.190em $ \sim$}}}
\newcommand{\gsim}{\mathrel{\mathop{\kern 0pt
      \rlap{\raise.2ex\hbox{$>$}}}\lower.9ex\hbox{\kern-.190em $\sim$}}}

\newcommand{\VectorTypefaceArrow}{
\let\oldvec\vec
\renewcommand{\vec}[1]{\oldvec{##1}} % use arrow
\newcommand{\uvec}[1]{\hat{##1}} % use a hat
}

\VectorTypefaceArrow

\newcommand{\op}[1]{\widehat{#1}} % use a hat

\newcommand{\curr}[1]{#1} % no special typeface
\newcommand{\mptype}[1]{{\rm #1}}

\newcommand{\mpL}{\mptype{L}}
\newcommand{\mpTE}{\mptype{TE}}
\newcommand{\mpTM}{\mptype{TM}}

\renewcommand{\Re}{\mathop{\rm Re}}
\renewcommand{\Im}{\mathop{\rm Im}}

\usepackage{upgreek} % nonslanted greek letters (for unit vectors in spherical coordinates)
\newcommand{\q}{{\widetilde{q}}}
\newcommand{\calO}{{\cal O}}
\newcommand{\tr}{\mathop{\rm tr}}
\newcommand{\rmfi}{{\rm f\,\!i}}

\usepackage{mathtools}
\newcommand{\myoverbracket}[1]{\overbracket[0.5pt][3pt]{#1}{}}

\newcommand{\lsymm}{[}
\newcommand{\rsymm}{]_{\rm sym}}

\newcommand{\plus}{\,+}

\usepackage{booktabs}
\usepackage{xcolor}
\usepackage{arydshln}

\allowdisplaybreaks

\title{The effective theory of nuclear scattering for a WIMP of arbitrary spin}

\author[a]{Paolo Gondolo,}
\affiliation[a]{Department of Physics, University of Utah, 115 South 1400
  East \#201, Salt Lake City, Utah 84112-0830}
\emailAdd{paolo.gondolo@utah.edu}

\author[b]{Sunghyun Kang,}
\author[b]{Stefano Scopel,}
\author[b, c]{Gaurav Tomar}
\affiliation[b]{Department of Physics, Sogang University, Seoul 121-742, South Korea}
\affiliation[c]{Physik-Department, Technische Universit\"at M\"unchen, James-Franck-Stra\ss e, 85748
Garching, Germany}
\emailAdd{francis735@naver.com}
\emailAdd{scopel@sogang.ac.kr}
\emailAdd{physics.tomar@tum.de}

\abstract{ We introduce a systematic approach to characterize the most
  general non-relativistic WIMP--nucleus interaction allowed by
  Galilean invariance for a WIMP of arbitrary spin $j_\chi$ in the
  approximation of one--nucleon currents. Five nucleon currents arise
  from the nonrelativistic limit of the free nucleon Dirac
  bilinears. Our procedure consists in (1) organizing the WIMP currents
  according to the rank of the $2 j_\chi+1$ irreducible operator
  products of up to $2 j_\chi$ WIMP spin vectors, and (2) coupling each
  of the WIMP
  currents to each of the five nucleon currents. The transferred momentum $q$ appears to a power fixed by
  rotational invariance. For a WIMP of spin $j_\chi$ we find a basis
  of 4+20$j_\chi$ independent operators that exhaust all the possible
  operators that drive elastic WIMP--nucleus scattering in the approximation
  of one--nucleon currents. By comparing our operator basis, which is complete, to the operators already
  introduced in the literature we show that some of the latter for
  $j_\chi=1$ were not independent and some were missing. We provide explicit formulas for the
  squared scattering amplitudes in terms of the nuclear response
  functions, which are available in the literature for most of the
  targets used in WIMP direct detection experiments.}

\begin{document}

\maketitle
\section{Introduction}
\label{sec:introduction}
In one of its most popular scenarios dark matter (DM) is believed to
be composed of Weakly Interacting Massive Particles (WIMPs) with a
mass in the GeV-TeV range and weak--type interactions with ordinary
matter. Such small but non vanishing interactions can drive WIMP
scattering off nuclear targets, and the measurement of the ensuing
nuclear recoils in low--background detectors (direct detection, DD)
represents the most straightforward way to detect them.

The most popular WIMP candidates are provided by extensions of the
Standard Model such as Supersymmetry or Large Extra Dimensions which
are in growing tension with the constraints from the Large Hadron
Collider (LHC). As a consequence, model-independent approaches have
become increasingly popular to interpret DM search
experiments~\cite{chang_momentum_dependence_2010, dobrescu_nreft,
  fan_2010, Hisano_vector_effective, Hisano_EW_DM, hill_solon_nreft,
  peter_nreft, cirelli_tools_2013, effective_wimps_2014, catena_nreft,
  catena_directionality_nreft_2015, Hisano_Wino_DM,
  Catena_Gondolo_global_fits, cerdeno_nreft,
  Catena_Gondolo_global_limits, nreft_bayesian, xenon100_nreft,
  cresst_nreft,matching_solon1,matching_solon2,chiral_eft,hoferichter_si,bishara_2017}.

In particular, since the DD process is non--relativistic (NR), on general grounds the WIMP-nucleon interaction can be
parameterized with an effective Hamiltonian ${\bf\mathcal{H}}$ that complies with
Galilean symmetry. 
The effective Hamiltonian ${\bf\mathcal{H}}$ to zero-th order in the
WIMP-nucleon relative velocity $\vec{v}$ and momentum transfer $\vec{q}$ has been known since at least
Ref.~\cite{GoodmanWitten}, and consists of the usual spin-dependent
(SD) and spin-independent (SI) terms.  To first order in $\vec{v}$, the
effective Hamiltonian ${\bf\mathcal{H}}$ has been systematically
described in~\cite{haxton1,haxton2} for WIMPs of spin 0 and 1/2, and
less systematically described
in~\cite{krauss_spin_1,catena_krauss_spin_1} for WIMPs of spin 1 and
in~\cite{barger_2008} for WIMPs of spin 3/2. An extension to spin-1/2 inelastic DM to first-order approximation in the WIMP mass difference can be found in~\cite{Barello:2014uda}.

In this paper we  systematically extend the WIMP-nucleon effective interaction approach to the case of a WIMP with arbitrary spin $j_\chi$. As in~\cite{haxton1,haxton2,krauss_spin_1,catena_krauss_spin_1,barger_2008}, we focus on elastic WIMP-nucleus scattering and include one-nucleon currents only \cite{haxton1,haxton2}. The effective Hamiltonian  ${\bf\mathcal{H}}$  is a sum of WIMP-nucleon operators $\mathcal{O}_{j}  t^{\tau}$, each multiplied by a coefficient $c_j^\tau$,
\begin{eqnarray}
{\bf\mathcal{H}}&=& \sum_{\tau=0,1} \sum_{j=1}^{N} c_j^{\tau} \mathcal{O}_{j}  t^{\tau}.
\label{eq:H}
\end{eqnarray}
Here $\tau$ is an isospin index (0 for isoscalar and 1 for isovector),  $t^0=1$, $t^1=\tau_3$ are nucleon isospin operators (the
$2\times2$ identity and the third Pauli matrix, respectively), and the $\mathcal{O}_{j}$'s ($j=1,N$) are operators in the space of WIMP-nucleon states. Alternatively, the sum over the isospin index $\tau$ can be replaced by a sum over protons and neutrons using the following relations between the isoscalar and isovector coupling constants $c^0_j$ and $c^{1}_j$ and the proton and neutron coupling constants $c^{p}_j$ and $c^{n}_j$,
\begin{align}
c^{p}_j=\frac{c^{0}_j+c^{1}_j}{2} , \qquad
c^{n}_j=\frac{c^{0}_j-c^{1}_j}{2} .
\end{align}

{ % begin Haxton operator table
\begin{table}[t]\centering
  \caption{Non-relativistic Galilean invariant operators discussed in
    the literature (\cite{haxton2, krauss_spin_1,
      catena_krauss_spin_1}) for a dark matter particle of spin $0$,
    $1/2$ and $1$, and their relation with the WIMP--nucleon operators
    $\calO_{X,s,l}$ defined in
    Eqs.~(\ref{eq:basis_WIMP_nucleon_operators}). Notice that the
    sign convention for the momentum transfer $\vec{q}$ used in this
    table and throughout the paper is opposite to that
    of Refs.~\cite{haxton2, krauss_spin_1, catena_krauss_spin_1}.}
\label{tab:Haxton_operators}
\renewcommand{\arraystretch}{1.5}
\addtolength{\tabcolsep}{2.0pt}
\vskip\baselineskip
\hspace{-4em}
\begin{minipage}{0.4\textwidth}
\begin{tabular}{@{}llr@{}}
\toprule
%Haxton et al.'s &Definition & Ours \\
%\midrule
$\calO_{1}$ & $1$ & $\calO_{M,0,0}$ \\
[0.5ex]\cdashline{1-3}
$\calO_{2}$ & $(\vec{v}{}^{\plus}_{\chi N})^2 $ & $N.A.$ \\
[0.5ex]\cdashline{1-3}
$\calO_{3}$ & $-i \vec{S}_N \cdot ( \vec{\q} \times \vec{v}{}^{\plus}_{\chi N} ) $ & $-\calO_{\Phi,0,1}$ \\
[0.5ex]\cdashline{1-3}
$\calO_{4}$ & $\vec{S}_\chi \cdot \vec{S}_N$ & $\calO_{\Sigma,1,0}$ \\
[0.5ex]\cdashline{1-3}
$\calO_{5}$ & $ - i \vec{S}_\chi \cdot ( \vec{\q} \times \vec{v}{}^{\plus}_{\chi N} )$ & $-\calO_{\Delta,1,1}$ \\
[0.5ex]\cdashline{1-3}
$\calO_{6}$ & $(\vec{S}_\chi\cdot \vec{\q})  ( \vec{S}_N \cdot \vec{\q} )$ & $-\calO_{\Sigma,1,2}$ \\
[0.5ex]\cdashline{1-3}
$\calO_{7}$ & $ \vec{S}_N \cdot \vec{v}{}^{\plus}_{\chi N} $ & $\calO_{\Omega,0,0}$ \\
[0.5ex]\cdashline{1-3}
$\calO_{8}$ & $\vec{S}_\chi \cdot \vec{v}{}^{\plus}_{\chi N}$ & $\calO_{\Delta,1,0}$ \\
[0.5ex]\cdashline{1-3}
$\calO_{9}$ & $-i \vec{S}_\chi \cdot (\vec{S}_N \times \vec{\q} )$ & $\calO_{\Sigma,1,1}$ \\
[0.5ex]\cdashline{1-3}
$\calO_{10}$ & $-i \vec{S}_N \cdot \vec{\q}$ & $-\calO_{\Sigma,0,1}$ \\
[0.5ex]\cdashline{1-3}
$\calO_{11}$ & $-i \vec{S}_\chi \cdot \vec{\q}$ & $-\calO_{M,1,1}$ \\
[0.5ex]\cdashline{1-3}
$\calO_{12}$ & $\vec{S}_\chi \cdot (\vec{S}_N \times \vec{v}{}^{\plus}_{\chi N} )$ & $-\calO_{\Phi,1,0}$ \\
\bottomrule
\end{tabular}
\end{minipage}
\hspace{1em}
\begin{minipage}{0.4\textwidth}
\begin{tabular}{@{}llr@{}}
\toprule
%Haxton et al.'s &Definition & Ours \\
%\midrule
$\calO_{13}$ & $\calO_{10}\calO_{8}$ & $-\calO_{\Phi,1,1}$ \\
[0.5ex]\cdashline{1-3}
$\calO_{14}$ & $\calO_{11}\calO_{7}$ & $-\calO_{\Omega,1,1}$ \\
[0.5ex]\cdashline{1-3}
$\calO_{15}$ & $-\calO_{11}\calO_{3}$ & $-\calO_{\Phi,1,2}$ \\
[0.5ex]\cdashline{1-3}
$\calO_{16}$ & $-\calO_{10}\calO_{5}$ &$-\calO_{\Phi,1,2}- \tilde{q}^2 \calO_{\Phi,1,0}$\\
[0.5ex]\cdashline{1-3}
$\calO_{17}$ & $-i\vec{\tilde{q}}\cdot{\bf \cal S}\cdot\vec{v}{}^{\plus}_{\chi N}$ & $\calO_{\Delta,2,1}$\\
[0.5ex]\cdashline{1-3}
$\calO_{18}$ & $-i\vec{\tilde{q}}\cdot{\bf \cal S}\cdot\vec{S}_N$ & $\calO_{\Sigma,2,1}-\frac{1}{3}\calO_{\Sigma,0,1}$\\
[0.5ex]\cdashline{1-3}
$\calO_{19}$ & $\vec{\tilde{q}}\cdot{\bf \cal S}\cdot\vec{\tilde{q}}$ & $\calO_{M,2,2}+\frac{1}{3}\tilde{q}^2\calO_{M,0,0}$\\
[0.5ex]\cdashline{1-3}
$\calO_{20}$ & $\left (\vec{S}_N \times \vec{\tilde{q}}\right )\cdot{\bf \cal S}\cdot\vec{\tilde{q}}$ & -$\calO_{\Sigma,2,2}$\\
[0.5ex]\cdashline{1-3}
$\calO_{21}$ & $\vec{v}{}^{\plus}_{\chi N}\cdot{\bf \cal S}\cdot\vec{S}_N$ & $\frac{1}{3}{\cal O}_{\Omega,0,0}$\\
[0.5ex]\cdashline{1-3}
$\calO_{22}$ & $\left (- i\vec{\tilde{q}}\times\vec{v}{}^{\plus}_{\chi N}\right )\cdot{\cal S}\cdot \vec{S}_N$ & $- {\cal O}_{\Phi,2,1}-\frac{1}{3}{\cal O}_{\Phi,0,1}$\\
[0.5ex]\cdashline{1-3}
$\calO_{23}$ & $- i\vec{\tilde{q}}\cdot{\cal S}\cdot\left (\vec{S}_N\times\vec{v}{}^{\plus}_{\chi N} \right )$ & $- {\cal O}_{\Phi,2,1}+\frac{1}{3}{\cal O}_{\Phi,0,1}$\\
[0.5ex]\cdashline{1-3}
$\calO_{24}$ & $- \vec{v}{}^{\plus}_{\chi N}\cdot{\cal S}\cdot\left (\vec{S}_N\times i\vec{\tilde{q}}\right)$ & $- {\cal O}_{\Phi,2,1}-\frac{1}{3}{\cal O}_{\Phi,0,1}$\\
\bottomrule
\end{tabular}
\vspace{1.5\baselineskip}
\end{minipage}
\end{table}
} % end Haxton operator table

The $\mathcal{O}_{j}$ operators introduced
in~\cite{haxton2,krauss_spin_1,catena_krauss_spin_1} are listed in the
first and second columns of Table~\ref{tab:Haxton_operators} (the third column shows their expressions in terms of the operators $\calO_{X,s,l}$ that we introduce systematically in Section~\ref{sec:basis_operators}). The
symbol $1_{\chi \N}$ denotes the identity operator, $\vec{q}$ is the momentum transferred
from the WIMP to the nucleus,\footnote{We use the long-standing
  convention for $\vec{q}$ in the dark matter direct detection
  literature instead of the convention
  in~\cite{haxton2,krauss_spin_1}. In the latter, $\vec{q}$ is the momentum
  lost by the nucleus and thus has the opposite sign to ours. This
  explains the signs in the definition of the $\vec{q}$-dependent
  operators in Table~\ref{tab:Haxton_operators}.} a tilde over $q$ denotes $\tilde{q}=q/m_N$ (and 
$\vec{\tilde{q}}$=$\vec{q}/m_N$), where $m_N$ is the nucleon mass, $\vec{S}_{\chi}$
and $\vec{S}_{\N}$ are the WIMP and nucleon spins, respectively, and
$%\begin{equation}
\mathcal{S}_{ij} = \delta_{ij} - \frac{1}{2} ( S_{\chi i} S_{\chi j} + S_{\chi j} S_{\chi i} ) 
% \mathcal{S}^{s s'}_{ij}=\frac{1}{2}(\epsilon^{s}_i\epsilon^{s'}_j+\epsilon^s_j\epsilon^{s'}_i)
  \label{eq:s}
$ %\end{equation}
\noindent is a DM spin--1 operator (see Section~\ref{sec:spin1} for
its identification with the symbol $\mathcal{S}$ used
in~\cite{krauss_spin_1,catena_krauss_spin_1}). Moreover, 
\begin{align}
\vec{v}_{\chi N}= \vec{v}_{\chi} - \vec{v}_{N}
\end{align}
is the
WIMP--nucleon relative velocity, and
\begin{align}
\vec{v}{}^{\plus}_{\chi N} = \vec{v}_{\chi N} - \frac{\vec{q}}{2\mu_{\chi N}},
\end{align}
where $\mu_{\chi N}$ is the reduced WIMP--nucleon mass.
The operators listed in Table~\ref{tab:Haxton_operators} are invariant under Galilean transformations.

The operators $\calO_1$ and $\calO_4$ are the only two operators to zero-th order in $\vec{v}_{\chi N}$ and $\vec{q}$.
If terms up to first order in  $\vec{v}_{\chi N}$
are included, ${\bf\mathcal{H}}$ in~\cite{haxton1,haxton2} contains 4
terms for a WIMP of spin 0 ($\calO_{1,3,7,10}$) and 15 terms for a WIMP of spin 1/2 ($\calO_{1,3,\ldots,16}$).
Earlier work on effective WIMP-nucleon interactions beyond the usual
SI and SD considered only operators independent of $\vec{v}_{\chi N}$
\cite{fan_2010}. Later work to include WIMPs of spin 1 enlarged the
effective Hamiltonian to a total of 18 terms in~\cite{krauss_spin_1} ($\calO_{1,\ldots,18}$)
and eventually 24 terms
in~\cite{catena_krauss_spin_1} ($\calO_{1,\ldots,24}$). Beyond spin 1,
Ref.~\cite{barger_2008} shows a particular example for a WIMP of spin
3/2. Our systematic treatment shows that some of these operators are not independent. Specifically, a look at the third column in Table~\ref{tab:Haxton_operators} reveals that $\calO_{7}$ and $\calO_{21}$ are multiples of the same operator, $\calO_{22}$ and $\calO_{24}$ are the same operator, and $\calO_{23}$ is a linear combination of $\calO_{3}$ and $\calO_{22}$. Details are given in Section~\ref{sec:op_counting}.

The expected DD scattering rate is obtained by evaluating the effective Hamiltonian $\mathcal{H}$ between initial and final nuclear states. 
The expected differential rate for WIMP--nucleus elastic scattering off a nuclear target $T$, differential in the energy deposited $E_R$, is given by
\begin{equation}
\left(\frac{dR}{dE_R}\right)_T=M N_T \int_{v_{T,min}(E_R)}^{v_{esc}} \frac{\rho_{\chi}}{m_{\chi}} \, 
v \, \frac{d\sigma_T}{dE_R} \, f(\vec{v},t) d^3v,
\label{eq:dr_der}
  \end{equation}

\noindent where $M$ is
the mass of the detector,  $N_T$ is the number of target nuclei per unit detector mass, $\rho_\chi$ is the
mass density of dark matter in the neighborhood of the Sun, $m_\chi$ is 
the WIMP mass, $f(v)$ is the WIMP speed distribution in the reference
frame of the Earth, and $v_{T,min}(E_R)$ is the minimal speed an incoming WIMP needs to have in the
target reference frame to deposit energy $E_R$. For elastic WIMP scattering,
\begin{equation}
  v_{T,\rm min}(E_R)=\sqrt{\frac{m_T E_R}{2 \mu_{\chi T}^2}},
  \label{eq:vmin}
\end{equation}

\noindent where $m_T$ is equal to the nuclear
target mass and ${\mu_{\chi T}}$ is equal to the WIMP--nucleus reduced mass. As shown in~\cite{haxton2}, the differential cross section ${d\sigma_T}/{d E_R}$ in Eq.~(\ref{eq:dr_der}) can be put into the form
\be
\frac{d\sigma_T}{d E_R}=\frac{2 m_T}{4\pi v^2}\sum_{\tau=0,1}\sum_{\tau^{\prime}=0,1}\sum_{k} R_k^{\tau\tau^{\prime}} \widetilde{F}_{T k}^{\tau\tau^{\prime}},
\label{eq:dsigma_de}
\ee
where the sums contain products of WIMP and nuclear response functions
$R_k^{\tau\tau^{\prime}}(v,q)$ and $\widetilde{F}_{T
  k}^{\tau\tau^{\prime}}(q)$ (the latter are the nuclear response
functions $W_{T k}^{\tau\tau'}(q)$ in~\cite{haxton1,haxton2} apart
from a multiplying factor).  In the expression above, WIMP and nuclear
physics are factorized in the product of the nuclear response
functions $\widetilde{F}_{T k}^{\tau\tau^{\prime}}$, which depend on
$q^2$, and the WIMP response functions $R_k^{\tau\tau^{\prime}}$,
which depend on $c^{\tau}_j$, $q^2$, and $ (\vec{v}{}^{\plus}_{\chi T}
)^2 = \big[ \vec{v}_{\chi T} -\vec{q}/(2\mu_{\chi T})\big]^2$, where
$\vec{v}_{\chi T}$ is the WIMP-nucleus relative velocity and
$\mu_{\chi T}$ is the WIMP-nucleus reduced mass. The index $k$ runs
over combinations of nucleon currents. This factorization holds if
two--nucleon effects~\cite{chiral_eft, two_body_1,two_body_2} are
neglected.

To generalize the expressions above for a
WIMP of arbitrary spin, the crucial observation is that, thanks to the
factorization between the WIMP and the nucleon currents, the latter are
unchanged and completely fixed irrespective of the WIMP spin. This
has two consequences: (i) the effective operators $\calO_j t^\tau$ for a
WIMP of arbitrary spin can be obtained in a systematic way by
saturating the nucleon current with increasing powers of the vectors
$\vec{q}$, $\vec{v}{}^{\plus}_{\chi N}$ and $\vec{S_{\chi}}$; (ii) the shell model
determinations of the nuclear response functions
$\widetilde{F}^{\tau\tau^{\prime}}_{T k}(q)$ available in the
literature~\cite{haxton2,catena} can also be used for WIMPs of spin
higher that 1.

The only new ingredients required to upgrade the cross section of
Eq.~(\ref{eq:dsigma_de}) to a WIMP of arbitrary spin are WIMP
response functions $R_k^{\tau\tau^{\prime}}$ that include the WIMP-nucleon operators for WIMPs of any spin. We compute their explicit expressions and give them in
Eqs.~(\ref{eq:wimp_response_functions_sent}). To obtain such expressions we
find it convenient to define WIMP--nucleon interaction Hamiltonian
operators in terms of tensors irreducible under the rotation
group. The corresponding operator basis $\calO_{X,s,l}$ is given in
Eqs.~(\ref{eq:basis_WIMP_nucleon_operators}) or, alternatively, in
Eqs.~(\ref{eq:basis_WIMP_nucleon_operators_alt}), and differs from that
of the $\mathcal{O}_{j}$ operators of Eq.~(\ref{eq:H}). The third column in
Table~\ref{tab:Haxton_operators} gives the dictionary between the two operator
bases, from which an analogous dictionary among the
corresponding Wilson coefficients $c^\tau_j$ can be obtained in a
straightforward way.

This paper is organized as follows. In
Section~\ref{sec:non_relativistic_nuclear_currents} we review the
nuclear currents that arise in the non-relativistic limit of nucleon
Dirac bilinears. In Section~\ref{sec:basis_operators} we introduce a
basis $\calO_j t^\tau$ of WIMP--nucleon interaction operators for the
Hamiltonian of Eq.~(\ref{eq:H}) and a WIMP of arbitrary spin. In Section~\ref{sec:effective_wimp_nucleus_hamiltonian} we ``put the nucleons inside the nucleus'' and present the
ensuing effective WIMP--nucleus Hamiltonian. In Section~\ref{sec:amplitude} we derive
the squared WIMP--nucleus scattering amplitude, resulting in
Eqs.~(\ref{eq:wimp_response_functions_sent}),
which are the main result of this paper. We discuss our findings in
Section~\ref{sec:discussion} and conclude in
Section~\ref{sec:conclusions}.

\section{Non-relativistic nucleon currents}
\label{sec:non_relativistic_nuclear_currents}
There is a standard procedure to find all possible non-relativistic one-nucleon current operators in a nucleus. First one finds the free-nucleon operators that appear in the non-relativistic limit of the free nucleon currents (the Dirac bilinears). Then one sums the corresponding density operators over the $A$ nucleons in the nucleus.\footnote{Notice that a free-nucleon operator acts in the space of one nucleon, while a one-nucleon operator acts in the space of many nucleons, and it equals the sum over all nucleons of the volume density of the free-nucleon operators, each multiplied by the identity operator in the subspace of the other nucleons.}

In the non-relativistic limit, the nucleon Dirac bilinears
$\overline{\psi}_{\rm f} \Gamma \psi_{\rm i}$, where $\Gamma$ is any
combination of Dirac $\gamma$ matrices and $\psi$ is the Dirac spinor for a relativistic free nucleon, reduce to linear combinations
of five non-relativistic bilinears $\chi^\dagger_{\rm f} \op{O}_X t_N^\tau
\chi^{}_{\rm i}$, where $\chi$ is a non-relativistic Pauli spinor for the nucleon, $t^\tau_N$ is the isospin
operator ($\tau=0,1$ for the isoscalar and isovector parts, respectively), and
$\op{O}_X$ is one of the free-nucleon operators
\begin{align}
\op{O}_{M} = 1,
\quad
\op{\vec{O}}_{\Sigma} = \vec{\sigma}_N ,
\quad
\op{\vec{O}}_{\Delta} = \op{\vec{v}}{}^{\plus}_N ,
\quad
\op{\vec{O}}_{\Phi} = \op{\vec{v}}{}^{\plus}_N \times \vec{\sigma}_N ,
\quad
\op{O}_{\Omega} = \op{\vec{v}}{}^{\plus}_N \cdot \vec{\sigma}_N  .
\label{eq:five_operators}
\end{align}
Here $\vec{\sigma}_N$ is the vector of Pauli spin matrices acting on the spin states of the nucleon $N$,  and $\op{\vec{v}}{}^{\plus}_N$ is the operator
\begin{align}
  \op{\vec{v}}{}^{\plus}_N = - \frac{i}{m_N} \left( \overrightarrow{\frac{\partial}{\partial\vec{r}_N}} - \overleftarrow{\frac{\partial}{\partial\vec{r}_N}}  \right) \label{eq:vperp}
  \end{align}
  \noindent  (in the position representation), where $\vec{r}_N$ and $m_N$ are the position vector and the mass of the nucleon $N$.

The operator $\op{\vec{v}}{}^{\plus}_N$ is defined so that its matrix elements between free nucleon states are
\begin{align}
\chi^\dagger_{\rm f} \, \op{\vec{v}}{}^{\plus}_N  \chi^{}_{\rm i} = \vec{v}{}^{\plus}_{N} \equiv \frac{ \vec{v}_{N,\rm i} + \vec{v}_{N,\rm f} }{2} ,
\label{eq:v_N_perp}
\end{align}
where $\vec{v}_{N,\rm i}$ and $\vec{v}_{N,\rm f}$ are the initial and final velocities of the nucleon. By contrast, the nucleon velocity operator is
\begin{align}
\op{\vec{v}}_{N} = - \frac{i}{m_N} \frac{\partial}{\partial\vec{r}_{N}} .
\end{align}

For a nucleon in a nucleus, one introduces one-nucleon current densities where a volume-density version of the operator $\op{\vec{v}}{}^{\plus}_N$ appears. Each of the five free-nucleon operators $\op{\calO}_X t_N^\tau$ ($X=M,\Sigma,\Delta,\Phi,\Omega$) has a corresponding one-nucleon current
density defined by
%\begin{subequations}
\begin{align}
\op{j}^{\tau}_{M}(\vec{r}) & = \sum_{N=1}^{A} \delta(\vec{r} - \op{\vec{r}}_N) \, t^{\tau}_N, 
\nonumber\\
\op{\vec{j}}{}^{\tau}_{\Sigma}(\vec{r}) & = \sum_{N=1}^{A}  \delta(\vec{r} - \op{\vec{r}}_N) \, \vec{\sigma}_N  t^{\tau}_N, 
\nonumber\\
\op{\vec{j}}{}^{\tau}_{\Delta, \rm sym}(\vec{r}) & = \sum_{N=1}^{A} \big\lsymm \delta(\vec{r}-\op{\vec{r}}_N) \, \op{\vec{v}}_N \big\rsymm t^{\tau}_N ,
\nonumber\\
\op{\vec{j}}{}^{\tau}_{\Phi, \rm sym}(\vec{r}) & = \sum_{N=1}^{A} \big\lsymm \delta(\vec{r}-\op{\vec{r}}_N) \, \op{\vec{v}}_N \big\rsymm \times \vec{\sigma}_N t^{\tau}_N,
\nonumber\\
\op{j}^{\tau}_{\Omega, \rm sym}(\vec{r}) & = \sum_{N=1}^{A} \big\lsymm \delta(\vec{r}-\op{\vec{r}}_N) \, \op{\vec{v}}_N \big\rsymm \cdot \vec{\sigma}_N  t^{\tau}_N .
\end{align}
%\end{subequations}
Here the index $N$ refers to the nucleon on which the operator acts
(we abuse the notation $N$ by using it to refer to a particular free nucleon
and also as a summation index over nucleons bound in a nucleus). Moreover, the symbol $\big\lsymm \cdots \big\rsymm$ stands
for the symmetrization operation
\begin{align}
 \big\lsymm \delta(\vec{r}-\op{\vec{r}}_N) \, \op{\vec{v}}_N \big\rsymm = \frac{1}{2} [  \delta(\vec{r}-\op{\vec{r}}_N) \, \op{\vec{v}}_N + \op{\vec{v}}_N  \delta(\vec{r}-\op{\vec{r}}_N) ]  .
\end{align}
This symmetrized operator is Hermitian and is the volume density version of the operator $\op{\vec{v}}{}^{\plus}_N$, in the sense that its free-nucleon matrix elements between nucleon wave functions $\psi_1(\vec{r}_N)$ and $\psi_2(\vec{r}_N)$ obey the relation
\begin{align}
\int \psi_1^*(\vec{r}_N) \big\lsymm \delta(\vec{r}-\op{\vec{r}}_N) \, \op{\vec{v}}_N \big\rsymm \psi_2^*(\vec{r}_N) \, d^3r_N = \int  \delta(\vec{r}-\vec{r}_N) \, \big[ \psi_1^*(\vec{r}_N)  \op{\vec{v}}{}^{\plus}_N \psi_2^*(\vec{r}_N) \big]  \, d^3 r_N .
\end{align}
This justifies the replacement of $\vec{v}{}^{\plus}_N$ with $\lsymm \delta(\vec{r}-\vec{r}_N) \vec{v}_N \rsymm$ in passing from a free-nucleon operator to a one-nucleon current in a nucleus. 

Hence the following correspondence applies between free-nucleon operators and one-nucleon currents,
%\begin{subequations}
\begin{eqnarray}
\op{O}_{M} t_N^\tau &\to& \op{j}^\tau_{M}(\vec{r}) ,
 \nonumber\\
\op{\vec{O}}_{\Sigma} t_N^\tau &\to&  \op{\vec{j}}{}^\tau_{\Sigma}(\vec{r}) ,
 \nonumber\\
 \op{\vec{O}}_{\Delta} t_N^\tau &\to&  \op{\vec{j}}{}^\tau_{\Delta,\rm sym}(\vec{r}) ,
 \nonumber\\
 \op{\vec{O}}_{\Phi} t_N^\tau &\to& \op{\vec{j}}{}^\tau_{\Phi,\rm sym}(\vec{r}) ,
 \nonumber\\
\op{O}_{\Omega} t_N^\tau &\to& \op{j}{}^\tau_{\Omega,\rm sym}(\vec{r}).
\end{eqnarray}
%\end{subequations}

%In the Born approximation, the effective WIMP--nucleus Hamiltonian contains the

In problems involving the transfer of a momentum $\vec{q}$ to the nucleus (such as the problem we are interested in, namely the scattering of WIMPs off nuclei), another variant of the one-nucleon currents appears. These are currents defined in the Breit frame of the nucleus, namely the reference frame in which  the nucleus momentum changes sign when the momentum $\vec{q}$ is transferred.\footnote{For elastic scattering, the energy transferred to the nucleus in the Breit frame is zero. The use of the Breit frame in the definition of form factors for particles of any spin has been discussed in~\cite{Yennie:1957}. The Breit frame is particularly relevant for nucleon form factors  (see, e.g.,~\cite{Friar:1973,Kelly:2002if,Perdrisat:2006hj}).} The velocity of the Breit frame is\footnote{In the notation of~\cite{haxton1,haxton2}, $\vec{v}{}^{\perp}_{T}$ is used in place of our $\vec{v}{}^{\plus}_{\chi T}$ in Eq.~\eqref{eq:v_chi_t}, and there is no $\vec{v}{}^{\plus}_{T}$. Moreover, $\vec{v}{}^{\perp}$ is used in place of our $\vec{v}{}^{\plus}_{\chi N}$ in Eq.~(\ref{eq:v_perp_chi_N}). To err in the direction of clarity, we have chosen to maintain the particle labels as subscripts and to use the different symbol $\!\plus$ in place of $\perp$ to distinguish $\vec{v}{}^{\perp}_{T}$ in~\cite{haxton1,haxton2} from our $\vec{v}{}^{\plus}_{T}$ in Eq.~\eqref{eq:v_T_perp}.\label{footnote:haxton}}
\begin{align}
\vec{v}{}^{\plus}_{T} = \frac{ \vec{v}_{T,\rm i} + \vec{v}_{T,\rm f} }{2} ,
\label{eq:v_T_perp}
\end{align}
where $\vec{v}_{T,\rm i}$ and $\vec{v}_{T,\rm f}$ are the initial and final velocities of the nucleus. Since the velocity of the nucleus $\vec{v}_T$ equals the velocity of the center of mass of the system of nucleons,
\begin{align}
\vec{v}_T = \frac{1}{A} \sum_{N} \vec{v}_N ,
\end{align}
Eqs.~\eqref{eq:v_N_perp} and~\eqref{eq:v_T_perp} imply
\begin{align}
\vec{v}{}^{\plus}_T = \frac{1}{A} \sum_{N} \vec{v}{}^{\plus}_N .
\end{align}
%Moreover, in terms of $\vec{q}$,
%\begin{align}
%\vec{v}{}^{\plus}_{T} =  \vec{v}_{T,\rm i}  + \frac{ \vec{q} }{2 m_T} .
%\end{align}
Let
\begin{align}
\vec{v}_{NT} = \vec{v}_{N} - \vec{v}{}^{\plus}_{T} 
\end{align}
be the nucleon velocity in the nucleus Breit frame corresponding to momentum transfer $\vec{q}$.
The Breit-frame currents are defined as the symmetrized currents with $\vec{v}_{N}$ replaced by $\vec{v}_{NT}$,
%\begin{subequations}
\begin{align}
%\op{j}^{\tau}_{M}(\vec{r}) & = \sum_{N=1}^{A} \delta(\vec{r} - \op{\vec{r}}_N) \, t^{\tau}_N, 
%\\
%\op{\vec{j}}{}^{\tau}_{\Sigma}(\vec{r}) & = \sum_{N=1}^{A}  \delta(\vec{r} - \op{\vec{r}}_N) \, \vec{\sigma}_N  t^{\tau}_N, 
%\\
\op{\vec{j}}{}^{\tau}_{\widetilde\Delta}(\vec{r}) & = \sum_{N=1}^{A} \lsymm \delta(\vec{r} - \op{\vec{r}}_N) \, \op{\vec{v}}_{NT} \rsymm t^{\tau}_N, 
 \nonumber\\
\op{\vec{j}}{}^{\tau}_{\widetilde\Phi}(\vec{r}) & = \sum_{N=1}^{A} \lsymm \delta(\vec{r} - \op{\vec{r}}_N) \, \op{\vec{v}}_{NT} \times \vec{\sigma}_N \rsymm t^{\tau}_N ,
\nonumber \\
\op{j}^{\tau}_{\widetilde\Omega}(\vec{r}) & = \sum_{N=1}^{A} \lsymm \delta(\vec{r} - \op{\vec{r}}_N) \, \op{\vec{v}}_{NT} \cdot \vec{\sigma}_N  \rsymm t^{\tau}_N.
\label{eq:Breitframe_nuclear_currents}
\end{align}
%\end{subequations}
%(We have added the argument $\vec{q}$ to $\op{j}^{\tau}_{M}$ and $\op{\vec{j}}{}^{\tau}_{\Sigma}$ for uniformity of notation, although they don't actually depend on $\vec{q}$.) 
The Breit-frame currents are related to the symmetrized currents via
%\begin{subequations}
\begin{align}
 \op{\vec{j}}{}^{\tau}_{\Delta,\rm sym}(\vec{r}) & = \op{\vec{j}}{}^{\tau}_{\widetilde\Delta}(\vec{r}) + \vec{v}^{\plus}_{T} \, \op{j}^{\tau}_{M}(\vec{r}) ,
\nonumber\\
 \op{\vec{j}}{}^{\tau}_{\Phi,\rm sym}(\vec{r}) & = \op{\vec{j}}{}^{\tau}_{\widetilde\Phi}(\vec{r}) + \vec{v}^{\plus}_{T} \times \op{j}^{\tau}_{\Sigma}(\vec{r}) ,
\nonumber\\
\op{\vec{j}}{}^{\tau}_{\Omega,\rm sym}(\vec{r}) & = \op{\vec{j}}{}^{\tau}_{\widetilde\Omega}(\vec{r}) + \vec{v}^{\plus}_{T} \cdot \op{j}^{\tau}_{\Sigma}(\vec{r}) .
\label{eq:j_sym_tilde}
\end{align}
%\end{subequations}

One also defines the non-symmetrized currents in the Breit frame
%\begin{subequations}
\begin{align}
\op{\vec{j}}{}^{\tau}_{\Delta}(\vec{r}) & = \sum_{N=1}^{A} \delta(\vec{r} - \op{\vec{r}}_N) \, \op{\vec{v}}_{NT} t^{\tau}_N, 
\nonumber \\
\op{\vec{j}}{}^{\tau}_{\Phi}(\vec{r}) & = \sum_{N=1}^{A} \delta(\vec{r} - \op{\vec{r}}_N) \, \op{\vec{v}}_{NT} \times \vec{\sigma}_N t^{\tau}_N ,
 \nonumber\\
\op{j}^{\tau}_{\Omega}(\vec{r}) & = \sum_{N=1}^{A} \delta(\vec{r} - \op{\vec{r}}_N) \, \op{\vec{v}}_{NT} \cdot \vec{\sigma}_N  t^{\tau}_N.
\label{eq:nonsymmetrized_nuclear_currents}
\end{align}
%\end{subequations}

When we later consider the scattering of WIMPs in the Born approximation, the plane wave WIMP wave functions contribute a factor $e^{i\vec{q} \cdot \vec{r}}$ to the amplitude, and the Fourier transform of the one-nucleon Breit-frame currents appears,
%\begin{subequations}
\begin{align}
j^\tau_{\curr{X}}(\vec{q})  & =  \int d^3 r \, e^{i\vec{q}\cdot\vec{r}} \, j^\tau_{\curr{X}}(\vec{r}) , \qquad \text{for $X=M,\Omega,\widetilde\Omega$,} 
 \nonumber\\
\vec{j}^\tau_{\curr{X}}(\vec{q})  & =  \int d^3 r \, e^{i\vec{q}\cdot\vec{r}} \, \vec{j}^\tau_{\curr{X}}(\vec{r}) , \qquad \text{for $X=\Sigma,\Delta,\Phi,\widetilde\Delta,\widetilde\Phi$.\label{eq:def_j_q}}
\end{align}
%\end{subequations}
Substituting Eqs.~(\ref{eq:Breitframe_nuclear_currents}) into Eqs.~(\ref{eq:def_j_q}), and
using the relation
\begin{align}
& \int \psi_1^*(\vec{r}_N) \big\lsymm e^{i\vec{q}\cdot\vec{r}_N} \, \op{\vec{v}}_N \big\rsymm \psi_2^*(\vec{r}_N) \, d^3r_N 
%\nonumber \\ &
%= - \frac{i}{2m_N} \int \psi_1^*(\vec{r}_N) \Big[ e^{i\vec{q}\cdot\vec{r}_N}\, \frac{\partial}{\partial r_N} +   \frac{\partial}{\partial r_N} \, e^{i\vec{q}\cdot\vec{r}_N} \Big]  \psi_2^*(\vec{r}_N) \, d^3r_N
%\nonumber \\ &
%= - \frac{i}{2m_N} \int \psi_1^*(\vec{r}_N) \Big[ 2 e^{i\vec{q}\cdot\vec{r}_N} \, \frac{\partial}{\partial r_N}+i \vec{q} e^{i\vec{q}\cdot\vec{r}_N} \Big]  \psi_2^*(\vec{r}_N) \, d^3r_N
\nonumber \\ &
=  \int e^{i\vec{q}\cdot\vec{r}_N} \, \Big[ \psi_1^*(\vec{r}_N) \Big( \op{\vec{v}}_{N} + \frac{\vec{q}}{2m_N} \Big) \psi_2^*(\vec{r}_N) \Big]  \, d^3r_N ,
\label{eq:def_vN_perp}
\end{align}
%\begin{align}
%\op{\vec{v}}{}^{\plus}_{N} =
% \op{\vec{v}}_N + \frac{ \vec{q}}{2m_N} ,
%\end{align}
one finds the following identities between the Fourier-transformed symmetrized and non-symmetrized one--nucleon currents in the Breit frame,
%\begin{subequations}
\begin{align}
j^\tau_{\widetilde\Delta}(\vec{q}) & = j^\tau_{\Delta}(\vec{q}) +
\frac{\vec{q}}{2m_N} \, j^\tau_{M}(\vec{q}),
\nonumber \\ j^\tau_{\widetilde\Phi}(\vec{q}) & = j^\tau_{\Phi}(\vec{q}) +
\frac{\vec{q}}{2m_N} \times \vec{j}^\tau_{\Sigma}(\vec{q}),
\nonumber \\ j^\tau_{\widetilde\Omega}(\vec{q}) & = j^\tau_{\Omega}(\vec{q}) +
\frac{\vec{q}}{2m_N}\cdot
\vec{j}^\tau_{\Sigma}(\vec{q}). 
\label{eq:symmetrized_nuclear_currents}
\end{align}
%\end{subequations}

\section{WIMP--nucleon operators}
\label{sec:basis_operators}

%\subsection{Basis WIMP--nucleon operators}

In this section we describe the effective interaction Hamiltonian of a
WIMP with a free nucleon.  The five free-nucleon operators
$\op{O}_{X}$ ($X=M,\Omega,\Sigma,\Delta,\Phi$) in
Eq.~(\ref{eq:five_operators}) depend on the nucleon velocity, which is
not invariant under Galilean boosts. Indeed, to comply with Galilean
invariance one must introduce five corresponding WIMP-nucleon
operators $\op{\calO}_{X}$ ($X=M,\Omega,\Sigma,\Delta,\Phi$) that
depend on the relative WIMP-nucleon velocity instead (in the following
we drop the hat on top of operators, unless it is needed for clarity)
\begin{align}
\vec{v}_{\chi N} = \vec{v}_{\chi}  - \vec{v}_{N} .
\label{eq:v_diff}
\end{align}
However from the non-relativistic limit of the nucleon Dirac
bilinears one knows that $\vec{v}_{N}$ appears in the combination
$\vec{v}{}^{\plus}_{N}$ of Eq.~(\ref{eq:def_vN_perp}).  If the WIMP
has spin--1/2 the same argument implies that the analogous combination
\begin{align}
\vec{v}{}^{\plus}_{\chi} = \vec{v}_{\chi} - \frac{ \vec{q}}{2m_\chi} 
\label{eq:v_plus_chi}
\end{align}
\noindent appears also from the non-relativistic limit of the WIMP Dirac
bilinear. Then combining Eqs.~(\ref{eq:v_diff}) and
(\ref{eq:v_plus_chi}) one concludes that the WIMP--nucleon operators
consistent to Eq.~(\ref{eq:five_operators}) must be:
\begin{align}
\op{\calO}_{M} = 1 ,
\quad 
\op{\vec{\calO}}_{\Sigma} = \vec{\sigma}_N ,
\quad
\op{\vec{\calO}}_{\Delta} = \vec{v}{}^{\plus}_{\chi N} ,
\quad
\op{\vec{\calO}}_{\Phi} = \vec{v}{}^{\plus}_{\chi N} \times \vec{\sigma}_N ,
\quad
\op{\calO}_{\Omega} =  \vec{v}{}^{\plus}_{\chi N}  \cdot \vec{\sigma}_N .
\label{eq:calO_X}
\end{align}

\noindent where:

\begin{align}
  \vec{v}{}^{\plus}_{\chi N} = \vec{v}{}^{\plus}_{\chi} - \vec{v}{}^{\plus}_{N} % =   \vec{v}_{\chi N} - \frac{\vec{q}}{2\mu_{\chi N}} .
  \label{eq:v_perp_chi_N}.
\end{align}

We now show that this conclusion holds also for a WIMP of
arbitrary spin. In order to do so one writes the non-relativistic Hamiltonian $\op{H}_{\chi N}$ for an interacting system made of a WIMP $\chi$ and a nucleon $N$,
\begin{align}
H_{\chi N} = \frac{\vec{p}_{\chi}^2}{2m_\chi} + \frac{\vec{p}_N^2}{2m_N} + V_{\chi N} .
\end{align}

The most general interaction Hamiltonian $V_{\chi N}$ depends on
the WIMP spin operator $\vec{S}_\chi$, the nucleon spin operator
$\vec{S}_N$, and, imposing Galilean invariance, on the relative
WIMP-nucleon position operator $\vec{r}_{\chi N}$ and its conjugate
relative momentum operator $\vec{p}_{\chi N}$.  Moreover, 
Eqs.~(\ref{eq:five_operators}) imply that the interaction
Hamiltonian is either independent of
$\vec{p}_{\chi N}$ or linear in $\vec{p}_{\chi N}$.  In the
latter case, since $V_{\chi N}$ must be Hermitian and it depends on
the non-commuting operators $\vec{r}_{\chi N}$ and $\vec{p}_{\chi
  N}$, a prescription needs to be set up on the order in which these
two operators appear. Any combination of the form $f_1(\vec{r}_{\chi
  N}) \, \vec{p}_{\chi N} \, f_2(\vec{r}_{\chi N})$, where
$f_1(\vec{r}_{\chi N})$ and $f_2(\vec{r}_{\chi N})$ are arbitrary
functions, can be rearranged with the $\vec{r}_{\chi N}$ dependence on
the left of the operator $\vec{p}_{\chi N}$ by commuting
$\vec{p}_{\chi N}$ and $f_2(\vec{r}_{\chi N})$ and regarding
their commutator as an extra term in the Hamiltonian. Thus there is no
loss of generality in assuming that the dependence on $\vec{r}_{\chi
  N}$ is on the left of $\vec{p}_{\chi N}$, as in
$f(\vec{r}_{\chi N}) \, \vec{p}_{\chi N}$. Then an Hermitian term
in the Hamiltonian is obtained by constructing the symmetric
combination
\begin{align}
\Big\lsymm f(\vec{r}_{\chi N}) \, \vec{p}_{\chi N} \Big]_{\rm sym} = \frac{1}{2} \Big( f(\vec{r}_{\chi N}) \, \vec{p}_{\chi N} + \vec{p}_{\chi N} \, f(\vec{r}_{\chi N}) \Big) .
\end{align}

Since the nucleon has spin 1/2, the interaction Hamiltonian
$V_{\chi N}$ can be split into terms independent of the nucleon
spin operator $\vec{S}_N$ and terms linear in $\vec{S}_N$ (notice that
the non-relativistic limit of the nucleon Dirac bilinears in
Section~\ref{sec:non_relativistic_nuclear_currents} shows that
symmetric tensor terms of the form $p_{\chi N,i} S_{N,j} +
p_{\chi N,j} S_{N,i} $ do not appear). So the interaction
Hamiltonian $V_{\chi N} $ must have the form
\begin{align}
V_{\chi N} & = V_{\chi N}^\tau \, t_N^\tau 
\end{align}
with
\begin{align}
V_{\chi N}^\tau & = 
V_M^\tau\big(\vec{r}_{\chi N}, \vec{S}_\chi\big) 
 + \vec{S}_N \cdot \vec{V}_\Sigma^\tau\big(\vec{r}_{\chi N}, \vec{S}_\chi\big)  
+ \Big[ \vec{V}_\Delta^\tau\big(\vec{r}_{\chi N}, \vec{S}_\chi\big) \cdot \vec{v}_{\chi N} \Big]_{\rm sym}
\nonumber \\
&
+ \vec{S}_N \cdot \Big[ \vec{V}_\Phi^\tau\big(\vec{r}_{\chi N}, \vec{S}_\chi\big) \times \vec{v}_{\chi N} \Big]_{\rm sym}
+ \vec{S}_N \cdot \Big[ V_\Omega^\tau\big(\vec{r}_{\chi N}, \vec{S}_\chi\big) \, \vec{v}_{\chi N} \Big]_{\rm sym}
.
\label{eq:V_chi_N}
\end{align}

\noindent Here we have introduced the relative WIMP-nucleon velocity operator $\vec{v}_{\chi N}$ defined by 
\begin{align}
\vec{v}_{\chi N}  = \frac{1}{\mu_{\chi N}} \, \vec{p}_{\chi N} .
\end{align}

\noindent The interaction amplitude for the WIMP--nucleon scattering process (in the Born approximation) is
then given by
\begin{align}
\langle f | V_{\chi N}| i \rangle 
& = \int d^3r_\chi d^3r_N e^{-i\vec{p}_{\chi,\rm f} \cdot \vec{r}_\chi - i \vec{p}_{N,\rm f} \cdot \vec{r}_N} \, V_{\chi N}  \, e^{i\vec{p}_{\chi,\rm i} \cdot \vec{r}_\chi + i \vec{p}_{N,\rm i} \cdot \vec{r}_N} 
\nonumber \\
& = \int d^3R d^3r_{\chi N} e^{-i\vec{p}_{\rm tot,f} \cdot \vec{R} - i \vec{p}_{\chi N,\rm f} \cdot \vec{r}_{\chi N}} \, V_{\chi N}  \, e^{i\vec{p}_{\rm tot,i} \cdot \vec{R} + i \vec{p}_{\chi N,\rm i} \cdot \vec{r}_{\chi N}} 
\nonumber \\
& = (2\pi)^3 \delta( \vec{p}_{\rm tot,f} - \vec{p}_{\rm tot,i} ) \, \int d^3r_{\chi N} e^{- i \vec{p}_{\chi N,\rm f} \cdot \vec{r}_{\chi N}} \, V_{\chi N}  \, e^{i \vec{p}_{\chi N,\rm i} \cdot \vec{r}_{\chi N}} 
\label{eq:f_V_chi_N_i}
\end{align}

\noindent where $\vec{p}_{\chi,\rm i}$, $\vec{p}_{\chi,\rm f}$,
$\vec{p}_{N,\rm i}$, $\vec{p}_{N,\rm f}$ are the initial and final momenta of the WIMP and the nucleon, and in the integral we have explicitly separated
the motion of the center of mass with coordinates
$(\vec{R},\vec{p}_{\rm tot})$.

The integral appearing in Eq.~\eqref{eq:f_V_chi_N_i} is a function of $\vec{q} = \vec{p}_{\chi N,\rm i} - \vec{p}_{\chi N,\rm f}$ and $\vec{v}{}^{\plus}_{\chi N} = (\vec{p}_{\chi N,\rm i} + \vec{p}_{\chi N,\rm f})/(2\mu_{\chi N})$. The dependence on $\vec{v}{}^{\plus}_{\chi N} $ gives the operators in Eq.~\eqref{eq:calO_X} multiplied by functions of $\vec{q}$, namely the Fourier transforms 
\begin{align}
V_X^\tau\big(\vec{q}, \vec{S}_\chi\big)  = \int d^3r_{\chi N} \, e^{i \vec{q} \cdot \vec{r}_{\chi N}} V_X^\tau\big(\vec{r}_{\chi N}, \vec{S}_\chi\big) 
\label{eq:V_X_q_S}
\end{align}
of the potentials in Eq.~\eqref{eq:V_chi_N}.

As a way of example, the explicit contribution to the amplitude from
$\vec{V}_\Delta$ is
\begin{align}
 \int & d^3r_{\chi N} e^{-
  i \vec{p}_{\chi N,\rm f} \cdot \vec{r}_{\chi N}} \, \Big[
  \vec{V}_\Delta^\tau\big(\vec{r}_{\chi N}, \vec{S}_\chi\big) \cdot
  \vec{v}_{\chi N} \Big]_{\rm sym}\, e^{i \vec{p}_{\chi N,\rm i} \cdot \vec{r}_{\chi N}}\nonumber \\
&= \frac{1}{2 \mu_{\chi N}}\int d^3r_{\chi N} e^{-
  i \vec{p}_{\chi N,\rm f} \cdot \vec{r}_{\chi N}} \, \Big[
  2 \vec{V}_\Delta^\tau\big(\vec{r}_{\chi N}, \vec{S}_\chi\big) \cdot
  \vec{p}_{\chi N} + [ \vec{p}_{\chi N},\vec{V}_\Delta(\vec{r}_{\chi N}, \vec{S}_\chi\big) ]\Big]\, e^{i \vec{p}_{\chi N,\rm i} \cdot \vec{r}_{\chi N}} \nonumber  \\
&= \int d^3r_{\chi N} e^{
  i \vec{q} \cdot \vec{r}_{\chi N}} \, \Big[
  \vec{V}_\Delta^\tau\big(\vec{r}_{\chi N}, \vec{S}_\chi\big) \cdot
  \vec{v}_{\chi N,\rm i} -\frac{i}{2 \mu_{\chi N}} \vec{\nabla}\cdot\vec{V}_\Delta(\vec{r}_{\chi N}, \vec{S}_\chi\big) \Big] \nonumber \\
&=\int d^3r_{\chi N} e^{
  i \vec{q} \cdot \vec{r}_{\chi N}} \, 
  \vec{V}_\Delta^\tau\big(\vec{r}_{\chi N}, \vec{S}_\chi\big) \cdot
  \Big[ \vec{v}_{\chi N,\rm i} -\frac{\vec{q}}{2 \mu_{\chi N}} \Big]\nonumber\\
  &=\int d^3r_{\chi N} e^{
  i \vec{q} \cdot \vec{r}_{\chi N}} \, 
  \vec{V}_\Delta^\tau\big(\vec{r}_{\chi N}, \vec{S}_\chi\big) \cdot
  \vec{v}{}^{\plus}_{\chi N} 
\nonumber \\
  &=
  \vec{\widetilde{V}}{}^\tau_\Delta\big(\vec{q}, \vec{S}_\chi\big) \cdot
  \vec{v}{}^{\plus}_{\chi N} .
\end{align}
Analogous steps show that also the
contributions from $\vec{V}_\Phi$ and $\vec{V}_\Omega$ are
proportional to the $\vec{v}{}^{\plus}_{\chi N}$ operator. This shows
that the effective operators of Eq.~(\ref{eq:calO_X}) written in terms
of $\vec{v}{}^{\plus}_{\chi N}$ must drive the WIMP--nucleon interaction
also for WIMPs of spin higher that 1/2. Notice that for elastic
WIMP-nucleon scattering,
\begin{align}
\vec{q} \cdot \vec{v}{}^{\plus}_{\chi N} = 0.
\end{align}

The WIMP-nucleon operators $\op{\calO}_{X}$ are related to the free-nucleon operators $\op{O}_{X}$ by means of the relations, obtained by using $\vec{v}{}^{\plus}_{\chi N} =  \vec{v}{}^{\plus}_{\chi} - \vec{v}{}^{\plus}_{N} $,
%\begin{subequations}
\begin{align}
& \op{\calO}_{M} =  \op{O}_{M} ,
\nonumber \\
& \op{\vec{\calO}}_{\Sigma} = \op{\vec{O}}_{\Sigma} ,
\nonumber\\
& \op{\vec{\calO}}_{\Delta} = \vec{v}{}^{\plus}_{\chi} \op{O}_{M} - \op{\vec{O}}_{\Delta} ,
\nonumber\\
& \op{\vec{\calO}}_{\Phi} = \vec{v}{}^{\plus}_{\chi} \times \op{\vec{O}}_{\Sigma} - \op{\vec{O}}_{\Phi} ,
\nonumber\\
& \op{\calO}_{\Omega} = \vec{v}{}^{\plus}_{\chi} \cdot \op{\vec{O}}_{\Sigma} - \op{O}_{\Omega}.
\label{eq:calO_O}
\end{align}
%\end{subequations}

The operators $\op{\calO}_{X}$
($X=M,\Omega,\Sigma,\Delta,\Phi$) are either invariant under rotations ($\op{\calO}_{M}$ and
$\op{\calO}_{\Omega}$) or transform as vectors ($\op{\vec{\calO}}_{\Sigma}$,
$\op{\vec{\calO}}_{\Delta}$, $\op{\vec{\calO}}_{\Phi}$). Therefore rotational
invariance of the interaction Hamiltonian term imposes that the scalar
operators $\op{\calO}_{M}$ and $\op{\calO}_{\Omega}$ multiply a scalar WIMP
operator $\op{o}$, and the vector operators $\op{\vec{\calO}}_{\Sigma}$,
$\op{\vec{\calO}}_{\Delta}$, $\op{\vec{\calO}}_{\Phi}$ multiply a vector WIMP
operator $\op{\vec{o}}$ as in $\op{\vec{o}} \cdot \op{\vec{\calO}}_{X}$.

On the other hand, the effective interaction term for a WIMP of spin
$j_\chi$ must contain up to the product of $2 j_\chi$ WIMP spin
vectors, in order to mediate transitions where the third component of
the WIMP spin changes from $\pm j_\chi$ to $\mp j_\chi$. Using index
notation $S_i$ for the $i$-th component of the vector $\vec{S}_\chi$ (we drop the subscript $\chi$ in $S_{\chi,i}$ for more readability), there
are interaction terms containing no $S_i$ or a product of $s$ factors
$S_i$ up to $s=2j_\chi$,
\begin{align}
1, \quad S_{i_1}, \quad S_{i_1} S_{i_2}, \quad S_{i_1}S_{i_2}S_{i_3}, \quad\dots, \quad S_{i_1} S_{i_2} \cdots S_{i_{2j_\chi}}.
\label{eq:products_of_S}
\end{align}
In other words, there are $2j_\chi+1$ possible products of the WIMP
spin operator for a WIMP of spin $j_\chi$. Each product can be labeled
by the number $s$ of WIMP spin factors $S_i$. An alternative way to
reach the same conclusion is to show that the $2j_\chi+1$ products in
Eq.~(\ref{eq:products_of_S}) are a basis in the space of spin
operators for spin $j_\chi$. Once the number of WIMP spin factors is
fixed to $s$, and the scalar or vector nature of the free-nucleon
operator $\op{O}_{X}$ is considered, the number of $\q$ factors is
constrained by rotational invariance. In particular, in the case of a
scalar nucleon operator $\op{O}_{X}$ ($X=M, {\Omega}$), the WIMP
operator $\op{o}$ must be a scalar, and all the indices $i_1\cdots
i_s$ in $S_{i_1} \cdots S_{i_s}$ must be saturated by terms
$\q_{i_1}\cdots \q_{i_s}$. The resulting WIMP operator is $S_{i_1}
\cdots S_{i_s}\q_{i_1}\cdots \q_{i_s}$. On the other hand, in the case
of a vector nucleon operator $\vec{\op{O}}_{X}$
($X=\Sigma,\Delta,\Phi$), a vector WIMP operator $\op{\vec{o}}$ is
needed, and the $s$ indices in $S_{i_1} \cdots S_{i_s}$ must be
saturated by an appropriate number of $\q$ factors in order to obtain
a vector. This can be achieved in three ways: (1) by using $s-1$
factors of $\q$ to produce $S_{i_1} \cdots S_{i_s}\q_{i_1}\cdots
\q_{i_{s-1}}$ with free index $i_s$, (2) by using $s$ factors of $\q$
to produce $\epsilon_{i_s l m} S_{i_1} \cdots S_{i_{s-1}} S_l
\q_{i_1}\cdots \q_{i_{s-1}}\q_m$, again with free index $i_s$, and (3)
by using $s+1$ factors of $\q$ to produce $S_{i_1} \cdots
S_{i_s}\q_{i_1}\cdots \q_{i_s}\q_{i_{s+1}}$, with free index
$i_{s+1}$.

A further consideration informs our choice of basis interaction
terms. In the calculation of the cross section for WIMP--nucleus
scattering, traces of the $S_{i_1} \cdots S_{i_s}$ operators are
needed. The latter are greatly simplified if for the products of WIMP
spin operators one uses irreducible tensors (i.e., belonging to
irreducible representations of the rotation group). Irreducible
tensors are completely symmetric under exchange of any two of their
indices and have zero trace under contraction of any number of pairs
of indices (they are symmetric traceless tensors). In addition, an
irreducible tensor of rank $s$ has $2s+1$ independent components, and
belongs to the irreducible representation of the rotation group of
spin $s$. Irreducible tensor operators of different rank are
independent, in the sense that the trace of their product is zero. As
a consequence, there are no interference terms in the cross section
between irreducible operators of different spin. Therefore we use the following $2j_\chi+1$ irreducible spin tensors
as a basis in the spin space of a WIMP of spin $j_\chi$,
\begin{align}
  1, \quad \myoverbracket{S_{i_1}}, \quad \myoverbracket{S_{i_1} S_{i_2}}, \quad \myoverbracket{S_{i_1}S_{i_2}S_{i_3}}, \quad\dots, \quad \myoverbracket{S_{i_1} S_{i_2} \cdots S_{i_{2j_\chi}}}.
  \label{eq:irreducible_products_of_S}
\end{align}
Here, borrowing the notation of~\cite{hess}, we use an overbracket over an expression containing a set of indices to indicate that the free indices under the bracket are completely symmetrized and all of their contractions are subtracted. For example,
\begin{align}
\myoverbracket{A_{ij}} = \frac{1}{2} \left( A_{ij} + A_{ji} \right) - \frac{1}{3} \, \delta_{ij} \, A^k_{~k} .
\end{align}
Notice that $\myoverbracket{1}=1$ and $\myoverbracket{A_i}=A_i$. More details are given in Appendices~\ref{sec:symmetric_symmetrictraceless} and \ref{sec:WIMP_spin_averages}.

%When the potentials $V_X(\vec{r}_{\chi N}, \vec{S}_{\chi})$ in Eq.~\eqref{eq:V_chi_N}, or their Fourier transforms~\eqref{eq:V_X_q_S}, are expanded onto the basis \eqref{eq:irreducible_products_of_S}, the coefficients of the expansion are tensor functions of ranks from 0 to $2j_\chi$ of the magnitude $r_{\chi N} = | \vec{r}_{\chi N} |$ (or $q=|\vec{q}|$, respectively), and these vector functions can be written as derivatives of scalar functions of $r_{\chi N}$ (or as scalar functions of $q$ multiplied by tensor products of $q_i$).

When the potentials $V_X(\vec{r}_{\chi N}, \vec{S}_{\chi})$ in Eq.~\eqref{eq:V_chi_N} are expanded onto the basis \eqref{eq:irreducible_products_of_S}, the coefficients of the expansion are tensor functions of ranks from 0 to $2j_\chi+1$ of the magnitude $r_{\chi N} = | \vec{r}_{\chi N} |$, These tensor functions can be written as derivatives of scalar functions of $r_{\chi N}$. For instance, introducing a factor $(-1)^s$ for our later convenience, 
\begin{align}
V_M^\tau\big(\vec{r}_{\chi N}, \vec{S}_\chi\big) = \sum_{s=0}^{2j_\chi} \, (-1)^s \, \myoverbracket{S_{i_1} S_{i_2} \cdots S_{i_s}} \, \partial_{i_1} \partial_{i_2} \cdots \partial_{i_s} V_{M,s,s}^\tau(r_{\chi N})  . 
\end{align}
When the same procedure is applied to the Fourier transforms $V_X^\tau(\vec{q}, \vec{S}_{\chi})$ in Eq.~\eqref{eq:V_X_q_S}, the coefficient functions are tensor products of the form $i q_{i_1} \, i q_{i_2} \, \cdots \, i q_{i_s}$ multiplied by scalar functions of the magnitude $q=|\vec{q}|$. For example,
\begin{align}
V_M^\tau\big(\vec{q}, \vec{S}_\chi\big) = \sum_{s=0}^{2j_\chi} \,  \myoverbracket{S_{i_1} S_{i_2} \cdots S_{i_s}} i^{s}q_{i_1}  q_{i_2} \cdots q_{i_s} V_{M,s,s}^\tau(q)  . 
\end{align}
The scalar functions $V_{X,s,l}^\tau(q)$ will give the $q$ dependence of the coefficients $c^\tau_{X,s,l}(q)$ in Eq.~\eqref{eq:O_chi_N} below.

Using the irreducible spin products in Eq.~\eqref{eq:irreducible_products_of_S}
%\begin{align}
%  1, \quad \myoverbracket{S_{i_1}}, \quad \myoverbracket{S_{i_1} S_{i_2}}, \quad \myoverbracket{S_{i_1}S_{i_2}S_{i_3}}, \quad\dots, \quad \myoverbracket{S_{i_1} S_{i_2} \cdots S_{i_{2j_\chi}}},
%  \label{eq:irreducible_products_of_S}
%\end{align}
in place of those in Eq.~(\ref{eq:products_of_S}), we are lead to introduce the scalar WIMP operators
\begin{align}
i^s \myoverbracket{ S_{i_1} \cdots S_{i_s} } \, \q_{i_1} \cdots \q_{i_s} ,
\end{align}
and the vector WIMP operators
%\begin{subequations}
\begin{align}
i^s \myoverbracket{S_{i_1} \cdots S_{i_s}}\q_{i_1}\cdots \q_{i_{s-1}} && \text{(free index $i_s$)},
\nonumber \\
i^s \epsilon_{ijk}  \myoverbracket{S_{i_1} \cdots S_{i_{s-1}} S_{j}}
\q_{i_1}\cdots \q_{i_{s-1}}\q_{k} && \text{(free index $i$)},
 \nonumber\\
i^{s+1} \myoverbracket{ S_{i_1} \cdots S_{i_s}} \q_{i_1}\cdots \q_{i_s}\q_{i_{s+1}} && \text{(free index $i_{s+1}$)}.
\end{align}
%\end{subequations}
The three vector operators correspond to the three possible combinations of angular momenta $s$ (the number of $S$ factors) and $l$ (the number of $\q$ factors) with total angular momentum 1.

Following the procedure outlined above we define the following basis
of WIMP--nucleon operators $\calO_{X,s,l}$, all of which are irreducible in WIMP spin space and Hermitian,
%\begin{subequations}
\begin{align}
\calO_{M,s,s} & = i^s \, \myoverbracket{ S_{i_1} \cdots S_{i_s} } \, \q_{i_1} \cdots \q_{i_s} && (s\ge0),\nonumber\\
\calO_{\Omega,s,s} & = i^s \, \myoverbracket{ S_{i_1} \cdots S_{i_s} } \, \q_{i_1} \cdots \q_{i_s} (\vec{v}{}^{\plus}_{\chi N} \cdot \vec{\sigma}_N)/2 && (s\ge0), \nonumber\\
\calO_{\Sigma,s,s-1} & = i^{s-1} \, \myoverbracket{ S_{i_1} \cdots S_{i_s} } \, \q_{i_1} \cdots \q_{i_{s-1}} (\vec{\sigma}_N)_{i_s}/2 && (s\ge1), \nonumber\\
\calO_{\Sigma,s,s} & = i^{s} \, \myoverbracket{ S_{i_1} \cdots S_{i_s} } \, \q_{i_1} \cdots \q_{i_{s-1}} (\vec{\q} \times \vec{\sigma}_N)_{i_s}/2 && (s\ge1), \nonumber\\
\calO_{\Sigma,s,s+1} & = i^{s+1} \, \myoverbracket{ S_{i_1} \cdots S_{i_s} } \, \q_{i_1} \cdots \q_{i_s} (\vec{\q} \cdot \vec{\sigma}_N)/2 && (s\ge0), \nonumber\\
\calO_{\Delta,s,s-1} & = i^{s-1} \, \myoverbracket{ S_{i_1} \cdots S_{i_s} } \, \q_{i_1} \cdots \q_{i_{s-1}} (\vec{v}{}^{\plus}_{\chi N})_{i_s} && (s\ge1), \nonumber \\
\calO_{\Delta,s,s} & = i^{s} \, \myoverbracket{ S_{i_1} \cdots S_{i_s} } \, \q_{i_1} \cdots \q_{i_{s-1}} (\vec{\q}\times\vec{v}{}^{\plus}_{\chi N})_{i_s} && (s\ge1),\nonumber \\
\calO_{\Delta,s,s+1} & = i^{s+1} \, \myoverbracket{ S_{i_1} \cdots S_{i_s} } \, \q_{i_1} \cdots \q_{i_s} (\vec{\q} \cdot \vec{v}{}^{\plus}_{\chi N}) && (s\ge0),\nonumber \\
\calO_{\Phi,s,s-1} & = i^{s-1} \, \myoverbracket{ S_{i_1} \cdots S_{i_s} } \, \q_{i_1} \cdots \q_{i_{s-1}} (\vec{v}{}^{\plus}_{\chi N} \times \vec{\sigma}_N)_{i_s}/2 && (s\ge1), \nonumber\\
\calO_{\Phi,s,s} & = i^{s} \, \myoverbracket{ S_{i_1} \cdots S_{i_s} } \, \q_{i_1} \cdots \q_{i_{s-1}} (\vec{\q} \times ( \vec{v}{}^{\plus}_{\chi N} \times \vec{\sigma}_N))_{i_s} /2 && (s\ge1),\nonumber \\
\calO_{\Phi,s,s+1} & = i^{s+1} \, \myoverbracket{ S_{i_1} \cdots S_{i_s} } \, \q_{i_1} \cdots \q_{i_s} (\vec{\q} \cdot \vec{v}{}^{\plus} _{\chi N} \times \vec{\sigma}_N)/2 && (s\ge0).
\label{eq:basis_WIMP_nucleon_operators}
\end{align}
%\end{subequations}
Each operator of Eqs.~(\ref{eq:basis_WIMP_nucleon_operators}) is to be multiplied by the isoscalar or isovector operator $t^0$ or $t^1=\tau_3$ to form $\calO^\tau_{X,s,l} = \calO_{X,s,l}  t^\tau $.

The basis operators in Eqs.~(\ref{eq:basis_WIMP_nucleon_operators})
can also be written in vector notation as follows, where the
overbrackets amount to taking the symmetric traceless part of the
product of WIMP spin matrices (in the following equation and in
Tables~\ref{tab:spin0}%, \ref{tab:spin1h}, \ref{tab:spin1}, \ref{tab:spin3h} and
--\ref{tab:spin2} we use the notation $\vec{S}_N$ and
$\vec{S}_\chi$ for the nucleon and WIMP spins, respectively)
%\begin{subequations}
\begin{align}
\calO_{M,s,s} & = \myoverbracket{ (i \, \vec{\q} \cdot \vec{S}_\chi)^s } && (s\ge0),\nonumber \\
\calO_{\Omega,s,s} & = \myoverbracket{ (i \, \vec{\q} \cdot \vec{S}_\chi)^s } \, (\vec{v}{}^{\plus}_{\chi N} \cdot \vec{S}_N) && (s\ge0),\nonumber \\
\calO_{\Sigma,s,s-1} & = \myoverbracket{ (i \,\vec{\q} \cdot \vec{S}_\chi)^{s-1} (\vec{S}_N \cdot \vec{S}_\chi) }  && (s\ge1), \nonumber\\
\calO_{\Sigma,s,s} & =\myoverbracket{ (i \,\vec{\q} \cdot \vec{S}_\chi)^{s-1} (i \, \vec{\q} \times \vec{S}_N \cdot \vec{S}_\chi) }  && (s\ge1),\nonumber \\
\calO_{\Sigma,s,s+1} & =\myoverbracket{ (i \, \vec{\q} \cdot \vec{S}_\chi)^s } \, (i \, \vec{\q} \cdot \vec{S}_N) && (s\ge0),\nonumber \\
\calO_{\Delta,s,s-1} & = \myoverbracket{ (i \,\vec{\q} \cdot \vec{S}_\chi)^{s-1} (\vec{v}{}^{\plus}_{\chi N} \cdot \vec{S}_\chi) }   && (s\ge1),\nonumber \\
\calO_{\Delta,s,s} & = \myoverbracket{ (i \,\vec{\q} \cdot \vec{S}_\chi)^{s-1} (i \, \vec{\q} \times  \vec{v}{}^{\plus}_{\chi N}  \cdot \vec{S}_\chi) }  && (s\ge1),\nonumber \\
\calO_{\Delta,s,s+1} & =\myoverbracket{ (i \, \vec{\q} \cdot \vec{S}_\chi)^s } (i\, \vec{\q} \cdot \vec{v}{}^{\plus}_{\chi N} )&& (s\ge0), \nonumber\\
\calO_{\Phi,s,s-1} & = \myoverbracket{ (i \,\vec{\q} \cdot \vec{S}_\chi)^{s-1} ( \vec{v}{}^{\plus}_{\chi N}  \times \vec{S}_N \cdot \vec{S}_\chi) }  && (s\ge1),\nonumber \\
\calO_{\Phi,s,s} & =  \myoverbracket{ (i \,\vec{\q} \cdot \vec{S}_\chi)^{s-1} (\vec{v}{}^{\plus}_{\chi N} \cdot \vec{S}_\chi) }  \, (i \, \vec{\q} \cdot \vec{S}_N) && (s\ge1), \nonumber\\
\calO_{\Phi,s,s+1} & = \myoverbracket{ (i \, \vec{\q} \cdot \vec{S}_\chi)^s } \, (i \, \vec{\q} \times \vec{v}{}^{\plus}_{\chi N} \cdot \vec{S}_N) && (s\ge0).
\label{eq:basis_WIMP_nucleon_operators_alt}
\end{align}
%\end{subequations}

The indices in the symbol of the operator $\calO_{\curr{X},s,l}$
follow the following scheme. The first index $X$ is the nucleon
current ($X=M$, $\Omega$, $\Sigma$, $\Delta$, and $\Phi$ for the
nucleon currents $1$, $\vec{v}{}^{\plus}_{\chi N} \cdot \vec{\sigma}_N$,
$\vec{\sigma}_N$, $\vec{v}{}^{\plus}_{\chi N}$, and $\vec{v}{}^{\plus}_{\chi N} \times
\vec{\sigma}_N$, respectively). The second index $s$ is the number of
WIMP spin operators $\vec{S}_{\chi}$ appearing in $\calO_{\curr{X},s,l}$. This
can be considered as the spin of the operator. It ranges from $s=0$ to
twice the WIMP spin $s=2j_\chi$. The third index $l$ is the power of
the momentum exchange vector $q_i$ in the operator
$\calO_{\curr{X},s,l}$. This can be considered as the angular momentum
of the operator. A factor of $i$ is introduced for every power of
$q$. We include the operator $\calO_{\Delta,s,s+1}$ in our list of
basis operators even if it is zero for elastic scattering because $
\vec{v}{}^{\plus}_{\chi N} \cdot \vec{q} = 0$; it may appear in
inelastic scattering in which the nucleus transitions to another
energy level.

The relation between our operators and those defined in~\cite{haxton1,
  haxton2} and~\cite{krauss_spin_1} is listed in
Table~\ref{tab:Haxton_operators} (see Section~\ref{sec:spin1} for the
case of WIMP spin 1). Notice that following common usage in the WIMP
dark matter community we define $\vec{q}$ as the momentum transferred
\emph{to} the nucleus, whereas~\cite{haxton1, haxton2} use $\vec{q}$
for the momentum \emph{lost} by the nucleus; thus our $\vec{q}$ and
that in~~\cite{haxton1, haxton2} have opposite signs.
Tables~\ref{tab:spin0}--\ref{tab:spin2} summarize the explicit forms
of the effective operators for WIMPs of spin 0, 1/2, 1, 3/2, and 2.

{ % begin operators spin 0
\begin{table}[t]\centering
\caption{Effective WIMP-nucleon operators appearing for WIMPs of spin $\ge0$.}
\label{tab:spin0}
\renewcommand{\arraystretch}{1.6}
\addtolength{\tabcolsep}{2.0pt}
\vskip\baselineskip
\begin{tabular}{@{}l@{\hspace{6em}}l@{}}
\toprule
\hbox to 4em{$\calO_{M,0,0} = \,$} 1 
& %[0.5ex]\cdashline{1-1}
\hbox to 4em{$\calO_{\Sigma,0,1}= \,$} $ i  \vec{\q} \cdot \vec{S}_N $  
\\[0.5ex]\cdashline{1-2}
\hbox to 4em{$\calO_{\Phi,0,1} = \,$} $ i  \vec{\q} \times \vec{v}{}^{\plus}_{\chi N} \cdot \vec{S}_N $ 
& %[0.5ex]\cdashline{1-1}
\hbox to 4em{$\calO_{\Omega,0,0} = \,$} $ \vec{v}{}^{\plus}_{\chi N} \cdot \vec{S}_N $ 
\\ 
\bottomrule
\end{tabular}
\end{table}
} % end operators spin 0

{ % begin operators spin 1/2
\begin{table}[t]\centering
\caption{Effective WIMP-nucleon operators for WIMPs of spin $\ge1/2$.}
\label{tab:spin1h}
\renewcommand{\arraystretch}{1.5}
\addtolength{\tabcolsep}{2.0pt}
\vskip\baselineskip
\begin{tabular}{@{}l@{\hspace{6em}}l@{}}
\toprule
\hbox to 4em{$\calO_{M,1,1}= \,$} $ i \vec{S}_\chi \cdot \vec{\q} $  
& % [0.5ex]\cdashline{1-1}
\hbox to 4em{$\calO_{\Sigma,1,0} = \,$} $ \vec{S}_\chi \cdot \vec{S}_N $ \\
[0.5ex]\cdashline{1-2}
\hbox to 4em{$\calO_{\Sigma,1,1}= \,$} $ i \vec{S}_\chi \cdot ( \vec{\q} \times \vec{S}_N ) $ 
& % [0.5ex]\cdashline{1-1}
\hbox to 4em{$\calO_{\Sigma,1,2} = \,$} $ - ( \vec{S}_\chi \cdot \vec{\q} ) ( \vec{\q} \cdot \vec{S}_N)  $ \\
[0.5ex]\cdashline{1-2}
\hbox to 4em{$\calO_{\Delta,1,0}= \,$} $ \vec{S}_\chi \cdot\vec{v}{}^{\plus}_{\chi N}  $  &
%[0.5ex]\cdashline{1-1}
\hbox to 4em{$\calO_{\Delta,1,1} = \,$} $ i \vec{S}_\chi  \cdot ( \vec{\q} \times \vec{v}{}^{\plus}_{\chi N} ) $ \\
[0.5ex]\cdashline{1-2}
\hbox to 4em{$\calO_{\Phi,1,0}= \,$} $ \vec{S}_\chi \cdot ( \vec{v}{}^{\plus}_{\chi N} \times \vec{S}_N ) $  &
%[0.5ex]\cdashline{1-1}
\hbox to 4em{$\calO_{\Phi,1,1}= \,$} $ i (\vec{S}_\chi \cdot \vec{v}{}^{\plus}_{\chi N} ) ( \vec{\q} \cdot \vec{S}_N  ) $  \\
[0.5ex]\cdashline{1-2}
\hbox to 4em{$\calO_{\Phi,1,2}= \,$} $ - ( \vec{S}_\chi \cdot \vec{\q} ) ( \vec{\q} \times \vec{v}{}^{\plus}_{\chi N} \cdot \vec{S}_N  ) $  &
%[0.5ex]\cdashline{1-1}
\hbox to 4em{$\calO_{\Omega,1,1}= \,$} $ i ( S_\chi \cdot \vec{\q} ) ( \vec{v}{}^{\plus}_{\chi N} \cdot \vec{S}_N  ) $  \\
\bottomrule
\end{tabular}
\end{table}
} % end operators spin 1/2

{ % begin operators spin 1
\begin{table}[t]\centering
\def\myabbr{(\vec{\q}\cdot\vec{S}_\chi)^2 \,\llap{$\myoverbracket{\phantom{t}\hspace{3.1em}}$}}
\def\myabbrev{(\vec{\q}\cdot\vec{S}_\chi) \vec{S}_\chi  \llap{$\myoverbracket{\phantom{t}\hspace{3.7em}}$}}
\caption{Effective WIMP-nucleon operators for WIMPs of spin $\ge1$. }
\label{tab:spin1}
\renewcommand{\arraystretch}{1.5}
\addtolength{\tabcolsep}{2.0pt}
\vskip\baselineskip
\begin{tabular}{@{}l@{\hspace{6em}}l@{}}
\toprule
\hbox to 4em{$\calO_{M,2,2}= \,$} $ - \, \myabbr  $  &
%[0.5ex]\cdashline{1-1}
\hbox to 4em{$\calO_{\Sigma,2,1} = \,$} $ i \, \myabbrev \cdot \vec{S}_N $ \\
[0.5ex]\cdashline{1-2}
\hbox to 4em{$\calO_{\Sigma,2,2}= \,$} $ -  \, \myabbrev   \times \vec{\q} \cdot \vec{S}_N  $  &
%[0.5ex]\cdashline{1-1}
\hbox to 4em{$\calO_{\Sigma,2,3} = \,$} $ - i \, \myabbr \, (\vec{\q} \cdot \vec{S}_N ) $ \\
[0.5ex]\cdashline{1-2}
\hbox to 4em{$\calO_{\Delta,2,1}= \,$} $ i  \, \myabbrev \cdot \vec{v}{}^{\plus}_{\chi N}  $   &
%[0.5ex]\cdashline{1-1}
\hbox to 4em{$\calO_{\Delta,2,2} = \,$} $ - \, \myabbrev  \times \vec{\q} \cdot \vec{v}{}^{\plus}_{\chi N}  $ \\
[0.5ex]\cdashline{1-2}
\hbox to 4em{$\calO_{\Phi,2,1}= \,$} $ i \, \myabbrev \cdot  \vec{v}{}^{\plus}_{\chi N} \times \vec{S}_N $   &
%[0.5ex]\cdashline{1-1}
\hbox to 4em{$\calO_{\Phi,2,2}= \,$} $ -  \, \myabbrev \cdot \vec{v}{}^{\plus}_{\chi N} \, (\vec{\q} \cdot \vec{S}_N)  $  \\
[0.5ex]\cdashline{1-2}
\hbox to 4em{$\calO_{\Phi,2,3}= \,$} $- i \, \myabbr \,  (\vec{\q}  \cdot \vec{v}{}^{\plus}_{\chi N} \times \vec{S}_N )  $  &
%[0.5ex]\cdashline{1-1}
\hbox to 4em{$\calO_{\Omega,2,2}= \,$} $ - \, \myabbr  \, ( \vec{v}{}^{\plus}_{\chi N} \cdot \vec{S}_N ) $  \\
\bottomrule\\[-4ex]
\multicolumn{2}{@{}l}{
\small $\myabbr = ( \vec{S}_\chi \cdot \vec{\q} )^2 - \tfrac{1}{3} j_\chi(j_\chi+1) \q^2 $ , \qquad
\small $\myabbrev = \frac{1}{2}  \big[  (\vec{S}_\chi\cdot\vec{\q}) \vec{S}_\chi + \vec{S}_\chi (\vec{S}_\chi\cdot\vec{\q})  \big] - \tfrac{1}{3} j_\chi(j_\chi+1) \, \vec{\q} $
}
\end{tabular}
\end{table}
} % end operators spin 1

{ % begin operators spin 3/2
\begin{table}[t]\centering
\def\myabbr{(\vec{\q}\cdot\vec{S}_\chi)^3 \,\llap{$\myoverbracket{\phantom{t}\hspace{3.1em}}$}}
\def\myabbrev{(\vec{\q}\cdot\vec{S}_\chi)^2 \vec{S}_\chi  \llap{$\myoverbracket{\phantom{t}\hspace{4.2em}}$}}
\caption{Effective WIMP-nucleon operators for WIMPs of spin $j_\chi\ge3/2$.}
\label{tab:spin3h}
\renewcommand{\arraystretch}{1.5}
\addtolength{\tabcolsep}{2.0pt}
\vskip\baselineskip
\begin{tabular}{@{}l@{\hspace{6em}}l@{}}
\toprule
\hbox to 4em{$\calO_{M,3,3}= \,$} $ - i \, \myabbr $  &
%[0.5ex]\cdashline{1-1}
\hbox to 4em{$\calO_{\Sigma,3,2} = \,$} $ - \, \myabbrev  \cdot \vec{S}_N$ \\
[0.5ex]\cdashline{1-2}
\hbox to 4em{$\calO_{\Sigma,3,3}= \,$} $ - i  \, \myabbrev   \times \vec{\q} \cdot \vec{S}_N $  &
%[0.5ex]\cdashline{1-1}
\hbox to 4em{$\calO_{\Sigma,3,4} = \,$} $ \myabbr  \,  (\vec{\q} \cdot \vec{S}_N ) $ \\
[0.5ex]\cdashline{1-2}
\hbox to 4em{$\calO_{\Delta,3,2}= \,$} $- \, \myabbrev   \cdot \vec{v}{}^{\plus}_{\chi N} $   &
%[0.5ex]\cdashline{1-1}
\hbox to 4em{$\calO_{\Delta,3,3} = \,$} $  - i \, \myabbrev  \times \vec{\q} \cdot \vec{v}{}^{\plus}_{\chi N} $ \\
[0.5ex]\cdashline{1-2}
\hbox to 4em{$\calO_{\Phi,3,2}= \,$} $  - \, \myabbrev  \cdot\vec{v}{}^{\plus}_{\chi N} \times\vec{S}_N $   &
%[0.5ex]\cdashline{1-1}
\hbox to 4em{$\calO_{\Phi,3,3}= \,$} $ - i \, \myabbrev   \cdot \vec{v}{}^{\plus}_{\chi N}  \, (\vec{\q} \cdot \vec{S}_N ) $  \\
[0.5ex]\cdashline{1-2}
\hbox to 4em{$\calO_{\Phi,3,4}= \,$} $  \myabbr \, (\vec{\q}  \cdot \vec{v}{}^{\plus}_{\chi N} \times \vec{S}_N )  $  &
%[0.5ex]\cdashline{1-1}
\hbox to 4em{$\calO_{\Omega,3,3}= \,$} $ - i \, \myabbr \, ( \vec{v}{}^{\plus}_{\chi N} \cdot \vec{S}_N ) $  \\
\bottomrule\\[-4ex]
\multicolumn{2}{@{}l}{
\small $\myabbr = (\vec{\q} \cdot \vec{S}_\chi)^3 - \frac{3}{5} \q^2 j_\chi(j_\chi+1) ( \vec{\q} \cdot \vec{S}_\chi) $
}
\\
\multicolumn{2}{@{}l}{
\small $\myabbrev =  \frac{1}{3} \big[ (\vec{\q}\cdot\vec{S}_\chi)^2 \vec{S}_\chi  + (\vec{\q}\cdot\vec{S}_\chi) \vec{S}_\chi (\vec{\q}\cdot\vec{S}_\chi) +  \vec{S}_\chi (\vec{\q}\cdot\vec{S}_\chi)^2 \big] 
%$ }
%\\[-0.7ex] 
%\multicolumn{2}{@{}l}{
%$ \phantom{myabbrev = } 
- \frac{2}{5} j_\chi(j_\chi+1) (\vec{\q}\cdot\vec{S}_\chi) \, \vec{\q} - \frac{1}{5}  j_\chi(j_\chi+1)  \q^2 \, \vec{S}_\chi $
}
\end{tabular}
\end{table}
} % end operators spin 3/2

{ % begin operators spin 2
\begin{table}[t]\centering
\def\myabbr{(\vec{\q}\cdot\vec{S}_\chi)^4 \,\llap{$\myoverbracket{\phantom{t}\hspace{3.2em}}$}}
\def\myabbrev{(\vec{\q}\cdot\vec{S}_\chi)^3 \vec{S}_\chi  \llap{$\myoverbracket{\phantom{t}\hspace{4.2em}}$}}
\caption{Effective WIMP-nucleon operators for WIMPs of spin $j_\chi\ge2$.}
\label{tab:spin2}
\newcommand{\ob}[2]{\myoverbracket{ \begin{minipage}[b][7.5pt][b]{#1}$#2$\end{minipage}}}
\renewcommand{\arraystretch}{1.5}
\addtolength{\tabcolsep}{2.0pt}
\vskip\baselineskip
\begin{tabular}{@{}l@{\hspace{6em}}l@{}}
\toprule
\hbox to 4em{$\calO_{M,4,4}= \,$} $ \myabbr $  &
%[0.5ex]\cdashline{1-1}
\hbox to 4em{$\calO_{\Sigma,4,3} = \,$} $ - i \, \myabbrev \cdot \vec{S}_N $ \\
[0.5ex]\cdashline{1-2}
\hbox to 4em{$\calO_{\Sigma,4,4}= \,$} $ \myabbrev   \times \vec{\q} \cdot \vec{S}_N  $  &
%[0.5ex]\cdashline{1-1}
\hbox to 4em{$\calO_{\Sigma,4,5} = \,$} $ i \, \myabbr \,  (\vec{\q} \cdot \vec{S}_N ) $ \\
[0.5ex]\cdashline{1-2}
\hbox to 4em{$\calO_{\Delta,4,3}= \,$} $  - i \, \myabbrev \cdot \vec{v}{}^{\plus}_{\chi N} $   &
%[0.5ex]\cdashline{1-1}
\hbox to 4em{$\calO_{\Delta,4,4} = \,$} $  \myabbrev \times \vec{\q} \cdot \vec{v}{}^{\plus}_{\chi N} $ \\
[0.5ex]\cdashline{1-2}
\hbox to 4em{$\calO_{\Phi,4,3}= \,$} $  - i \, \myabbrev \cdot \vec{v}{}^{\plus}_{\chi N} \times \vec{S}_N  $  &
%[0.5ex]\cdashline{1-1}
\hbox to 4em{$\calO_{\Phi,4,4}= \,$} $\myabbrev \cdot \vec{v}{}^{\plus}_{\chi N} \, (\vec{\q} \cdot \vec{S}_N)  $  \\
[0.5ex]\cdashline{1-2}
\hbox to 4em{$\calO_{\Phi,4,5}= \,$} $  i \, \myabbr \, (\vec{\q}  \cdot \vec{v}{}^{\plus}_{\chi N} \times \vec{S}_N )  $  &
%[0.5ex]\cdashline{1-1}
\hbox to 4em{$\calO_{\Omega,4,4}= \,$} $\myabbr \, ( \vec{v}{}^{\plus}_{\chi N} \cdot \vec{S}_N ) $  \\
\bottomrule\\[-4ex]
\multicolumn{2}{@{}l}{
\small $\myabbr = (\vec{\q} \cdot \vec{S}_\chi)^4 - \frac{6}{7} j_\chi(j_\chi+1) \q^2 ( \vec{\q} \cdot \vec{S}_\chi)^2 + \frac{3}{35} j_\chi^2(j_\chi+1)^2 \q^4  $ 
}
\\
\multicolumn{2}{@{}l}{
\small $\myabbrev =  \frac{1}{4} \big[  (\vec{\q} \cdot \vec{S}_\chi)^3 (\vec{S}_N\cdot\vec{S}_\chi) +  (\vec{\q} \cdot \vec{S}_\chi)^2 (\vec{S}_N\cdot\vec{S}_\chi)  (\vec{\q} \cdot \vec{S}_\chi) +  (\vec{\q} \cdot \vec{S}_\chi) (\vec{S}_N\cdot\vec{S}_\chi)  (\vec{\q} \cdot \vec{S}_\chi)^2 +  (\vec{S}_N\cdot\vec{S}_\chi) (\vec{\q} \cdot \vec{S}_\chi)^3 \big] $ 
} 
\\[-0.7ex]
\multicolumn{2}{@{}l}{
$ \phantom{myabbrev = } 
- \frac{3}{14} j_\chi(j_\chi+1) \q^2 \big[  (\vec{\q} \cdot \vec{S}_\chi) (\vec{S}_N\cdot\vec{S}_\chi) + (\vec{S}_N\cdot\vec{S}_\chi) (\vec{\q} \cdot \vec{S}_\chi) \big] $
} 
\\[-0.7ex]
\multicolumn{2}{@{}l}{
$ \phantom{myabbrev = }  - \frac{3}{7}  j_\chi (j_\chi+1) (\vec{S}_N\cdot\vec{\q})  ( \vec{\q} \cdot \vec{S}_\chi)^2  + \frac{3}{35} j_\chi^2(j_\chi+1)^2 \q^2  (\vec{\q} \cdot \vec{S}_N )  $
}
\end{tabular}
\end{table}
} % end operators spin 2

%\subsection{WIMP--nucleon Hamiltonian}

A general WIMP--nucleon operator $\calO_{\chi N}$ is a linear
combination of the basis WIMP--nucleon operators in
Eqs.~(\ref{eq:basis_WIMP_nucleon_operators}),
\begin{align}
& \op{\calO}_{\chi N} = \sum_{X\tau s\,l} c^{\tau}_{X,s,l}(q) \, \op{\calO}_{X,s,l} \, t^\tau_N.
\label{eq:O_chi_N}
\end{align}
The coefficients $c^{\tau}_{X,s,l}(q) $ are in principle functions of the magnitude $q$ of the momentum transfer, determined by the Fourier transforms of the potentials in Eq.~\eqref{eq:V_chi_N} as $c^{\tau}_{X,s,l}(q)  = m_N^l  V_{X,s,l}^\tau(q)$.  In some phenomenological studies they have been taken as constants.

We
can group the basis operators according to the five nucleon currents
$X=M,\Omega,\Sigma,\Delta,\Phi$ as
\begin{align}
& \op{\calO}_{\chi N} =  \sum_{\tau} t^\tau \Big( \ell_M^\tau \, \op{\calO}_{M} + \vec{\ell}_\Sigma^\tau  \cdot \op{\vec{\calO}}_{\Sigma} + \vec{\ell}_\Delta^\tau \cdot   \op{\vec{\calO}}_{\Delta} + \vec{\ell}_{\Phi}^\tau \cdot   \op{\vec{\calO}}_{\Phi}  + \ell_\Omega^\tau \, \op{\calO}_{\Omega}  \Big ).
\label{eq:O_chi_N_ell}
\end{align}
Here the operators $\op{\calO}_{M}, \op{\vec{\calO}}_{\Sigma}, \op{\vec{\calO}}_{\Delta} , \op{\vec{\calO}}_{\Phi}, \op{\calO}_{\Omega} $ are those appearing in Eq.~\eqref{eq:calO_X}, and the WIMP currents $\ell_M$, $\ell_\Omega$, $\vec{\ell}_{\Sigma}$, $\vec{\ell}_{\Delta}$, $\vec{\ell}_{\Phi}$ can be obtained by substituting Eqs.~(\ref{eq:basis_WIMP_nucleon_operators}) into Eq.~\eqref{eq:O_chi_N},
%\begin{subequations}
\begin{align}
\ell_M^\tau & = \sum_{s=0}^{2j_\chi}  i^s \, \myoverbracket{ S_{i_1} \cdots S_{i_s} } \, \q_{i_1} \cdots \q_{i_s} \, c^{\tau}_{M,s,s}, 
\nonumber\\ 
\ell_\Omega^\tau & = \frac{1}{2} \sum_{s=0}^{2j_\chi}  i^s \, \myoverbracket{ S_{i_1} \cdots S_{i_s} } \, \q_{i_1} \cdots \q_{i_s} \, c^{\tau}_{\Omega,s,s},
\nonumber\\
\ell_{\Sigma,i}^\tau & =  \frac{1}{2} i c^{\tau}_{\Sigma,0,1} \, \q_i \nonumber\\ & + \frac{1}{2} \sum_{s=1}^{2j_\chi} i^{s-1} \, \myoverbracket{ S_{i_1} \cdots S_{i_s} } \, \q_{i_1} \cdots \q_{i_{s-1}} \Big ( 
c^{\tau}_{\Sigma,s,s-1} \, \delta_{i_si} - i c^{\tau}_{\Sigma,s,s} \, \epsilon_{i_sij} \q_j - c^{\tau}_{\Sigma,s,s+1} \, \q_{i_s} \q_i 
\Big ) , 
\nonumber\\
\ell_{\Delta,i}^\tau & = i c^{\tau}_{\Delta,0,1} \, \q_i \nonumber \\ & + \sum_{s=1}^{2j_\chi} i^{s-1} \, \myoverbracket{ S_{i_1} \cdots S_{i_s} } \, \q_{i_1} \cdots \q_{i_{s-1}} \Big ( c^{\tau}_{\Delta,s,s-1} \, \delta_{i_si} - i c^{\tau}_{\Delta,s,s} \, \epsilon_{i_s i j} \q_j - c^{\tau}_{\Delta,s,s+1} \, \q_{i_s} \q_i \Big ) ,
\nonumber\\
\ell_{\Phi,i}^\tau & =  \frac{1}{2} i c^{\tau}_{\Phi,0,1} \, \q_i \nonumber \\ & + \frac{1}{2} \sum_{s=1}^{2j_\chi} i^{s-1} \, \myoverbracket{ S_{i_1} \cdots S_{i_s} } \, \q_{i_1} \cdots \q_{i_{s-1}} \Big ( c^{\tau}_{\Phi,s,s-1} \, \delta_{i_si} - i c^{\tau}_{\Phi,s,s} \epsilon_{i_sij} \q_j- c^{\tau}_{\Phi,s,s+1} \, \q_{i_s} \q_i \Big ) .\label{eq:l_phi} 
\end{align}
%\end{subequations}

\noindent Eq.~\eqref{eq:O_chi_N_ell} applies to WIMP interactions with a free nucleon. 

\section{Effective WIMP--nucleus Hamiltonian}
\label{sec:effective_wimp_nucleus_hamiltonian}

We now pass from the Hamiltonian describing the interaction of a WIMP with a free nucleon to the effective Hamiltonian that describes the interaction of the WIMP with the whole nucleus. Under the approximation that the WIMP interacts only with one nucleon at a time (the one-nucleon approximation), what we need to do is to ``put the nucleon inside the nucleus'' and use the relative velocity of the WIMP with respect to the nucleus (i.e., the center of mass of the system of nucleons).

Let $\vec{v}_{\chi T}$ be the WIMP velocity in the
reference frame of the nucleus center of mass. Introduce $\vec{v}{}^{\plus}_{\chi T}$ as
\begin{align}
  \vec{v}{}^{\plus}_{\chi T} = \vec{v}{}^{\plus}_{\chi} - \vec{v}{}^{\plus}_{T}=\vec{v}_{\chi T}-\frac{\vec{q}}{2\mu_{\chi T}}.
  \label{eq:v_chi_t}
\end{align}
In the notation of~\cite{haxton1,haxton2},
\begin{align}
 \vec{v}{}^{\plus}_{\chi T} = \vec{v}{}^{\perp}_{T} 
\end{align}
(see footnote~\ref{footnote:haxton}).

For elastic WIMP-nucleus scattering,
\begin{align}
\vec{q} \cdot \vec{v}{}^{\plus}_{\chi T} = 0,
\end{align}
and
\begin{align}
( \vec{v}{}^{\plus}_{\chi T} )^2 = v_{\chi T}^2 - \frac{q^2}{4\mu_{\chi T}^2}.
\end{align}

The recipe to ``put the nucleon inside the nucleus''  is to replace the free-nucleon operators $\op{O}_{X} t^\tau$  by their respective symmetrized nucleon current densities $\op{j}^\tau_{X}$. In more detail, using
\begin{align}
\vec{v}{}^{\plus}_{\chi N} = \vec{v}{}^{\plus}_{\chi T} - \vec{v}{}^{\plus}_{NT},
\end{align}
%where
%\begin{align}
%  \vec{v}{}^{\plus}_{\chi T} = \vec{v}{}^{\plus}_{\chi} - \vec{v}{}^{\plus}_{T}=\vec{v}_{\chi T}-\frac{\vec{q}}{2\mu_{\chi T}},
%  \label{eq:v_chi_t}
%\end{align}
%\noindent  $\vec{v}_{\chi T}$ the WIMP velocity in the
%reference frame of the nuclear center of mass, 
Eqs.~(\ref{eq:j_sym_tilde})
and~(\ref{eq:calO_O}) imply the following
replacements
%\begin{subequations}
\begin{align}
& \op{\calO}_{M} t_N^\tau \to \op{j}^\tau_{M}  ,
\nonumber \\
& \op{\vec{\calO}}_{\Sigma} t_N^\tau \to \op{\vec{j}}{}^\tau_{\Sigma} ,
\nonumber\\
& \op{\vec{\calO}}_{\Delta} t_N^\tau \to  \vec{v}{}^{\plus}_{\chi T} \op{j}{}^\tau_{M} -  \op{\vec{j}}{}^\tau_{\widetilde\Delta} ,
\nonumber\\
& \op{\vec{\calO}}_{\Phi} t_N^\tau \to \vec{v}{}^{\plus}_{\chi T} \times \op{\vec{j}}{}^\tau_{\Sigma} - \op{\vec{j}}{}^\tau_{\widetilde\Phi} ,
\nonumber\\
& \op{\calO}_{\Omega} t_N^\tau \to \vec{v}{}^{\plus}_{\chi T} \cdot  \op{\vec{j}}{}^\tau_{\Sigma} - \op{j}{}^\tau_{\widetilde\Omega} .
\end{align}
%\end{subequations}

A Fourier transform (which applies for WIMP wave functions that are plane waves)
leads to the WIMP-nucleus effective Hamiltonian
\begin{align}
%\op{H}  = \sum_{\tau} \Big [ \widetilde{\ell}_M^\tau \,  {j}^\tau_M + \vec{\widetilde{\ell}}{}_\Sigma^\tau \cdot {\vec{j}}^\tau_{\Sigma} - \vec{\ell}_\Delta^\tau \cdot  {\vec{j}}^\tau_{\widetilde\Delta} - \vec{\ell}_{\Phi}^\tau \cdot  {\vec{j}}^\tau_{\widetilde\Phi}  - \ell_\Omega \, {j}^\tau_{\widetilde\Omega}  \Big ] ,
\op{H}(\vec{q})  = \sum_{\tau} \Big [ \widetilde{\ell}_M^\tau \, \op{j}^\tau_M(\vec{q}) + \vec{\widetilde{\ell}}{}_\Sigma^\tau \cdot \op{\vec{j}}{}^\tau_\Sigma(\vec{q}) - \vec{\ell}_\Delta^\tau \cdot  \op{\vec{j}}{}^\tau_{\widetilde\Delta}(\vec{q}) - \vec{\ell}_{\Phi}^\tau \cdot \op{\vec{j}}{}^\tau_{\widetilde\Phi}(\vec{q})  - \ell_\Omega \, \op{j}^\tau_{\widetilde\Omega}(\vec{q})   \Big ] ,
\label{eq:H_in_terms_of_ell_and_j}
\end{align}
%where we wrote $j^\tau_X$ for $\op{j}^\tau_X(\vec{q})$ and introduced
where 
%\begin{subequations}
\begin{align}
\widetilde{\ell}_M^\tau & = \ell_M^\tau + \vec{\ell}_\Delta^\tau \cdot  \vec{v}{}^{\plus}_{\chi T} , 
\nonumber\\
\vec{\widetilde{\ell}}{}_\Sigma^\tau & = \vec{\ell}_\Sigma^\tau + \ell_\Omega \, \vec{v}{}^{\plus}_{\chi T} + \vec{\ell}_{\Phi}^\tau \times \vec{v}{}^{\plus}_{\chi T} .
\label{eq:tilde_ell}
\end{align}

\section{Scattering amplitude squared}
\label{sec:amplitude}

In this Section we outline the procedure to calculate the square of
the amplitude for the scattering process driven by the effective
Hamiltonian of Eq.~(\ref{eq:H_in_terms_of_ell_and_j}). As already
pointed out, the factorization between the nuclear currents
$j^{\tau}_X$, $\vec{j}^{\tau}_X$ and the WIMP currents $l^{\tau}_X$,
$\vec{l}^{\tau}_X$ implies that, compared to the results in the
literature for a WIMP of
spin $\le$1~\cite{haxton1,haxton2,krauss_spin_1} , the nuclear part of
the calculation will not change when the currents
(\ref{eq:l_phi}) are used to describe the interaction
of a WIMP with arbitrary spin.  As a consequence, part of the
procedure has already been described
elsewhere~\cite{haxton1,haxton2}. Nevertheless, for completeness, in
this Section we review the full calculation, albeit focusing on
how to obtain the WIMP spin averages from the currents of
Eqs.~(\ref{eq:l_phi}). In the latter derivation the
convenience of assuming irreducible representations of the rotation
group for the basis WIMP--nucleon operators introduced in
Section~\ref{sec:basis_operators} becomes apparent, as all the
results are obtained by using the two master equations~(\ref{eq:WIMP_spin_average_master_equation_a}--\ref{eq:WIMP_spin_average_master_equation_b}) for traces of products of irreducible spin operators.
The proof of some of the derivations used in this Section, including those of 
Eqs.~(\ref{eq:WIMP_spin_average_master_equation_a}--\ref{eq:WIMP_spin_average_master_equation_b}),
are provided in the Appendices.

\subsection{Sum/average over nuclear spins}
\label{sec:nuclear_part}
Nuclear targets in direct dark matter detection experiments are
usually unpolarized, thus the cross section is summed over final
nuclear spins and averaged over initial nuclear spins. Let $H_\rmfi = \langle {\rm f} | \op{H} | {\rm i} \rangle $ indicate the transition matrix element of the effective Hamiltonian between an initial WIMP--nucleus state $| {\rm i} \rangle $ and a final WIMP--nucleus state $| {\rm f} \rangle$. The sum/average over nuclear polarizations is defined as a sum over final nuclear azimuthal quantum numbers $M_f$ and an average over initial nuclear azimuthal quantum numbers $M_i$,
\begin{align}
\overline{H_\rmfi^* H_\rmfi^{}}& =  \frac{1}{2J_i+1} \sum_{M_i=-J_i}^{J_i} \sum_{M_f=-J_f}^{J_f}
H_\rmfi^* H_\rmfi .
\end{align}
Here $J_i$ and $J_f$ denote the initial and final total angular momentum of the nucleus.

As far as the nuclear part is concerned, the calculation requires to
expand the nuclear currents $j^{\tau}_X$, $\vec{j}^{\tau}_X$ in
spherical and vector spherical harmonics, and to obtain the sums over
initial and final nuclear spins for each nuclear current multipole
operator making use of the Wigner--Eckart theorem.

When the Fourier transform of the non-symmetrized nucleon currents in
Eqs.~(\ref{eq:nonsymmetrized_nuclear_currents}) is expanded into multipoles one obtains 
\begin{align}
& \op{j}_{M}^\tau(\vec{q}) = \sum_{JM} 4 \pi i^J \, Y_{JM}^{*}(\uvec{q}) \, \op{M}_{JM}^\tau(q) ,
\nonumber \\
& \op{j}_{\Omega}^\tau(\vec{q}) = \sum_{JM} 4 \pi i^J \, Y_{JM}^{*}(\uvec{q}) \, \op{\Omega}_{JM}^\tau(q) 
\label{eq:nucleon_current_multipole_expansion_a}
\end{align}
for the scalar currents, and 
\begin{align}
& \op{\vec{j}}{}^\tau_{\Sigma}(q) = \sum_{JM} 4 \pi i^J \big[ - i \vec{Y}_{JM}^{(\mpL)*}(\uvec{q}) \, \op{\Sigma}_{JM}^{\prime\prime\,\tau}(q) - i \vec{Y}_{JM}^{(\mpTE)*}(\uvec{q}) \, \op{\Sigma}_{JM}^{\prime\,\tau}(q) + i \vec{Y}_{JM}^{(\mpTM)*}(\uvec{q}) \, \op{\Sigma}_{JM}^\tau(q) \big ] , 
\nonumber \\
& \op{\vec{j}}{}^\tau_{\Delta}(q) = - \frac{iq}{m_N} \sum_{JM} 4 \pi i^J \big[ - i \vec{Y}_{JM}^{(\mpL)*}(\uvec{q}) \, \op{\Delta}_{JM}^{\prime\prime\,\tau}(q)  + i \vec{Y}_{JM}^{(\mpTE)*}(\uvec{q}) \, \op{\Delta}_{JM}^{\prime\,\tau}(q) + i \vec{Y}_{JM}^{(\mpTM)*}(\uvec{q}) \, \op{\Delta}_{JM}^\tau(q) \big ] ,
\nonumber \\
& \op{\vec{j}}{}^\tau_{\Phi}(q) = - \frac{iq}{m_N} \sum_{JM} 4 \pi i^J \big[  \vec{Y}_{JM}^{(\mpL)*}(\uvec{q}) \, \op{\Phi}_{JM}^{\prime\prime\,\tau}(q) + \vec{Y}_{JM}^{(\mpTE)*}(\uvec{q}) \, \op{\Phi}_{JM}^{\prime\,\tau}(q) - \vec{Y}_{JM}^{(\mpTM)*}(\uvec{q}) \, \op{\Phi}_{JM}^\tau(q) \big ] 
%\label{eq:nucleon_current_multipole_expansion_b}
\end{align}
\noindent for the vector currents. In the expression above, which is obtained using the
multipole expansion of the scalar and vector plane waves provided in
Appendix \ref{sec:multipole_expansion}, the one-nucleon operators
$\op{X}^{\tau}_{JM}$, $\op{X}^{\prime\tau}_{JM}$ and
$\op{X}^{\prime\prime\tau}_{JM}$ (with $X$=$M, \Sigma, \Delta,
\Phi,\Omega$) arise~\cite{deforest_walecka, donnelly_walecka,
  walecka}. We provide them explicitly in
Eq.~(\ref{eq:one_nucleon_multipoles}).  For the vector operators
$X$=$\Sigma, \Delta, \Phi$, we follow the standard notation
that double--primed quantities indicate a longitudinal multipole (L),
single--primed quantities correspond to a transverse--electric
multipole (TE) and unprimed quantities indicates a transverse--magnetic multipole
(TM). Moreover, in the expressions above $\vec{Y}_{JM}^{(\mpL)}$,
$\vec{Y}_{JM}^{(\mpTE)}$ and $\vec{Y}_{JM}^{(\mpTM)}$ are
longitudinal, transverse electric, and transverse magnetic spherical
harmonics defined in terms of the vector spherical harmonics $\vec{Y}_{JLM}(\uvec{q})$. We
provide them explicitly in Eqs.~(\ref{eq:jeiqrB1})--(\ref{eq:jeiqrB3})
and (\ref{eq:vector_spherical}).

The operators $\op{\vec{j}}{}^{\tau}_{\Delta}$,
$\op{\vec{j}}{}^{\tau}_{\Phi}$ and $\op{j}_{\Omega}^\tau$ in
Eq.~(\ref{eq:one_nucleon_multipoles}) correspond to the
non-symmetrized nuclear currents of
Eqs.~(\ref{eq:nonsymmetrized_nuclear_currents}). As explained in
Section~\ref{sec:non_relativistic_nuclear_currents} the WIMP--nucleus
scattering process is driven by the symmetrized currents in
Eqs.~(\ref{eq:symmetrized_nuclear_currents}). So after symmetrization
one obtains
%\begin{subequations}
\begin{align}
& \op{j}_{\Omega,\rm sym}^\tau(\vec{q}) = \sum_{JM} 4 \pi i^J \, Y_{JM}^{*}(\uvec{q}) \, \op{\widetilde\Omega}{}_{JM}^\tau(q) ,\nonumber \\
& \op{\vec{j}}{}^\tau_{\Delta,\rm sym}(q) = - \frac{iq}{m_N} \sum_{JM} 4 \pi i^J \big[  - i \vec{Y}_{JM}^{(\mpL)*}(\uvec{q}) \, \op{\widetilde\Delta}{}_{JM}^{\prime\prime\,\tau}(q)  + i \vec{Y}_{JM}^{(\mpTE)*}(\uvec{q}) \, \op{\Delta}_{JM}^{\prime\,\tau}(q) + i \vec{Y}_{JM}^{(\mpTM)*}(\uvec{q}) \, \op{\Delta}_{JM}^\tau(q) \big ] ,\nonumber \\
& \op{\vec{j}}{}^\tau_{\Phi,\rm sym}(q) = - \frac{iq}{m_N} \sum_{JM} 4 \pi i^J \big[  \vec{Y}_{JM}^{(\mpL)*}(\uvec{q}) \, \op{\Phi}_{JM}^{\prime\prime\,\tau}(q) + \vec{Y}_{JM}^{(\mpTE)*}(\uvec{q}) \, \op{\widetilde\Phi}{}_{JM}^{\prime\,\tau}(q) - \vec{Y}_{JM}^{(\mpTM)*}(\uvec{q}) \, \op{\widetilde\Phi}{}_{JM}^{\tau}(q) \big ] ,
\label{eq:nucleon_current_multipole_expansion_c}
\end{align}
%\end{subequations}
with the symmetrized operators, indicated by a tilde, given in
Eqs.~(\ref{eq:basis_free_nucleon_multipole_operators_b}).

When the multipole expansions of the nucleon currents
(\ref{eq:nucleon_current_multipole_expansion_a},\ref{eq:nucleon_current_multipole_expansion_c})
are inserted into the effective WIMP--nucleon Hamiltonian $\op{H}$ in
Eq.~\eqref{eq:H_in_terms_of_ell_and_j}, one obtains the multipole
expansion of $\op{H}$,
\begin{align}
\op{H} = \sum_{JM} 4 \pi i^J \bigg [ Y_{JM}^{\tau*}(\uvec{q}) \, \op{H}_{JM} + \vec{Y}_{JM}^{(\mpTE)}(\uvec{q}) \cdot \op{\vec{H}}{}_{JM}^{(\mpTE)} + \vec{Y}_{JM}^{(\mpTM)}(\uvec{q}) \cdot \op{\vec{H}}{}_{JM}^{(\mpTM)} \bigg ] ,
\label{eq:effective_hamiltonian_multipole_expansion}
\end{align}
with
%\begin{subequations}
\begin{align}
& \op{H}_{JM} = \sum_{\tau} \bigg( \widetilde{\ell}_M^\tau \, \op{M}_{JM}^\tau - i \vec{\widetilde{\ell}}{}_{\Sigma} \cdot \uvec{q} \, \op{\Sigma}_{JM}^{\prime\prime\,\tau} - \frac{q}{m_N} \vec{\ell}_{\Delta}^\tau \cdot \uvec{q} \, \op{\widetilde\Delta}{}_{JM}^{\prime\prime\,\tau}- \frac{iq}{m_N} \vec{\ell}_{\Phi}^\tau \cdot \uvec{q} \, \op{\Phi}_{JM}^{\prime\prime\,\tau} - \frac{iq}{m_N} \ell_{\Omega}^\tau \, \op{\widetilde\Omega}{}_{JM}^\tau \bigg ) ,\nonumber \\
& \op{\vec{H}}{}^{(\mpTE)}_{JM} = \sum_{\tau} \bigg( - i \vec{\widetilde{\ell}}{}_{\Sigma}^\tau \, \op{\Sigma}_{JM}^{\prime\,\tau} + \frac{q}{m_N} \vec{\ell}_{\Delta}^\tau \, \op{\Delta}_{JM}^{\prime\,\tau} - \frac{iq}{m_N} \vec{\ell}_\Phi \op{\widetilde\Phi}_{JM}^{\prime\,\tau} \bigg ) ,\nonumber \\
& \op{\vec{H}}{}^{(\mpTM)}_{JM} = \sum_{\tau} \bigg( i \vec{\widetilde{\ell}}{}_{\Sigma}^\tau \, \op{\Sigma}_{JM}^{\tau} + \frac{q}{m_N} \vec{\ell}_{\Delta}^\tau \, \op{\Delta}_{JM}^{\tau} + \frac{iq}{m_N} \vec{\ell}_\Phi \op{\widetilde\Phi}_{JM}^{\tau} \bigg ) .
\label{eq:HJM}
\end{align}
%\end{subequations}

We provide the
details of the rest of the calculation of the sum/average over nuclear spins in
Appendix~\ref{app:sum_nuclear_spins}. The result is
\begin{align}
\overline{H_\rmfi^* H_\rmfi^{}}& = \sum_{\tau\tau^\prime} \bigg \{ 
\widetilde{\ell}_{M}^{\tau} \widetilde{\ell}_M^{\tau^\prime*} \,
F^{\tau\tau^\prime}_{M} + \widetilde{\ell}_{\Sigma i}^{\tau}
\widetilde{\ell}_{\Sigma j}^{\tau^\prime*} \hat{q}_i\hat{q}_j \,
F^{\tau\tau^\prime}_{\Sigma''} + \frac{1}{2} \widetilde{\ell}_{\Sigma
  i}^{\tau} \widetilde{\ell}_{\Sigma j}^{\tau^\prime*} \big(
\delta_{ij} - \hat{q}_i\hat{q}_j \big) \,
F^{\tau\tau^\prime}_{\Sigma'} \nonumber \\ & + \frac{2q}{m_N} \Im \big
( \hat{q}_i \ell_{\Phi i}^{\tau} \widetilde{\ell}_{M}^{\tau^\prime*}
\, F^{\tau\tau^\prime}_{\Phi'' M} \big ) + \frac{q}{m_N} \Im \big (
\epsilon_{ijk} \widetilde{\ell}_{\Sigma,i}^{\tau} \ell_{\Delta
  j}^{\tau^\prime*} \hat{q}_k F^{\tau\tau^\prime}_{\Sigma'\Delta} \big
) \nonumber \\ & + \qm^2 \ell_{\Phi i}^{\tau} \ell_{\Phi
  j}^{\tau^\prime*} \hat{q}_i\hat{q}_j \, F^{\tau\tau^\prime}_{\Phi''}
+ \frac{1}{2} \qm^2 \ell_{\Phi i}^{\tau} \ell_{\Phi j}^{\tau^\prime*}
\big( \delta_{ij} - \hat{q}_i\hat{q}_j \big) \,
F^{\tau\tau^\prime}_{\widetilde\Phi'} \nonumber \\ & + \frac{1}{2}
\qm^2 \ell_{\Delta i}^{\tau} \ell_{\Delta j}^{\tau^\prime*} \big(
\delta_{ij} - \hat{q}_i\hat{q}_j \big) \, F^{\tau\tau^\prime}_{\Delta}
\bigg \} .
\label{eq:HfiHfi_groundstate}
\end{align}
Here we use the notation of~\cite{haxton1}, where the nuclear
response functions $F^{\tau\tau^\prime}_X$ are defined by
\begin{align}
F^{\tau\tau^\prime}_{\curr{X}\curr{Y}}(q) = \frac{4\pi}{2J_i+1} \sum_{J_f} \langle J_f || \op{X}_{J}^\tau(q) || J_i \rangle \, \langle J_f || \op{Y}_{J}^\tau(q) || J_i \rangle^* , 
\label{eq:FXYAB}
\end{align}
\noindent with $ \langle J_f || \op{\curr{X}}^{\tau}_{J}(q) ||
J_i\rangle $ being the reduced matrix elements of the one-nucleon multipole operator $\op{X}^\tau_{JM}$ defined in
Eq.~\eqref{eq:one_nucleon_multipoles}. Ref.~\cite{haxton2} uses the notation
\begin{align}
F^{\tau\tau^\prime}_{\curr{X}\curr{Y}}(q)  = \frac{4\pi}{2J_i+1} \sum_{J_f} \, W^{\tau\tau^\prime}_{\curr{X}\curr{Y}} (q).
\end{align}
We
write $F^{\tau\tau'}_{X}(q)$ for $F^{\tau\tau'}_{XX}(q)$. In
Eq.~(\ref{eq:HfiHfi_groundstate}) only the multipole operators
$X$=$M$, $\Sigma^{\prime}$, $\Sigma^{\prime\prime}$,
$\Delta$, $\Phi^{\prime\prime}$ and $\widetilde\Phi'$ appear,
which correspond to $P$ and $T$ invariant nuclear ground states.
These are the only allowed responses under the assumption that the
nuclear ground state is an eigenstate of P and CP. The
parity of the nucleon currents and their multipoles under
space-reflection $P$ and time-reversal $T$ are collected in Table
\ref{tab:nucleon_current_parities}.

{ % begin nucleon current parities table
\begin{table}[t]\centering
\caption{Parity of the nucleon currents under space reflection $P$ and time reversal $T$. Columns $P_J$ and $T_J$ list the parities of their $J$-th multipole moments (the notation $\mpL$, $\mpTE$, and $\mpTM$ stands for longitudinal, transverse electric, and transverse magnetic multipole, respectively). The last column lists the allowed $J$'s in a ground state that is $P$ and $T$ (or $CP$) invariant.}
\label{tab:nucleon_current_parities}
\renewcommand{\arraystretch}{1.5}
\addtolength{\tabcolsep}{2.0pt}
\vskip\baselineskip
\begin{tabular}{@{}llllllll@{}}
\toprule
$X$ & Operator & $P$ & $T$ & Mult.: & $P_J$ & $T_J$ & Ground state  \\
\midrule
$M$ & $1$ & $+1$ & $+1$ & & $(-1)^{J}$ & $(-1)^{J}$ & even $J$ \\
[0.5ex]\cdashline{1-8}
$\widetilde\Omega$ & $\vec{v}{}^{\plus}_N\cdot\vec{\sigma}_N$ & $-1$ & $+1$ & & $(-1)^{J+1}$ & $(-1)^{J}$ & forbidden \\
[0.5ex]\cdashline{1-8}
$\Sigma$ & $\vec{\sigma}_N$ & $+1$ & $-1$ & $\mpL$: & $(-1)^{J+1}$ & $(-1)^{J+1}$ & odd $J$ \\
&&&& $\mpTE$: & $(-1)^{J+1}$ & $(-1)^{J+1}$ & odd $J$ \\
&&&& $\mpTM$: & $(-1)^{J}$ & $(-1)^{J+1}$ & forbidden \\
[0.5ex]\cdashline{1-8}
$\widetilde\Delta$ & $\vec{v}{}^{\plus}_N$ & $-1$ & $-1$ & $\mpL$: & $(-1)^{J}$ & $(-1)^{J+1}$ & forbidden \\
&&&& $\mpTE$: & $(-1)^{J}$ & $(-1)^{J+1}$ & forbidden \\
&&&& $\mpTM$: & $(-1)^{J+1}$ & $(-1)^{J+1}$ & odd $J$ \\
[0.5ex]\cdashline{1-8}
$\widetilde\Phi$ & $\vec{v}{}^{\plus}_N\times\vec{\sigma}_N$ & $-1$ & $+1$ & $\mpL$: & $(-1)^{J}$ & $(-1)^{J}$ & even $J$ \\
&&&& $\mpTE$: & $(-1)^{J}$ & $(-1)^{J}$ & even $J$ \\
&&&& $\mpTM$: & $(-1)^{J+1}$ & $(-1)^{J}$ & forbidden \\
\bottomrule
\end{tabular}
\end{table}
} % end nucleon current parities table

\subsection{Sum/average over WIMP spins}
\label{sec:wimp_part}

The sum/averages over the nuclear spins
Eqs.~\eqref{eq:HfiHfi_groundstate} contain products of the WIMP
currents $\ell^{\tau}_{\curr{X}}$ and
$\vec{\ell}^{\tau}_{\curr{X}}$. The average of these products over the
initial WIMP spins and their sum over the final WIMP spins defines the
unpolarized WIMP response functions
$R^{\tau\tau^\prime}_{\curr{X}\curr{Y}}$, apart from
conventional factors. We indicate the sum/average over WIMP spins with
an overline over the product of WIMP currents. (The context makes it
clear if the overline denotes a sum/average over nuclear spins or
WIMP spins; a double overline denotes a sum/average over both.) Thinking of the WIMP currents $\ell^{\tau}_{\curr{X}}$
and $\vec{\ell}^{\tau}_{\curr{X}}$ as matrices in WIMP spin space, and
thus of $\ell^{\tau*}_{\curr{X}}$ as the Hermitian conjugate of the
matrix $\ell^{\tau}_{\curr{X}}$, we have
\begin{align}
\overline{ \ell^{\tau}_{\curr{X}} \ell^{\tau^\prime*}_{\curr{Y}} } \equiv \frac{1}{2j_\chi+1} \tr \big( \ell^{\tau}_{\curr{X}} \ell^{\tau^\prime*}_{\curr{Y}} \big )
\end{align}
and similar relations for the vector WIMP currents. 

In particular, taking the average over nuclear and WIMP spins of
Eq.~(\ref{eq:HfiHfi_groundstate}) yields 
\begin{align}
\overline{\overline{H_\rmfi^* H_\rmfi^{}}} =&
 \sum_{\tau\tau^\prime}  \bigg \{ 
R^{\tau\tau^\prime}_{M} \, F^{\tau\tau^\prime}_{M} 
+ R^{\tau\tau^\prime}_{\Sigma''} \, F^{\tau\tau^\prime}_{\Sigma''}
+ R^{\tau\tau^\prime}_{\Sigma'} \, F^{\tau\tau^\prime}_{\Sigma'}
\nonumber \\ &
+ \qm^2 \Big [ R^{\tau\tau^\prime}_{\Phi'' M} \, F^{\tau\tau^\prime}_{\Phi'' M} 
+ R^{\tau\tau^\prime}_{\Sigma' \Delta} \,  F^{\tau\tau^\prime}_{\Sigma'\Delta} 
+ R^{\tau\tau^\prime}_{\Phi''} \, F^{\tau\tau^\prime}_{\Phi''}
+ R^{\tau\tau^\prime}_{\widetilde\Phi'}  \, F^{\tau\tau^\prime}_{\widetilde\Phi'}
+R^{\tau\tau^\prime}_{\Delta} \, F^{\tau\tau^\prime}_{\Delta} 
\Big ]  \bigg \},
\label{eq:HfiHfi_final}
\end{align}
\noindent where, matching the notation of~\cite{haxton2},
%\begin{subequations}
\begin{align}
R^{\tau\tau^\prime}_{M} &=  \overline{ \widetilde{\ell}^{\tau}_{M} \widetilde{\ell}^{\tau^\prime*}_{M} } ,\nonumber \\
R^{\tau\tau^\prime}_{\Sigma'} &= \frac{1}{2} \big( \delta_{ij} - \hat{q}_i\hat{q}_j \big)   \overline{  \widetilde{\ell}_{\Sigma i}^{\tau} \widetilde{\ell}_{\Sigma j}^{\tau^\prime*} } ,\nonumber \\
R^{\tau\tau^\prime}_{\Sigma''} &= \hat{q}_i \hat{q}_j  \overline{ \widetilde{\ell}_{\Sigma i}^{\tau} \widetilde{\ell}_{\Sigma j}^{\tau^\prime*} } , \nonumber\\
R^{\tau\tau^\prime}_{\Delta} &=\frac{1}{2} \big( \delta_{ij} - \hat{q}_i\hat{q}_j \big)   \overline{ \ell_{\Delta i}^{\tau} \ell_{\Delta j}^{\tau^\prime*} } ,\nonumber \\
R^{\tau\tau^\prime}_{\widetilde\Phi'} &= \frac{1}{2} \big( \delta_{ij} - \hat{q}_i\hat{q}_j \big)   \overline{  \ell_{\Phi i}^{\tau} \ell_{\Phi j} ^{\tau^\prime*} } ,\nonumber \\
R^{\tau\tau^\prime}_{\Phi''} &= \hat{q}_i \hat{q}_j  \overline{ \ell_{\Phi i}^{\tau} \ell_{\Phi j}^{\tau^\prime*} } , \nonumber\\
R^{\tau\tau^\prime}_{\Phi'' M} &=  \frac{2m_N}{q} \Im \big ( \hat{q}_i  \overline{ \ell_{\Phi i}^{\tau} \widetilde{\ell}_{M}^{\tau^\prime*} } \big ), \nonumber\\
R^{\tau\tau^\prime}_{\Sigma'\Delta} &= \frac{m_N}{q} \Im \big ( \epsilon_{ijk} \overline{ \widetilde{\ell}_{\Sigma' i}^{\tau} \ell_{\Delta j}^{\tau^\prime*} } \hat{q}_k \big ) .
\end{align}
%\end{subequations}

We now use Eqs.~(\ref{eq:tilde_ell}) and the
fact that the $\overline{\ell^{}_X \ell^*_Y}$ are functions of the
vector $\vec{q}$ only. Thus, for example, $\overline{ \ell^{\tau}_M
  \ell^{\tau'*}_{X,i} }$ is proportional to $\uvec{q}_i$,
\begin{align}
\overline{ \ell^{\tau}_M \ell^{\tau'*}_{X,i} } = L^{\tau\tau'}_{MX} \, \uvec{q}_i 
\end{align}
with coefficient given by
\begin{align}
L^{\tau\tau'}_{MX} = \uvec{q}_i \overline{ \ell^{\tau}_M \ell^{\tau'*}_{X,i} }.
\end{align}
On the other hand
 $\overline{ \ell^{\tau}_{X,i} \ell^{\tau'*}_{Y,j} }$ is the sum of a term in $\delta_{ij}-\uvec{q}_i\uvec{q}_j$, a term in $\uvec{q}_i \uvec{q}_j$, and a term in $\epsilon_{ijk} \uvec{q}_k$,
 \begin{align}
 \overline{ \ell^{\tau}_{X,i} \ell^{\tau'*}_{Y,j} } = L^{\perp\tau\tau'}_{XY} \, ( \delta_{ij}-\uvec{q}_i\uvec{q}_j ) + L^{||\tau\tau'}_{XY} \, \uvec{q}_i \uvec{q}_j + L^{\times\tau\tau'}_{XY} \, \epsilon_{ijk} \uvec{q}_k ,
 \end{align}
with respective coefficients given by
%\begin{subequations}
\begin{align}
L^{\perp\tau\tau'}_{XY} & = \tfrac{1}{2} (\delta_{ij}-\uvec{q}_i\uvec{q}_j)\overline{ \ell^{\tau}_{X,i} \ell^{\tau'*}_{Y,j} } ,\nonumber \\
L^{||\tau\tau'}_{XY} & = \uvec{q}_i\uvec{q}_j \, \overline{ \ell^{\tau}_{X,i} \ell^{\tau'*}_{Y,j} } , \nonumber\\
L^{\times\tau\tau'}_{XY} & = \tfrac{1}{2} \epsilon_{ijk} \uvec{q}_k \overline{ \ell^{\tau}_{X,i} \ell^{\tau'*}_{Y,j} } . 
\end{align}
%\end{subequations}

We can express the WIMP response functions $R^{\tau\tau^\prime}_{XY}$ in terms of the coefficients $L^{\tau\tau^\prime}_{XY}$. Writing $L^{\tau\tau'}_{XX} = L^{\tau\tau'}_{X}$ and introducing 
\begin{align}
L_M^{\tau\tau'} = \overline{ \ell_M^\tau \ell_M^{\tau^\prime*} }, \qquad L_\Omega^{\tau\tau'}= \overline{ \ell_\Omega^\tau \ell_\Omega^{\tau^\prime*} } ,
\end{align}
we obtain
%\begin{subequations}
\begin{align}
R^{\tau\tau^\prime}_{M} &= 
L^{\tau\tau'}_{M} +  (v{}^{\plus}_{\chi T})^2 \, L^{\perp\tau\tau^\prime}_{\Delta}  ,\nonumber \\
R^{\tau\tau^\prime}_{\Sigma'} &= L^{\perp\tau\tau'}_{\Sigma}+ \frac{1}{2} (v{}^{\plus}_{\chi T})^2 \, \big ( L^{\tau\tau'}_{\Omega}  + L^{||\tau\tau^\prime}_{\Phi}  \big ) , \nonumber\\
R^{\tau\tau^\prime}_{\Sigma''} &= L^{||\tau\tau'}_{\Sigma}  +  (v{}^{\plus}_{\chi T})^2 \, L^{\perp\tau\tau^\prime}_{\Phi} , \nonumber\\
R^{\tau\tau^\prime}_{\Delta} &=L^{\perp\tau\tau'}_{\Delta}  , \nonumber\\
R^{\tau\tau^\prime}_{\widetilde\Phi'} &= L^{\perp\tau\tau'}_{\Phi} , \nonumber\\
R^{\tau\tau^\prime}_{\Phi''} &= L^{||\tau\tau'}_{\Phi} , \nonumber\\
R^{\tau\tau^\prime}_{\Phi'' M} &=  \frac{2m_N}{q} \Im L^{\tau\tau'}_{\Phi M} ,\nonumber \\
R^{\tau\tau^\prime}_{\Sigma'\Delta} &= \frac{2m_N}{q} \Im L^{\times \tau\tau'}_{\Sigma \Delta}.
\label{eq:R_of_L}
\end{align}
%\end{subequations}

The last step is the calculation of the traces of the WIMP currents
contained in the coefficients $L^{\tau\tau'}_{XY}$. In
Section~\ref{sec:basis_operators} we chose to write the effective
Hamiltonian in terms of irreducible tensors $\myoverbracket{
  S_{i_1} \cdots S_{i_s} }$ of products of WIMP spin operators. As a
consequence, all the traces can be calculated by making use of the two
following master equations
\begin{align}
\label{eq:WIMP_spin_average_master_equation_a}
\frac{1}{2j_\chi+1} \tr 
& \Big (\myoverbracket{S_{i_1} \cdots S_{i_{s}}} 
\uvec{q}_{i_1} \cdots \uvec{q}_{i_s}
\myoverbracket{S_{j_1} \cdots S_{j_{s'}}} 
\uvec{q}_{j_1} \cdots \uvec{q}_{j_{s'}} 
\Big )
= 
\delta_{ss'} \, B_{j_\chi,s} ,
\intertext{and}
\frac{1}{2j_\chi+1} \tr 
& \Big (\myoverbracket{S_{i_1} \cdots S_{i_{s}}} 
\uvec{q}_{i_1} \cdots \uvec{q}_{i_{s-1}} a_{i_s} 
\myoverbracket{S_{j_1} \cdots S_{j_{s'}}} 
\uvec{q}_{j_1} \cdots \uvec{q}_{j_{s'-1}} b_{j_{s'}} 
\Big )
= \nonumber \\ & 
\Big [ \, \uvec{q}_i \uvec{q}_j 
+ \frac{s+1}{2s} ( \delta_{ij} - \uvec{q}_i \uvec{q}_j ) \Big ] a_i b_j  
\, \delta_{ss'} \, B_{j_\chi,s} 
\qquad (s \ge 1) .
\label{eq:WIMP_spin_average_master_equation_b}
\end{align}
Here
\begin{equation}
  B_{j_\chi,s}=\frac{s!}{(2s+1)!!}\frac{s!}{(2s-1)!!} \, K_{j_\chi,0} \cdots K_{j_\chi,s-1} 
%  B_{j_\chi,s}=\frac{1}{2s+1} \left( \frac{s!}{(2s-1)!!} \right)^2 \, K_{j_\chi,0} \cdots K_{j_\chi,s-1} 
  \label{eq:define_Bs} ,
\end{equation}
with
\begin{equation}
  K_{j_\chi,i}=j_\chi\left(j_\chi+1\right)-\frac{i}{2}\left(\frac{i}{2}+1\right).
\end{equation}
The first few values of $B_{j_\chi,s}$ are
\begin{align}
& B_{j_\chi,0}=1, \qquad B_{j_\chi,1}=\frac{j_\chi(j_\chi+1)}{3}, \qquad B_{j_\chi,2}=\frac{4}{45} j_\chi(j_\chi+1) \Big( j_\chi(j_\chi+1) - \frac{3}{4} \Big ), \nonumber \\ & B_{j_\chi,3}=\frac{4}{175} j_\chi(j_\chi+1) \Big( j_\chi(j_\chi+1) - \frac{3}{4} \Big ) \Big( j_\chi(j_\chi+1) - 2 \Big ) , \nonumber \\ &  B_{j_\chi,4}=\frac{64}{11025}  j_\chi(j_\chi+1) \Big( j_\chi(j_\chi+1) - \frac{3}{4} \Big ) \Big( j_\chi(j_\chi+1) - 2 \Big ) \Big( j_\chi(j_\chi+1) - \frac{15}{4} \Big ) .
\end{align}
\noindent A proof of the equations above is provided in
Appendix~\ref{sec:WIMP_spin_averages}.

Let us start with the scalar currents, which are readily obtained. For example,
\begin{align}
L^{\tau\tau'}_{M} &  = \frac{1}{2j_\chi+1} \tr \Big(  \ell^{\tau}_M \ell^{\tau'*}_{M} \big) 
\nonumber \\
& = \sum_{s=0}^{2j_\chi}\sum_{s'=0}^{2j_\chi}\frac{1}{2j_\chi+1} \tr \Big( 
\myoverbracket{ S_{i_1} \cdots S_{i_s} } \, \q_{i_1} \cdots \q_{i_s} \, c^{\tau}_{M,s,s}
\myoverbracket{ S_{j_1} \cdots S_{j_s} } \, \q_{j_1} \cdots \q_{j_s} \, c^{\tau'*}_{M,s,s}
\Big)
\nonumber \\
& = \sum_{s=0}^{2j_\chi} B_{\j_\chi,s} c^{\tau}_{M,s,s} \, c^{\tau'*}_{M,s,s} \,\q^{2s} .
\end{align}
And similarly
\begin{align}
%L^{\tau\tau'}_{M} & = \sum_{s=0}^{2j_\chi}  B_{j_\chi,s} c^{\tau}_{M,s,s} \, c^{\tau'*}_{M,s,s} \, \q^{2s} , 
%\\
L^{\tau\tau'}_{\Omega} & = \frac{1}{4} \sum_{s=0}^{2j_\chi}  B_{j_\chi,s} c^{\tau}_{\Omega,s,s} \, c^{\tau'*}_{\Omega,s,s} \, \q^{2s} .
\end{align}
The vector currents $2 \ell_{\Sigma,i}^\tau$, $\ell_{\Delta,i}^\tau$,
and $2 \ell_{\Phi,i}^\tau$ have similar expressions, and we give
details about the calculation of $L^{\tau\tau'}_{\Sigma}$ only. We
need
\begin{align}
\frac{1}{2j_\chi+1} \tr \Big(  
\myoverbracket{ S_{i_1} \cdots S_{i_s} } \, \q_{i_1} \cdots \q_{i_{s-1}} a^{\tau}_{\Sigma,i_si} 
\myoverbracket{ S_{j_1} \cdots S_{j_s} } \, \q_{j_1} \cdots \q_{j_{s-1}} a^{\tau'*}_{\Sigma,j_sj} 
\Big)
,
\end{align}
where
\begin{align}
a^{\tau}_{\Sigma,i_si} =
c^{\tau}_{\Sigma,s,s-1} \, \delta_{i_si} - i c^{\tau}_{\Sigma,s,s} \, \epsilon_{i_sik} \q_k - c^{\tau}_{\Sigma,s,s+1} \, \q_{i_s} \q_i .
\end{align}
Split $a^{\tau}_{\Sigma,i_si}$ into a part parallel to $\uvec{q}_{i_s}$ and a part perpendicular to $\uvec{q}_{i_s}$,
\begin{align}
a^{\tau}_{\Sigma,i_si}  = a^{||\tau}_{\Sigma,i} \uvec{q}_{i_s} + a^{\perp\tau}_{\Sigma,i_si},
\end{align}
where 
\begin{align} 
a^{||\tau}_{\Sigma,i} & = c^{||\tau}_{\Sigma,s}  \, \uvec{q}_i  ,
\nonumber \\
a^{\perp\tau}_{\Sigma,i_si} & = c^{\tau}_{\Sigma,s,s-1} \big( \delta_{i_si} - \uvec{q}_{i_s} \uvec{q}_i \big) -  i c^{\tau}_{\Sigma,s,s} \, \epsilon_{i_sij} \q_j ,
\end{align}
with
\begin{align}
c^{||\tau}_{\Sigma,s} &  = c^{\tau}_{\Sigma,s,s-1} - c^{\tau}_{\Sigma,s,s+1} \, \q^2.
\end{align}
Then
\begin{align}
& \frac{1}{2j_\chi+1} \tr \Big( 
\myoverbracket{ S_{i_1} \cdots S_{i_s} } \, \q_{i_1} \cdots \q_{i_{s-1}} a^{\tau}_{\Sigma,i_si} 
\myoverbracket{ S_{j_1} \cdots S_{j_s} } \, \q_{j_1} \cdots \q_{j_{s-1}} a^{\tau'*}_{\Sigma,j_sj} 
\Big)
\nonumber \\ & 
= B_{j_\chi,s} \, \q^{2s-2} \Big[ a^{||\tau}_{\Sigma,i} \, a^{||\tau'*}_{\Sigma,j} + \frac{s+1}{2s} \big( \delta_{mn} - \uvec{q}_{m} \uvec{q}_{n} \big) \, a^{\perp\tau}_{\Sigma,mi} \, a^{\perp\tau'*}_{\Sigma,nj} \Big] 
\nonumber \\ & 
= B_{j_\chi,s} \, \q^{2s-2} \Big[ \uvec{q}_i \uvec{q}_j \, c^{||\tau}_{\Sigma} c^{||\tau'*}_{\Sigma} + \frac{s+1}{2s} \big( \delta_{ij} - \uvec{q}_{i} \uvec{q}_{j} \big)  \, \big( c^{\tau}_{\Sigma,s,s-1} c^{\tau'*}_{\Sigma,s,s-1} + c^{\tau}_{\Sigma,s,s} c^{\tau'*}_{\Sigma,s,s} \, \q^2 \big )  \Big] .
\end{align}
Here we used 
\begin{align}
\big( \delta_{mn} - \uvec{q}_{m} \uvec{q}_{n} \big) \, a^{\perp\tau}_{\Sigma,mi} \, a^{\perp\tau'*}_{\Sigma,nj} 
& =
c^{\tau}_{\Sigma,s,s-1} c^{\tau'*}_{\Sigma,s,s-1} \, \big( \delta_{ij} - \uvec{q}_{i} \uvec{q}_{j} \big)
+ c^{\tau}_{\Sigma,s,s} c^{\tau'*}_{\Sigma,s,s} \, \epsilon_{mik} \q_k \,  \epsilon_{mjl} \q_l 
\nonumber \\ & =
\big( c^{\tau}_{\Sigma,s,s-1} c^{\tau'*}_{\Sigma,s,s-1} + c^{\tau}_{\Sigma,s,s} c^{\tau'*}_{\Sigma,s,s} \, \q^2 \big ) \, \big( \delta_{ij} - \uvec{q}_{i} \uvec{q}_{j} \big) .
\end{align}
Therefore,
\begin{align}
4 \overline{ \ell_{\Sigma,i}^{\tau} \ell_{\Sigma,j}^{\tau'*} } & = c^{\tau}_{\Sigma,0,1} \, c^{\tau'*}_{\Sigma,0,1}  \q_i \q_j + 
\sum_{s=1}^{2j_\chi}  B_{j_\chi,s} \, \q^{2s-2} \Big[ \uvec{q}_i \uvec{q}_j \, c^{||\tau}_{\Sigma} c^{||\tau'*}_{\Sigma} 
\nonumber \\ & \qquad \qquad + \frac{s+1}{2s} \big( \delta_{ij} - \uvec{q}_{i} \uvec{q}_{j} \big)  \, \big( c^{\tau}_{\Sigma,s,s-1} c^{\tau'*}_{\Sigma,s,s-1} + c^{\tau}_{\Sigma,s,s} c^{\tau'*}_{\Sigma,s,s} \, \q^2 \big )  \Big] .
\end{align}
Then
\begin{align}
L^{||\tau\tau'}_{\Sigma} & = \frac{1}{4} c^{\tau}_{\Sigma,0,1} \, c^{\tau'*}_{\Sigma,0,1} \q^2  + 
\frac{1}{4} \sum_{s=1}^{2j_\chi}  B_{j_\chi,s} \, \q^{2s-2} \big( c^{\tau}_{\Sigma,s,s-1} - c^{\tau}_{\Sigma,s,s+1} \, \q^2 \big)  \, \big( c^{\tau'*}_{\Sigma,s,s-1} - c^{\tau'*}_{\Sigma,s,s+1} \, \q^2 \big) ,
\\
L^{\perp\tau\tau'}_{\Sigma} & = 
\frac{1}{4} \sum_{s=1}^{2j_\chi}  B_{j_\chi,s} \, \frac{s+1}{2s}  \, \q^{2s-2 }\, \big( c^{\tau}_{\Sigma,s,s-1} c^{\tau'*}_{\Sigma,s,s-1} + c^{\tau}_{\Sigma,s,s} c^{\tau'*}_{\Sigma,s,s} \, \q^2 \big ) .
\end{align}
Similar calculations for the other vector currents give
\begin{align}
L^{\perp\tau\tau'}_{\Delta} & = 
\sum_{s=1}^{2j_\chi}  B_{j_\chi,s} \, \frac{s+1}{2s}  \, \q^{2s-2 }\, \big( c^{\tau}_{\Delta,s,s-1} c^{\tau'*}_{\Delta,s,s-1} + c^{\tau}_{\Delta,s,s} c^{\tau'*}_{\Delta,s,s} \, \q^2 \big ) ,
\\
L^{||\tau\tau'}_{\Phi} & = \frac{1}{4} c^{\tau}_{\Phi,0,1} \, c^{\tau'*}_{\Phi,0,1} \q^2  + 
\frac{1}{4} \sum_{s=1}^{2j_\chi}  B_{j_\chi,s} \, \q^{2s-2} \big( c^{\tau}_{\Phi,s,s-1} - c^{\tau}_{\Phi,s,s+1} \, \q^2 \big)  \, \big( c^{\tau'*}_{\Phi,s,s-1} - c^{\tau'*}_{\Phi,s,s+1} \, \q^2 \big) ,
\\
L^{\perp\tau\tau'}_{\Phi} & = 
\frac{1}{4} \sum_{s=1}^{2j_\chi}  B_{j_\chi,s} \, \frac{s+1}{2s}  \, \q^{2s-2 }\, \big( c^{\tau}_{\Phi,s,s-1} c^{\tau'*}_{\Phi,s,s-1} + c^{\tau}_{\Phi,s,s} c^{\tau'*}_{\Phi,s,s} \, \q^2 \big ) .
\end{align}
The quantities $L^{\tau\tau'}_{\Phi M}$ and $L^{\tau\tau'}_{\Sigma \Delta}$ are obtained as follows
\begin{align}
L^{\tau\tau'}_{\Phi M}  & = \uvec{q}_i \overline{\ell^{\tau}_{\Phi,i} \ell^{\tau'*}_{M} } \nonumber\\
& = \frac{i}{2} \Big [ c^{\tau}_{\Phi,0,1} \, c^{\tau'*}_{M,0,0} \, \q - \sum_{s=1}^{2j_\chi} \, B_{j_\chi,s} \,  \q^{2s-1} \, ( c^{\tau}_{\Phi,s,s-1} - c^{\tau}_{\Phi,s,s+1} \q^2 ) c^{\tau'*}_{M,s,s} \Big ]
,
\\
L^{\times\tau\tau'}_{\Sigma \Delta} & = \frac{1}{2} \epsilon_{ijk} \uvec{q}_k \overline{ \ell^{}_{\Sigma,i} \ell^*_{\Delta,j} } 
%\\
%& = \frac{1}{4} \epsilon_{ijk} \uvec{q}_k \Big [ \sum_{s=1}^{2j_\chi} i^{s-1} \, \myoverbracket{ S_{i_1} \cdots S_{i_s} } \, \q_{i_1} \cdots \q_{i_{s-1}} \Big ( 
%c^{\tau}_{\Sigma,s,s-1} \, ( \delta_{i_si}-\uvec{q}_{i_s} \uvec{q}_i ) - i c^{\tau}_{\Sigma,s,s} \, \epsilon_{i_sii'} \q_{i'}  \Big )
%\Big ] 
%\nonumber \\ & \qquad \qquad
%\Big [ \sum_{s=1}^{2j_\chi} i^{1-s} \, \myoverbracket{ S_{i_1} \cdots S_{i_s} } \, \q_{i_1} \cdots \q_{i_{s-1}} \Big ( 
%c^{\tau'*}_{\Delta,s,s-1} \, ( \delta_{i_sj}-\uvec{q}_{i_s} \uvec{q}_j ) + i c^{\tau'*}_{\Delta,s,s} \, \epsilon_{i_sjj'} \q_{j'} \Big )
%\Big ] 
%\\
%& = \frac{1}{4} \epsilon_{ijk} \uvec{q}_k \sum_{s=1}^{2j_\chi} \, B_{j_\chi,s} \, \q^{2s-2} \Big ( 
%c^{\tau}_{\Sigma,s,s-1} c^{\tau'*}_{\Delta,s,s-1} \, ( \delta_{ij}-\uvec{q}_{i} \uvec{q}_j ) + i c^{\tau}_{\Sigma,s,s-1} c^{\tau'*}_{\Delta,s,s} \, \epsilon_{ijj'} \q_{j'} 
%\nonumber \\ & \qquad\qquad\qquad
%- i c^{\tau}_{\Sigma,s,s} c^{\tau'*}_{\Delta,s,s-1} \, \epsilon_{jij'} \q_{i'} + c^{\tau}_{\Sigma,s,s} c^{\tau'*}_{\Delta,s,s} \q^2 ( \delta_{ij}-\uvec{q}_{i} \uvec{q}_j )
%\Big )
\nonumber\\
& = \frac{i}{2} \sum_{s=1}^{2j_\chi} B_{j_\chi,s} \, \frac{s+1}{2s} \, \q^{2s-1} \big ( c^{\tau}_{\Sigma,s,s-1} c^{\tau'*}_{\Delta,s,s} + c^{\tau}_{\Sigma,s,s} c^{\tau'*}_{\Delta,s,s-1} \big ) .
\end{align}

Finally, inserting the expressions for $L^{\tau\tau'}_{XY}$ into
\eqref{eq:R_of_L}, the explicit
expressions in the next subsection are obtained for the eight response functions $R^{\tau\tau'}_{XY}$
with $\curr{X}$=$M$, $\Phi^{\prime \prime}$, $\Phi^{\prime \prime} M$, $\tilde{\Phi}^\prime$,
$\Sigma^{\prime \prime}$, $\Sigma^\prime$, $\Delta$ and $\Delta
\Sigma^\prime$.

\subsection{Results}

The unpolarized differential cross section for WIMP-nucleus scattering is given by the expression (our $v^{\plus 2}_{\chi T} \equiv (\vec{v}^{\plus}_{\chi T})^2$ is equal to $v_T^{\perp 2}$ in the notation of~\cite{haxton1})
\be
\frac{d\sigma_T}{d E_R}=\frac{2 m_T}{4\pi v^2} \sum_{\tau=0,1}\sum_{\tau^{\prime}=0,1} \sum_X R_X^{\tau\tau^{\prime}}\big(v^{\plus 2}_{\chi T}, \qmsq\big) \widetilde{F}_{TX}^{\tau\tau^{\prime}}(\q)  ,
\ee
where the sum is over $X=M, \Phi^{\prime \prime}, \Phi^{\prime \prime} M, \tilde{\Phi}^\prime,
\Sigma^{\prime \prime},\Sigma^\prime,\Delta,\Delta\Sigma^\prime$. The functions $\widetilde{F}_{TX}^{\tau\tau^{\prime}}(\q)$ are given in terms of the nuclear response functions in Eq.~(\ref{eq:FXYAB}) and available in the literature by the expressions
\begin{align}
& \widetilde{F}_{TX}^{\tau\tau^{\prime}}(\q) = F_{X}^{\tau\tau^{\prime}}(\q) , && \text{for $X=M,\Sigma',\Sigma''$} , 
\nonumber \\
& \widetilde{F}_{TX}^{\tau\tau^{\prime}}(\q) = \q^2 \, F_{X}^{\tau\tau^{\prime}}(\q) ,&& \text{for $X=\Delta,\widetilde\Phi',\Phi'',\Sigma'\Delta,\Phi''M$} .
\end{align}
The functions $R^{\tau\tau'}_{k}\big(v^{\plus 2}_{\chi T}, \qmsq\big)$ are the WIMP response functions, given for WIMPs of any spin by
%\begin{subequations}
\begin{align}
R_{M}^{\tau \tau^\prime}&\big(v^{\plus 2}_{\chi T}, \qmsq\big) =
   v^{\plus 2}_{\chi T} \, R_{\Delta}^{\tau \tau^\prime}\big(v^{\plus 2}_{\chi T}, \qmsq\big) 
   + \sum_{s=0}^{2j_\chi} B_{j_\chi,s} c_{M,s,s}^\tau c_{M,s,s}^{\tau^\prime *}\qm^{2s} \nonumber
\\
R_{\Phi^{\prime \prime}}^{\tau \tau^\prime}&\big(v^{\plus 2}_{\chi T}, \qmsq\big) =
   \frac{1}{4}c_{\Phi,0,1}^{\tau} c_{\Phi,0,1}^{\tau^\prime *} \qm^{2} 
   \nonumber \\ &
   + \frac{1}{4}\sum_{s=1}^{2j_\chi} B_{j_\chi,s}  \qm^{2s-2} 
   \big ( c_{\Phi,s,s-1}^{\tau} - c_{\Phi,s,s+1}^{\tau} \qm^{2} \big )
   \big ( c_{\Phi,s,s-1}^{\tau^\prime *} - c_{\Phi,s,s+1}^{\tau^\prime *} \qm^{2} \big )
 \nonumber\\
 R_{\Phi^{\prime \prime} M}^{\tau \tau^\prime}&\big(v^{\plus 2}_{\chi T}, \qmsq\big) = 
   -  c_{\Phi,0,1}^{\tau} c_{M,0,0}^{\tau^\prime *} 
   + \sum_{s=1}^{2j_\chi} B_{j_\chi,s}  \qm^{2s-2}  
   \big ( c_{\Phi,s,s-1}^{\tau} - c_{\Phi,s,s+1}^{\tau} \qm^{2} \big ) c_{M,s,s}^{\tau^\prime *} ,
\nonumber\\
R_{\tilde{\Phi}^\prime}^{\tau \tau^\prime}&\big(v^{\plus 2}_{\chi T}, \qmsq\big) =
   \sum_{s=1}^{2j_\chi} B_{j_\chi,s} \frac{s+1}{8s}  \qm^{2s-2}\big (
   c_{\Phi,s,s-1}^{\tau} c_{\Phi,s,s-1}^{\tau^\prime *}
   + c_{\Phi,s,s}^{\tau} c_{\Phi,s,s}^{\tau^\prime *} \qm^{2}
  \big ) ,
\nonumber\\
R_{\Sigma^{\prime \prime}}^{\tau \tau^\prime}&\big(v^{\plus 2}_{\chi T}, \qmsq\big)  =
   v^{\plus 2}_{\chi T} \, R_{\tilde{\Phi}^\prime}^{\tau \tau^\prime}\left(v^{\plus 2}_{\chi T}, \qmsq\right) 
   + \frac{1}{4} c_{\Sigma,0,1}^{\tau} c_{\Sigma,0,1}^{\tau^\prime *} \qm^{2} 
   \nonumber \\ &
   + \sum_{s=1}^{2j_\chi} \frac{1}{4} B_{j_\chi,s}  \qm^{2s-2} 
   \big ( c_{\Sigma,s,s-1}^{\tau} - c_{\Sigma,s,s+1}^{\tau} \qm^{2} \big )
   \big ( c_{\Sigma,s,s-1}^{\tau^\prime *} - c_{\Sigma,s,s+1}^{\tau^\prime *} \qm^{2} \big )
   ,
\nonumber\\
R_{\Sigma^\prime}^{\tau \tau^\prime}&\big(v^{\plus 2}_{\chi T}, \qmsq\big)  =
   \frac{1}{2} \, v^{\plus 2}_{\chi T} \, R_{\Phi^{\prime \prime}}^{\tau \tau^\prime}\left(v^{\plus 2}_{\chi T}, \qmsq\right) 
   + \sum_{s=0}^{2j_\chi} \frac{1}{8} B_{j_\chi,s} \,
   c_{\Omega,s,s}^{\tau} c_{\Omega,s,s}^{\tau^\prime *}  v^{\plus 2}_{\chi T} \qm^{2s}
   \nonumber \\ & 
   + \sum_{s=1}^{2j_\chi}  \frac{1}{8} B_{j_\chi,s}  \frac{s+1}{s} \qm^{2s-2} 
   \big (c_{\Sigma,s,s-1}^{\tau} c_{\Sigma,s,s-1}^{\tau^\prime *} + c_{\Sigma,s,s}^{\tau} c_{\Sigma,s,s}^{\tau^\prime *} \qm^{2}  \big ) ,
\nonumber\\      
R_{\Delta}^{\tau \tau^\prime}&\big(v^{\plus 2}_{\chi T}, \qmsq\big) =
    \sum_{s=1}^{2j_\chi} 
    B_{j_\chi,s} \frac{s+1}{2s}  \qm^{2s-2} \big (
    c_{\Delta,s,s-1}^{\tau} c_{\Delta,s,s-1}^{\tau^\prime *}
    + c_{\Delta,s,s}^{\tau} c_{\Delta,s,s}^{\tau^\prime *} \qm^{2}
    \big ) ,
\nonumber \\
R_{\Delta \Sigma^\prime}^{\tau \tau^\prime}&\big(v^{\plus 2}_{\chi T}, \qmsq\big) =
    -\sum_{s=1}^{2j_\chi} B_{j_\chi,s} \frac{s+1}{2s} \qm^{2s-2} \big (
    c_{\Delta,s,s}^{\tau} c_{\Sigma,s,s-1}^{\tau^\prime *} 
    + c_{\Delta,s,s-1}^{\tau} c_{\Sigma,s,s}^{\tau^\prime *} 
    \big ),
    \label{eq:wimp_response_functions_sent}
\end{align}
%\end{subequations}
\noindent We recall that
\begin{align}
v^{\plus 2}_{\chi T}  = v_{\chi T}^2 - \frac{q^2}{4 \mu_{\chi T}^2}
\end{align}
\noindent (see Eq.~(\ref{eq:v_chi_t}) with $\vec{q} \cdot \vec{v}{}^{\plus}_{\chi T}=0$) and
\begin{equation}
  B_{j_\chi,s}=\frac{s!}{(2s+1)!!}\frac{s!}{(2s-1)!!} \, K_{j_\chi,0} \cdots K_{j_\chi,s-1} 
\end{equation}
with
\begin{equation}
  K_{j_\chi,i}=j_\chi\left(j_\chi+1\right)-\frac{i}{2}\left(\frac{i}{2}+1\right)
\end{equation}
(see Eq.~\ref{eq:define_Bs}). 
The equations above are valid for a WIMP of arbitrary spin $j_{\chi}$
and are the main result of the present paper. In particular, the
adoption of the irreducible tensors in
Eq.~(\ref{eq:irreducible_products_of_S}) implies that for a given value of $s$=$2
j_\chi$ a different set of WIMP response functions $R_{X}^{\tau
  \tau^\prime}$ arises for each set of the operators $\calO_{X,s,l}$
introduced in Section~\ref{sec:basis_operators}. For a WIMP of spin
$j_\chi$ all the operators $\calO_{X,s,l}$ with $s\le 2 j_\chi$
contribute to the cross section.

\section{Discussion}
\label{sec:discussion}
In this Section we discuss some of the consequences of the results obtained in
the previous Sections.

\subsection{The case of spin 1}
\label{sec:spin1}

In Section~\ref{sec:basis_operators} we expressed the WIMP--nucleon
interaction Hamiltonian operators in terms of tensors irreducible
under the rotation group.  The case $j_\chi =1$ has already been
discussed in the literature in terms of reducible
operators~\cite{krauss_spin_1, catena_krauss_spin_1}, so it is
instructive to compare the two approaches.

The authors of Ref.~\cite{krauss_spin_1} introduce a symbol
$\mathcal{S}$ in expressions of the kind $\vec{a} \cdot \mathcal{S}
\cdot \vec{b}$, where $\vec{a}$ and $\vec{b}$ are vectors (see, e.g.,
their Eq.~(4)). They call it the symmetric combination of
polarization vectors $\epsilon_i$. In their Appendix they give the
expression
\begin{align}
S_{ij} = \frac{1}{2} \left( \epsilon_i^\dagger \epsilon_j + \epsilon_j^\dagger \epsilon_i \right) .
\end{align}
%One way to read these equations is that for any two vectors $\vec{a}$ and $\vec{b}$ one has
%\begin{align}
%a \cdot \mathcal{S} \cdot b = a_i S_{ij} b_j = \frac{1}{2} a_i b_j \left( \epsilon_i^\dagger \epsilon_j + \epsilon_j^\dagger \epsilon_i \right) .
%\end{align}
%

We want to identify the symbol $\mathcal{S}$ with an operator $\op{\mathcal{S}}$ in WIMP spin space (in this section we keep the hat over WIMP spin operators). We find the definitions of $\mathcal{S}$
and $S_{ij}$ as operators in Ref.~\cite{krauss_spin_1} a little obscure. We
interpret them as definitions in a particular basis, and then
translate them to basis-independent definition in terms of the WIMP spin
operators $S_i$ (where $i=1,2,3$). In particular, we identify the
quantities $\epsilon_i^s$ in~\cite{krauss_spin_1} with the components
of the WIMP spin eigenstate $|1,s\rangle$ in the linear polarization
basis $|e_i\rangle$, i.e.,
\begin{align}
\epsilon_i^s = \langle e_i | 1,s\rangle .
\end{align} 
As standard, the linear polarization states in the $x$, $y$, and $z$ directions $|e_i\rangle$ (with $i=1,2,3$) are given in terms of the angular momentum eigenstates $|1,m\rangle$ (with $m=+1,0,-1$) by
%\begin{subequations}
\begin{align}
& | e_1 \rangle = - \frac{1}{\sqrt{2}} \, |1,1\rangle + \frac{1}{\sqrt{2}} \, |1,-1\rangle ,
\nonumber\\
& | e_2 \rangle = \frac{i}{\sqrt{2}} \, |1,1\rangle + \frac{i}{\sqrt{2}} \, |1,-1\rangle ,
\nonumber\\
& | e_3 \rangle = |1,0\rangle .
\end{align}
%\end{subequations}
Notice that $\langle e_i | e_j \rangle = \delta_{ij}$. The
coefficients in the definition of the states $|e_i\rangle$ are the
same as in the expressions of the Cartesian unit vectors $e_x$, $e_y$,
$e_z$ in terms of the spherical basis vectors
$e_{+1}=(-e_x-ie_y)/\sqrt{2}$, $e_0=e_z$, and
$e_{-1}=(e_x-ie_y)/\sqrt{2}$.

The matrix elements of the spin matrices $\op{S}_k$ (with $k=1,2,3$) in the $|e_i\rangle$ and $|1,s\rangle$ bases are respectively
\begin{align}
& \langle e_i | \op{S}_k | e_j \rangle = i \epsilon_{ikj},
\\
& \langle 1,s'| \op{S}_k | 1, s \rangle = \langle 1,s'| e_i \rangle \, \langle e_i |S_k | e_j \rangle \, \langle e_j| 1, s \rangle = i \epsilon_{ikj} \epsilon_i^{s'*} \epsilon_j^s .
\end{align}
The latter expression matches the formula $iS_k = \epsilon_{ijk}
\epsilon_i^\dagger \epsilon_j$ after Eq.~(B4) in~\cite{krauss_spin_1}
if it is interpreted as $i \op{S}_k = \epsilon_{ijk} | e_j \rangle \,
\langle e_i |$, i.e., if the following identifications are made: $\epsilon_i \to |e_i\rangle$ and
$\epsilon_i^\dagger \to \langle e_i |$.  This motivates our
interpretation of the definition of $S_{ij}$ in the Appendix of
Ref.~\cite{krauss_spin_1}, namely $ S_{ij} = \frac{1}{2} (
\epsilon_i^\dagger \epsilon_j + \epsilon_j^\dagger \epsilon_i )$, as
\begin{align}
\op{S}_{ij} = \frac{1}{2} \big( |e_j \rangle \, \langle e_i |  + |e_i \rangle \, \langle e_j |  \big) .
\label{eq:def_Sij}
\end{align}

Our goal is to write the operator $\op{S}_{ij}$ so identified in terms of products of the spin operators $\op{S}_i$ (where $i=1,2,3$). In the $|e_i\rangle$ basis, from Eq.~(\ref{eq:def_Sij}),
\begin{align}
\langle e_m | \op{S}_{ij} | e_n \rangle = \frac{1}{2} \big( \delta_{in}\delta_{jm} + \delta_{jn} \delta_{im} \big) .
\end{align}
Also,
\begin{align}
\langle e_m | \op{S}_i \op{S}_j |e _n\rangle = i \epsilon_{mik} \, i \epsilon_{kjn} = \delta_{ij} \delta_{mn} - \delta_{in} \delta_{jm} .
\end{align}
Therefore
\begin{align}
\langle e_m | ( \op{S}_i \op{S}_j + \op{S}_j \op{S}_i ) |e _n\rangle & = 2 \delta_{ij} \delta_{mn} - (\delta_{in}\delta_{jm} + \delta_{jn} \delta_{im} ) 
\nonumber \\ & = 2 \, \langle e_m | \delta_{ij} \op{1} - \op{S}_{ij} | e_n \rangle .
\end{align}
Hence
\begin{align}
\op{S}_{ij} & = \delta_{ij} \op{1} - \frac{1}{2} ( \op{S}_i \op{S}_j + \op{S}_j \op{S}_i ) .
\label{eq:Sij}
\end{align}
Using the symmetrization symbol $\{..\}$ and $j_\chi=1$ in the relation
\begin{equation}
\myoverbracket{\op{S}_i\op{S}_j}=\{\op{S}_i \op{S}_j\}-\frac{j_{\chi}(j_\chi+1)}{3}\delta_{ij} \op{1},
\end{equation}
 $\op{S}_{ij}$ can also be written as
\begin{align}
\op{S}_{ij} = \delta_{ij} \op{1} - \{ \op{S}_i \op{S}_j \} 
 = \frac{1}{3} \delta_{ij} \op{1} - \myoverbracket{\op{S}_i\op{S}_j} .
\label{eq:s_operator}
\end{align}

The substitutions $\op{S}_{ij} \rightarrow
-\myoverbracket{\op{S}_i\op{S}_j}+\frac{1}{3}\delta_{ij}\op{1}$ and
$\vec{\tilde{q}}\rightarrow -\vec{\tilde{q}}$ produce the
relations in Table~\ref{tab:Haxton_operators} between the spin--1
operators $\calO_{17,\ldots,20}$ and the operators $\calO_{X,s,l}$
introduced in Section~\ref{sec:basis_operators}.

Similarly, we find the definition of the ${\cal S}_{ij}$ in Ref.~\cite{catena_krauss_spin_1} as operator also a little confusing. The definition in their equation (3.4) is consistent with the operator $\op{S}_{ij}$ that we identify in Eq.~(\ref{eq:Sij}) if their equation (3.4) is interpreted as the transition amplitude of the operator $\op{S}_{ij}$ between initial and final helicity eigenstates. Let the initial and final helicity eigenstates for a spin-1 particle be
\begin{align}
| h, s \rangle, \qquad | h', s' \rangle,
\end{align}
respectively. We identify the quantities $e_{si}$ and $e'_{s'i}$ in~\cite{catena_krauss_spin_1} with
\begin{align}
e_{si} =  \langle e_i |h, s \rangle ,
\qquad
e'_{s'i} = \langle h', s' | e_i \rangle .
\end{align}
Then from Eq.~\eqref{eq:def_Sij} we have
\begin{align}
\langle h',s' | \op{S}_{ij} | h,s \rangle = \frac{1}{2} \big( e_{si} e'_{s'j}  + e_{sj} e'_{s'i}  \big) ,
\end{align}
which equals ${\cal S}_{ij}^{s's}$ in~\cite{catena_krauss_spin_1} and reproduces their equation (3.4).

This clarifies that the symbols ${\cal S}$ in
Dent et al.~\cite{krauss_spin_1} and Catena et
al.~\cite{catena_krauss_spin_1} can be identified with the operators
\begin{align}
\op{S}_{ij} & = \delta_{ij} \op{1} - \frac{1}{2} ( \op{S}_i \op{S}_j + \op{S}_j \op{S}_i ) .
\end{align}

We now show that the additional operators $\calO_{21,\ldots,24}$ of order $q$ introduced in~\cite{catena_krauss_spin_1} are not independent in the one-nucleon approximation. These operators do not
arise from the non--relativistic limit of a high energy amplitude. They are obtained 
by combining ${\cal S}$ in rotationally invariant combinations with
$\vec{S}_N$, $\vec{q}$ and $\vec{v}{}^{\plus}_{\chi N}$. Consider for
example the operator $\calO_{21}=\vec{v}{}^{\plus}_{\chi N}\cdot{\cal S}\cdot
\vec{S}_N$. Using Eq.~(\ref{eq:s_operator}) one obtains
\begin{equation}
{\cal O}_{21}=-\myoverbracket{S_iS_j}v^{\plus}_{\chi N,i} S_{N,j}+\frac{1}{3}\vec{v}^{\plus}_{\chi N}\cdot
\vec{S}_N.  
\label{eq:O21_1}
%=\frac{1}{3}{\cal O}_7=\frac{1}{3}{\cal O}_{\Omega,0,0},
\end{equation}
Since $\myoverbracket{S_iS_j}v^{\plus}_{\chi N,i} S_{N,j}=\myoverbracket{S_iS_j}\myoverbracket{v^{\plus}_{\chi N,i}S_{N,j}}$,
and in one--nucleon approximation
$\myoverbracket{v^{\plus}_{\chi N,i}S_{N,j}}$ does not contribute to the
scattering process, i.e., it is not included among the currents in
Eq.~(\ref{eq:five_operators}), the first term in the right hand side of Eq.~(\ref{eq:O21_1}) vanishes. Thus in the one-nucleon--scattering approximation,
\begin{equation}
{\cal O}_{21}=\frac{1}{3}\vec{v}^{\plus}_{\chi N}\cdot
\vec{S}_N = \frac{1}{3}{\cal O}_7=\frac{1}{3}{\cal O}_{\Omega,0,0}.
\label{eq:O21}
\end{equation}

In general any interaction term
depending on $\vec{v}^{\plus}_{\chi N}$ and $\vec{S}_N$ must be projected onto the
currents of Eq.~(\ref{eq:five_operators}) using the decomposition
\begin{equation}
v{}^{\plus}_{\chi N,i} S_{N,j}=\frac{1}{3}\delta_{ij} \left (\vec{v}^{\plus}_{\chi N}\cdot
\vec{S}_N \right )+ \frac{1}{2} \epsilon_{ijk}\left (\vec{v}^{\plus}_{\chi N}
\times \vec{S}_N \right )_k+\myoverbracket{v^{\plus}_{\chi N,i} S_{N,j}}.
 \label{eq:dyad_decomposition}
\end{equation}
In this way, for the additional operators ${\cal
  O}_{22,\ldots,24}$ defined in~\cite{catena_krauss_spin_1}, we
obtain
%
%\begin{subequations}
\begin{eqnarray}
  {\cal O}_{22}&=&\left (i\frac{\vec{q}}{m_N}\times\vec{v}{}^{\plus}_{\chi N}\right )\cdot{\cal S}\cdot \vec{S}_N=-{\cal O}_{\Phi,2,1}-\frac{1}{3}{\cal O}_{\Phi,0,1},\nonumber\\
  {\cal O}_{23}&=&i\frac{\vec{q}}{m_N}\cdot{\cal S}\cdot\left (\vec{S}_N\times\vec{v}{}^{\plus}_{\chi N} \right )=-{\cal O}_{\Phi,2,1}+\frac{1}{3}{\cal O}_{\Phi,0,1}={\cal O}_{22}-\frac{2}{3}{\cal O}_{3},\nonumber\\
  {\cal O}_{24}&=&\vec{v}{}^{\plus}_{\chi N}\cdot{\cal S}\cdot\left (\vec{S}_N\times i\frac{\vec{q}}{m_N}\right)=-{\cal O}_{\Phi,2,1}-\frac{1}{3}{\cal O}_{\Phi,0,1}={\cal O}_{22}.
  \end{eqnarray}
%  \end{subequations}
%
We conclude that in one--nucleon--scattering approximation,
${\cal O}_{22}$ and ${\cal O}_{24}$ correspond to the same operator, while ${\cal O}_{23}$ is a linear combination of ${\cal O}_{22}$ and ${\cal O}_{3}$.

\subsection{The counting of independent operators}
\label{sec:op_counting}

The procedure outlined in Section~\ref{sec:basis_operators} consists
in coupling one of the five nucleon currents of
Eq.~(\ref{eq:five_operators}) to WIMP currents ordered according to the
rank of the irreducible operators $\myoverbracket{S_{i_s}
  \cdots S_{i_s}}$ ($s=0,1,2,...$). The power $l$ of the transferred
momentum $q$ descends from rotational invariance.  For elastic
WIMP-nucleus scattering, it is $l=s$ for the scalar nucleon operators
$\calO_M$ and $\calO_\Omega$, $l=s,s\pm 1$ for the vector operators
$\calO_\Sigma$ and $\calO_\Phi$, and $l=s,s-1$ for the vector operator
$\calO_\Delta$. Taking this into account, we can count the number of
basis WIMP-nucleon operators as follows. For $s=0$, there are two
operators $\calO_{M,0,0}$ and $\calO_{\Omega,0,0}$, and three
operators $\calO_{\Sigma,0,1}$, $\calO_{\Phi,0,1}$ and
$\calO_{\Delta,0,1}$ (with the exception that for elastic scattering
$\calO_{\Delta,0,1}$ vanishes and is not counted). Thus for $s=0$
there is a total of five operators (four for elastic scattering). For
$s>0$, Eqs.~(\ref{eq:basis_WIMP_nucleon_operators}) show that at a
fixed value of $s$ there is one operator for each scalar nucleon
current ($\calO_{X,s,s}$ for $X=M,\Omega$) and there are three
operators for each vector nucleon current ($\calO_{X,s,s-1}$,
$\calO_{X,s,s}$, $\calO_{X,s,s+1}$ for $X=\Sigma, \Delta,\Phi$, with
the exception that for elastic scattering $\calO_{\Delta,s,s+1}$
vanishes). This implies that each value of $s>0$ contributes
$2+3\times3=11$ new operators (10 for elastic scattering).  Since $s$
ranges from 0 to $2j_\chi$, the total number of independent operators
for a WIMP of spin $j_\chi$ is $4+10\times 2j_\chi=4 + 20 j_\chi$ for
elastic scattering ($5+11\times2j_\chi=5+22 j_\chi$ for inelastic
scattering). If we restrict the counting to operators that are independent of the WIMP-nucleon relative velocity, we keep only $X=M,\Sigma$, and find that at $s=0$ there are two operators and that each $s>0$ contributes 4 operators (one with $X=M$ and three with $X=\Sigma$). This gives a total of $2+8j_\chi$ velocity-independent basis operators. The number of linearly-independent operators for WIMPs of spin 0, 1/2, 1, 3/2, and 2 are collected in Table~\ref{tab:number_of_operators}.

{ % begin number of independent operators
\begin{table}[t]\centering
\caption{Number of linearly-independent operators in the one-nucleon approximation.}
\label{tab:number_of_operators}
\renewcommand{\arraystretch}{1.5}
\addtolength{\tabcolsep}{2.0pt}
\vskip\baselineskip
\begin{tabular}{@{}rrrr@{}}
\toprule
WIMP spin & Elastic scattering & Inelastic scattering & Velocity-independent \\
\midrule
$0$ & 4 & 5 & 2 \\
$\tfrac{1}{2}$ & 14 & 16 & 10 \\
$1$ & 24 & 27 & 18 \\
$\tfrac{3}{2}$ & 34 & 38 & 26 \\
$2$ & 44 & 49 & 34 \\
\bottomrule
\end{tabular}
\end{table}
} % end number of independent operators

The number of operators introduced so far in the literature for WIMP spin 
$j_{\chi}\le1$ is 24,  as shown in Table~\ref{tab:Haxton_operators}.
This number coincides with our counting of 24 basis operators for elastic scattering of WIMPs of spin
$j_{\chi}\le1$. This is only a coincidence. The total number of independent operators that have appeared in the literature so far is actually 19, as 1 of those in Table~\ref{tab:Haxton_operators} is of order $v^2$ (namely, $\calO_2$) and 4 are  linearly dependent on the other 19 (namely, $\calO_{16}$, $\calO_{21}$, and two among $\calO_{22}$, $\calO_{23}$, and $\calO_{24}$). The 5 linearly-independent operators that have so far been missing in the literature for $j_{\chi}\le$1 are 
\begin{align}
\calO_{\Omega,2,2},
\quad
\calO_{\Sigma,2,3}, 
\quad 
\calO_{\Delta,2,2},
\quad
\calO_{\Phi,2,2},
\quad
\calO_{\Phi,2,3}
\label{eq:additional_operators}
\end{align}
(see their absence from Table~\ref{tab:Haxton_operators} and their presence in Table~\ref{tab:spin1}). In addition, for inelastic scattering, one should add the linearly independent operators,
\begin{align}
\calO_{\Delta,0,1} ,
\quad
\calO_{\Delta,1,2} ,
\quad
\calO_{\Delta,2,3} .
\end{align}
Ref.~\cite{haxton2} introduced 14
independent operators for $j_{\chi}\le$ 1/2, in agreement to our
counting for elastic scattering: the 16 operators $\calO_{1,\ldots,16}$, minus the two operators $\calO_2$ and
$\calO_{16}$, the former being quadratic in $v{}^{\plus}$ and the
latter being a linear combination of $\calO_{12}$ and $\calO_{15}$.
Ref.~\cite{krauss_spin_1} introduced two additional operators for $j_\chi=1$,
$\calO_{17}$ and $\calO_{18}$, accounting for 16 of the 24 independent operators for $j_\chi=1$. Ref.~\cite{catena_krauss_spin_1} introduced six additional operators $\calO_{19,\ldots,24}$, but only three of them are linearly independent, bringing the number of independent operators for $j_\chi=1$ to 19 out of 24. Our addition of the operators in Eq.~\eqref{eq:additional_operators} completes the 24 linearly independent operators for elastic scattering of WIMPs of spin $j_\chi=1$.

% Taking this into account, at linear order in
%$\vec{v}^\perp$ there are $2+2\times3+2=10$ new operators for each
%value of $s$, with the exception of $s=0$, for which only the four
%operators ${\cal O}_{M,0,0}$, ${\cal O}_{\Omega,0,0}$, ${\cal
%  O}_{\Sigma,0,1}$ and ${\cal O}_{\Phi,0,1}$ appear. This ordering
%method nicely explains why 14 operators have been discussed for a spin
%1/2 particle in the literature~\cite{haxton2}, and exhausts all the
%possible operators that drive WIMP--nucleus scattering in the
%approximation of one--nucleon currents.

\subsection{Examples of differential scattering rates}
\label{sec:diff_rate}

\begin{figure}
\begin{center}
  \includegraphics[width=0.7\textwidth]{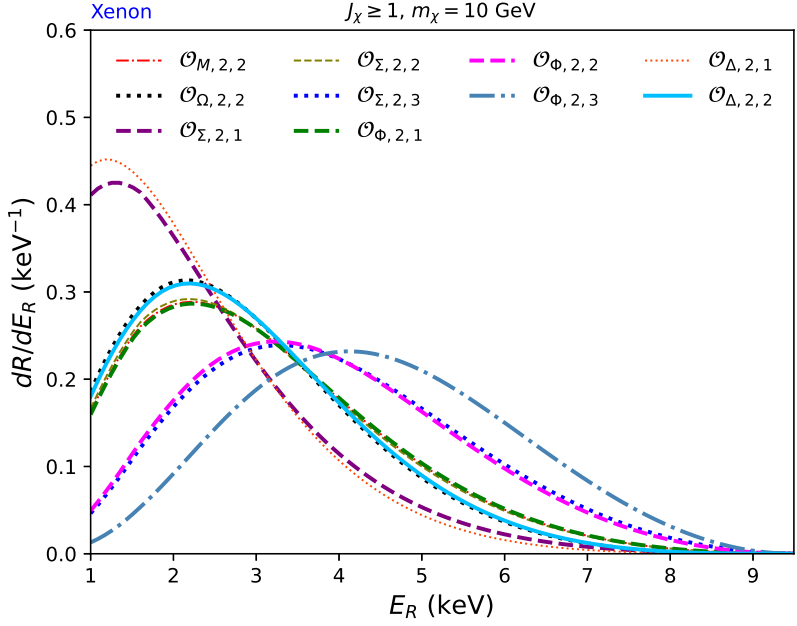}
\end{center}
\caption{Expected differential scattering rate (\ref{eq:dr_der})
  normalized to 1 event for a 10 GeV mass WIMP with a xenon target and
  for the 10 irreducible effective operators $\calO_{X,2,l}$ defined
  in Eqs.~(\ref{eq:basis_WIMP_nucleon_operators}), assuming a WIMP of
  spin $j_\chi\ge$1.
\label{fig:diff_rate_xe_1}}
\end{figure}

\begin{figure}
\begin{center}
  \includegraphics[width=0.7\textwidth]{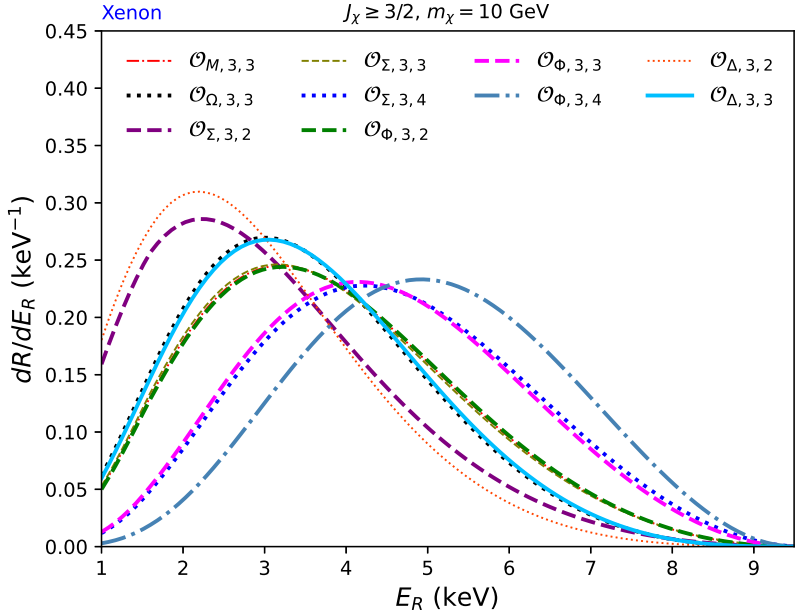}   
\end{center}
\caption{Same as Fig.~\ref{fig:diff_rate_xe_1} but for the 10
  irreducible effective operators $\calO_{X,3,l}$ and assuming a WIMP
  of spin $j_\chi\ge$3/2.
\label{fig:diff_rate_xe_3_2}}
\end{figure}

\begin{figure}
\begin{center}
  \includegraphics[width=0.7\textwidth]{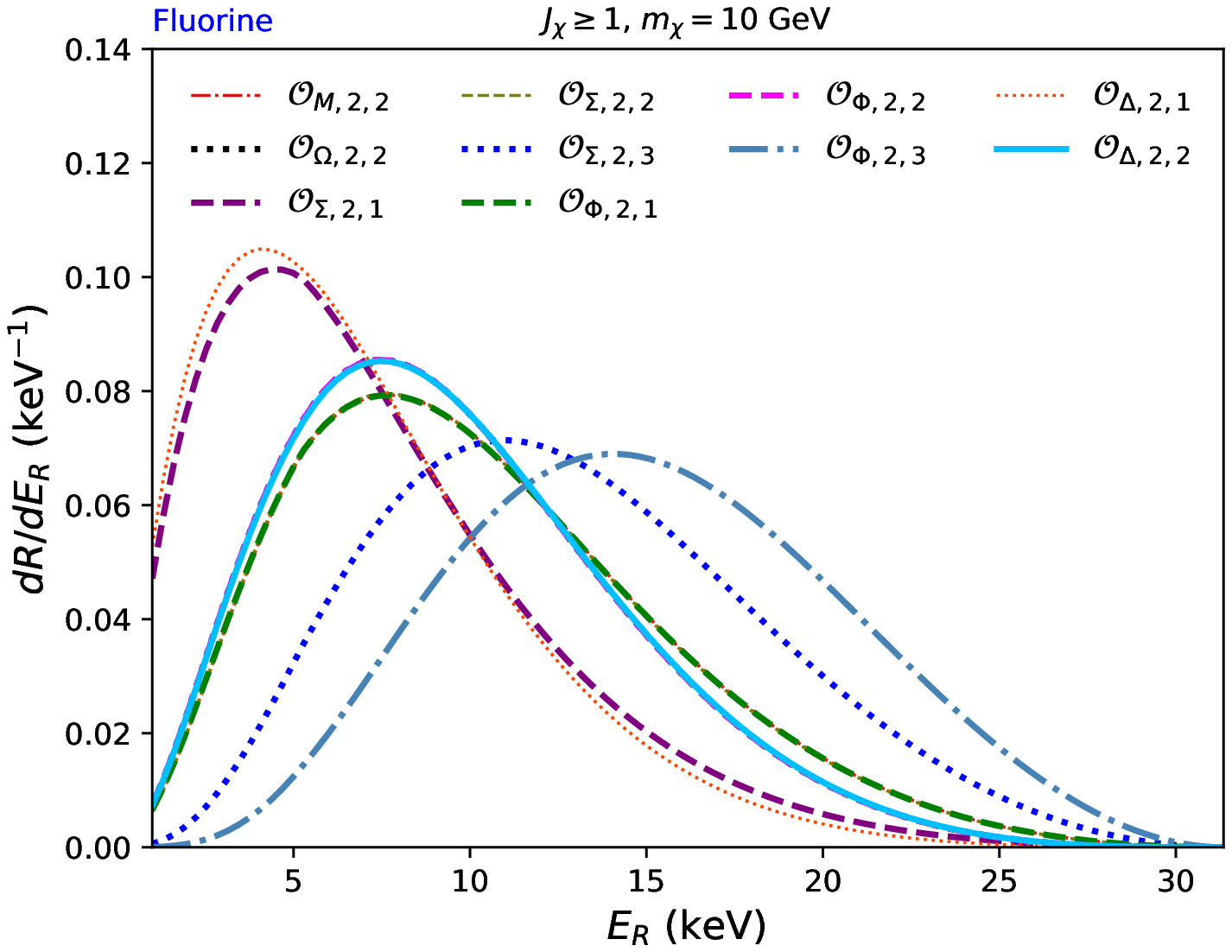}   
\end{center}
\caption{Same as Fig.~\ref{fig:diff_rate_xe_1} but for a fluorine target.
\label{fig:diff_rate_f_1}}
\end{figure}

\begin{figure}
\begin{center}
  \includegraphics[width=0.7\textwidth]{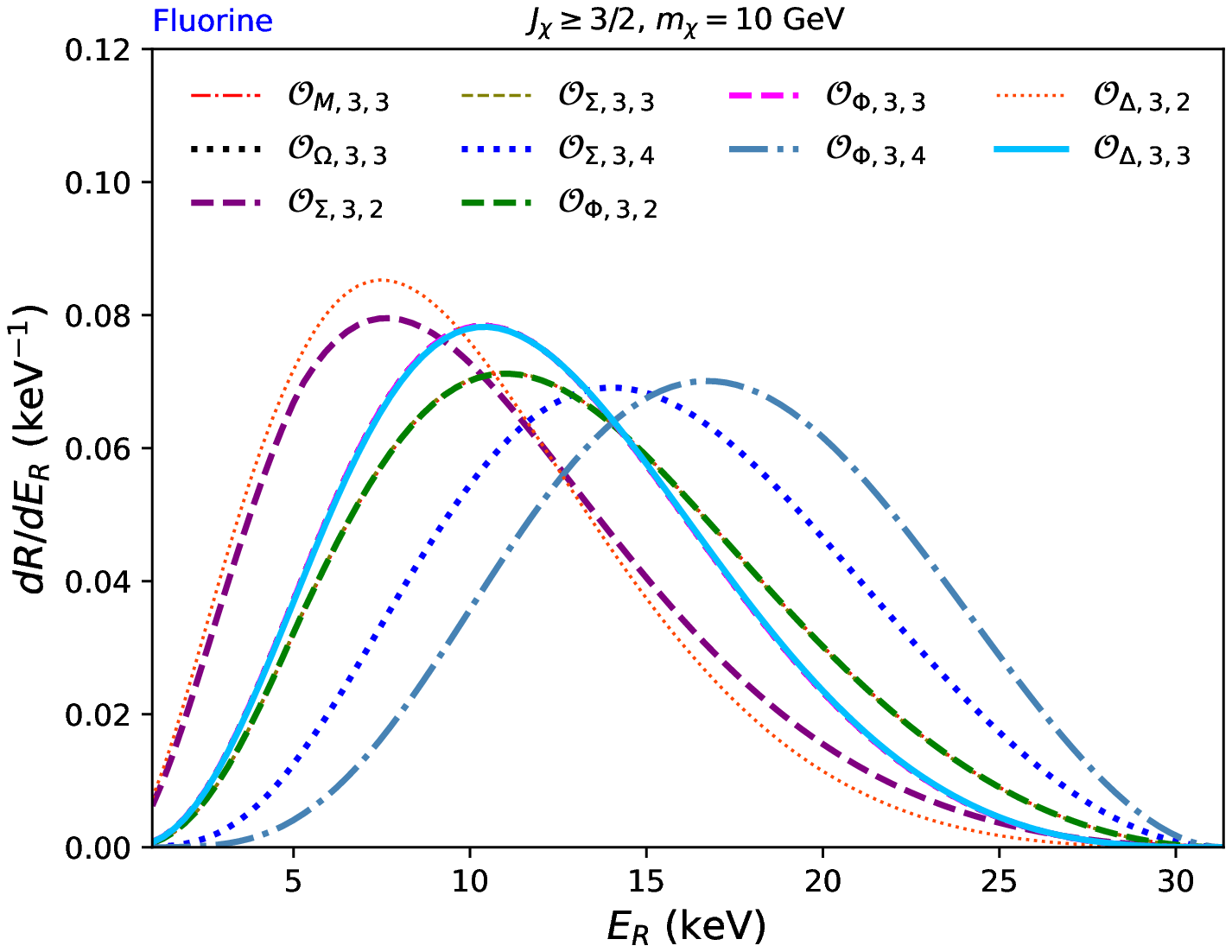}   
\end{center}
\caption{Same as in Fig.~\ref{fig:diff_rate_xe_3_2} but for a fluorine target.
\label{fig:diff_rate_f_3_2}}
\end{figure}

In Figs.~\ref{fig:diff_rate_xe_1}--\ref{fig:diff_rate_f_3_2} we
provide a few examples of the expected spectrum of the differential
rate in Eq~(\ref{eq:dr_der}) as driven by some of the irreducible
effective operators introduced in
Eqs.~(\ref{eq:basis_WIMP_nucleon_operators}). In particular
Fig.~\ref{fig:diff_rate_xe_1} shows the differential rate for a 10 GeV
mass WIMP on xenon and for the 10 irreducible effective operators
$\calO_{X,2,l}$ that arise for a WIMP with $j_\chi\ge$1. 
Fig.~\ref{fig:diff_rate_xe_3_2} shows the differential rate for the
operators $\calO_{X,3,l}$ arising for a WIMP with
$j_\chi\ge$3/2. Figs.~\ref{fig:diff_rate_f_1} and
\ref{fig:diff_rate_f_3_2} show the analogous cases for a fluorine
nuclear target. All the spectra are normalized to 1 event. For the
WIMP velocity distribution $f(v)$, a truncated Maxwellian with escape
velocity 550 km/s and rms velocity 270 km/s in the Galactic rest frame
is adopted.  In these plots one can observe how the spectra shift to
larger recoil energies $E_R$ for growing $j_\chi$ due to the correlation
between $E_R$ and the power of $q/m_N$ in the squared
amplitude. Such correlation implies also a suppression of the
contribution of higher--rank operators  compared to lower--rank operators
when their couplings are of the same order of magnitude. It must be remarked that from the point of view of a
non--relativistic effective theory, one cannot rule out the
possibility that the scattering rate of a WIMP with
spin $j_\chi$ is driven by one of the higher--rank operators. We expect
this to lead to non--standard phenomenological consequences.

%%%%%%%%%%%%%%%%%%%%%%%%%%%%%%%%%%%%%%%%%%%%%%%%%%%%%%%
%%% 
%%%%%%%%%%%%%%%%%%%%%%%%%%%%%%%%%%%%%%%%%%%%%%%%%%%%%%%

%%%%%%%%%%%%%%%%%%%%%%%%%%%%%%%%%%%%%%%%%%%%%%%%%%%%%%%
%%% 
%%%%%%%%%%%%%%%%%%%%%%%%%%%%%%%%%%%%%%%%%%%%%%%%%%%%%%%

%%%%%%%%%%%%%%%%%%%%%%%%%%%%%%%%%%%%%%%%%%%%%%%%%%%%%%%
%%% 
%%%%%%%%%%%%%%%%%%%%%%%%%%%%%%%%%%%%%%%%%%%%%%%%%%%%%%%

%%%%%%%%%%%%%%%%%%%%%%%%%%%%%%%%%%%%%%%%%%%%%%%%%%%%%%%
%%% 
%%%%%%%%%%%%%%%%%%%%%%%%%%%%%%%%%%%%%%%%%%%%%%%%%%%%%%%

\section{Conclusions}
\label{sec:conclusions}

In the present paper we have introduced a systematic approach that, in
the one--nucleon approximation, describes the most
general non-relativistic WIMP--nucleus interaction allowed by Galilean
invariance for a WIMP of arbitrary spin.  The resulting squared
scattering amplitudes depend on the WIMP response functions of
Eqs.~(\ref{eq:wimp_response_functions_sent}),
which are the main result of our paper, and on the same nuclear
response functions as for WIMPs of spin $\le$ 1.  Many nuclear
response functions are available in the literature for most of the
targets used in WIMP direct detection
experiments~\cite{haxton2,catena}.

In particular, we have expressed the WIMP--nucleon interaction
Hamiltonian operators in terms of tensors irreducible under the
rotation group.  This has several advantages:
\begin{itemize}
  \item  it includes all the operators allowed by symmetry,
including those that do not arise as the low--energy limit of standard
point--like particle interactions with spin $\le$1 mediators;
\item it avoids double counting, allowing to show that some of the
  operators introduced in the literature for the spin-1 WIMP case are
  not independent (see Section~\ref{sec:spin1});
\item it greatly simplifies the calculation of the cross section, that
  was obtained from the two master equations~(\ref{eq:WIMP_spin_average_master_equation_a}-\ref{eq:WIMP_spin_average_master_equation_b}) for the traces of WIMP spin operators;
\item for a given WIMP spin $j_\chi$ the scattering cross section is given by
  a sum of cleanly separated contributions from irreducible operators of 
  ranks 0, 1, 2, 3, \ldots, up to $2 j_\chi$, without interference terms (since irreducible operators of
  different rank do not interfere).
\end{itemize}

All the Wilson coefficients $c_{X,s,l}$ are defined up to arbitrary
functions of the transferred momentum $q^2$.  Moreover, as shown in
Table~\ref{tab:Haxton_operators}, in some cases the change of basis
from reducible to irreducible operators involves momentum--dependent
coefficients.

From the phenomenological point of view, contributions from
irreducible operators of higher rank are shifted to larger recoil
energies compared with contributions from operators of lower rank. It
may happen that lower rank operators vanish and the WIMP scattering
rate is dominated by a higher rank operator. We expect this to lead to
non--standard phenomenological consequences.

\section*{Acknowledgements}
The research of S.K, S.S and G.T. was supported by the National
Research Foundation of Korea(NRF) funded by the Ministry of Education
through the Center for Quantum Space Time (CQUeST) with grant number
2020R1A6A1A03047877 and by the Ministry of Science and ICT with grant
number 2019R1F1A1052231. GT is also supported by a TUM University
Foundation Fellowship. The work of P.G. has been partially supported
by NSF award PHY-1720282 at the University of Utah.

\appendix

\section{Multipole expansion of a vector plane wave}
\label{sec:multipole_expansion}

It is well known that a plane wave $e^{i \vec{q} \cdot \vec{r}}$ can be expanded into spherical harmonics according to the equation
\begin{align}
e^{i \vec{q} \cdot \vec{r}} = \sum_{L=0}^{\infty} \sum_{M=-L}^{L} 4 \pi i^L j_{L}(qr) Y_{LM}^*(\uvec{q}) Y_{LM}(\uvec{r}).
\label{eq:eiqr}
\end{align}
Here $j_L(qr)$ is the spherical Bessel function of order $L$.
Among equivalent forms of this expansion, Eq.~(\ref{eq:eiqr}) is a vector equation valid in any coordinate system that shows the explicit separate dependence on the unit vectors $\uvec{q}$ and $\uvec{r}$.

For a vector plane wave $\vec{\ell} \, e^{i \vec{q} \cdot \vec{r}}$,
it is not hard to find in the literature expressions of its expansion
into vector spherical harmonics in specific coordinate systems, for
the most part with the $z$ axis chosen along the direction of the
vector $\vec{q}$. Here we establish the following vector relation
valid in all coordinate systems, showing the explicit separate
dependence on the unit vectors $\uvec{q}$ and $\uvec{r}$ in analogy to
Eq.~(\ref{eq:eiqr}).
\begin{align}
e^{i \vec{q} \cdot \vec{r}} \, \vec{j}(\vec{r}) =  \sum_{J=0}^{\infty} \sum_{M=-J}^{J} 4 \pi i^{J} \bigg[ 
&  - i  \, \vec{Y}_{JM}^{(\mpL)*}(\uvec{q}) \,\, \bigg ( \frac{1}{q} \frac{\partial M_{JM}(q,\vec{r})}{\partial \vec{r}} \bigg ) \cdot  \vec{j}(\vec{r}) \nonumber \\
&  + i \, \vec{Y}_{JM}^{(\mpTE)*}(\uvec{q}) \,\, \bigg ( \! -\frac{i}{q} \frac{\partial}{\partial \vec{r}} \times  \vec{M}^{M}_{JJ}(q,\vec{r}) \bigg ) \cdot  \vec{j}(\vec{r}) \nonumber \\
&  + i \, \vec{Y}_{JM}^{(\mpTM)*}(\uvec{q}) \,\, \vec{M}^{M}_{JJ}(q,\vec{r}) \cdot \vec{j}(\vec{r})
 \bigg].
\label{eq:jeiqr}
\end{align}
Here L, TE, and TM stand for longitudinal, transverse electric and transverse magnetic, respectively; the $\mpTE$ and $\mpTM$ terms start at $J=1$;
\begin{align}
& M_{JM}(q,\vec{r}) = j_J(qr) Y_{JM}(\uvec{r}) , \\
& \vec{M}^{M}_{JL}(q,\vec{r}) = j_J(qr) \vec{Y}_{JLM}(\uvec{r}) .
\end{align}
Moreover,
\begin{align}
\label{eq:jeiqrB1}
\vec{Y}_{JM}^{(\mpL)}(\uvec{q}) & = \sqrt{\frac{J}{2J+1}} \vec{Y}_{J\,J-1\,M}(\uvec{q}) - \sqrt{\frac{J+1}{2J+1}} \vec{Y}_{J\,J+1\,M}(\uvec{q})  = \uvec{q} \, Y_{JM}(\uvec{q}) , \\
\vec{Y}_{JM}^{(\mpTE)}(\uvec{q}) & = \sqrt{\frac{J+1}{2J+1}} \vec{Y}_{J\,J-1\,M}(\uvec{q}) + \sqrt{\frac{J}{2J+1}} \vec{Y}_{J\,J+1\,M}(\uvec{q}) = \frac{q}{\sqrt{J(J+1)}} \frac{\partial Y_{JM}(\uvec{q}) }{\partial \vec{q}}, \\
\vec{Y}_{JM}^{(\mpTM)}(\uvec{q}) & = i \, \vec{Y}_{JJM}(\uvec{q}) = \uvec{q} \times \vec{Y}_{JM}^{(\mpTE)}(\uvec{q})  ,
\label{eq:jeiqrB3}
\end{align}
are the longitudinal, transverse electric and transverse magnetic spherical harmonics, defined in terms of the vector spherical harmonics
\begin{align}
  & \vec{Y}_{JLM}(\uvec{q}) = \sum_{\alpha=-L}^{L} \sum_{\beta=-1}^{1} C^{JM}_{L\alpha1\beta} \, Y_{L\alpha}(\uvec{q}) \, \uvec{e}_{\beta} ,
  \label{eq:vector_spherical}
\end{align}
where $C^{JM}_{L\alpha1\beta} $ is the Clebsch-Gordan coefficient for coupling angular momenta $L\alpha$ and $1\beta$ into $JM$, and 
$ \uvec{e}_{\beta} $  is the standard spherical basis 
\begin{align}
\uvec{e}_{+1} = - (\uvec{x} + i \uvec{y})/\sqrt{2} , \qquad \uvec{e}_0 = \uvec{z} , \qquad \uvec{e}_{-1} = (\uvec{x} - i \uvec{y})/\sqrt{2} .
\end{align}

Eq.~(\ref{eq:jeiqr}) can be obtained as follows. Write
\begin{align}
\vec{\ell} \cdot \vec{j} \,  e^{i \vec{q} \cdot \vec{r}} =  \sum_{JM} \bigg[ 
c^{(\mpL)}_{JM} \, \frac{1}{q} \frac{\partial M_{JM}(q,\vec{r})}{\partial \vec{r}}
- i c^{(\mpTE)}_{JM}  \, \frac{1}{q} \frac{\partial}{\partial \vec{r}} \times  \vec{M}^{M}_{JJ}(q,\vec{r}) 
+ c^{(\mpTM)}_{JM} \,   \vec{M}^{M}_{JJ}(q,\vec{r}) 
 \bigg]  \cdot \vec{j} .
\label{eq:jeiqrD1}
\end{align}
By using the relations
\begin{align}
&  \frac{\partial M_{JM}(q,\vec{r})}{\partial \vec{r}}  = q \bigg( \sqrt{\frac{J}{2J+1}} \vec{M}^{M}_{J\,J-1} + \sqrt{\frac{J+1}{2J+1}} \vec{M}^{M}_{J\,J+1} \bigg) , \\
\intertext{and}
& \frac{\partial}{\partial \vec{r}} \times  \vec{M}^{M}_{JJ}(q,\vec{r}) = i q \bigg( \sqrt{\frac{J+1}{2J+1}} \vec{M}^{M}_{J\,J-1} - \sqrt{\frac{J}{2J+1}} \vec{M}^{M}_{J\,J+1} \bigg) ,
\end{align}
express Eq.~(\ref{eq:jeiqrD1}) in the form
\begin{align}
\vec{\ell}  \cdot \vec{j} \,  e^{i \vec{q} \cdot \vec{r}} =  \sum_{JLM} c_{JLM} \, j_{L}(qr) \, \vec{Y}_{JLM}(\uvec{r})  \cdot \vec{j} ,
\end{align}
where $L=J-1,J,J+1$ and
\begin{align}
& c^{(\mpL)}_{JM} =  \sqrt{\frac{J}{2J+1}}c_{J\,J-1\,M} + \sqrt{\frac{J+1}{2J+1}} c_{J\,J+1\,M} , \\
& c^{(\mpTE)}_{JM} =  \sqrt{\frac{J+1}{2J+1}}c_{J\,J-1\,M} + \sqrt{\frac{J}{2J+1}} c_{J\,J+1\,M} , \\
& c^{(\mpTM)}_{JM} =  c_{JJM} . 
\end{align}
The orthogonality relation for the vector spherical harmonics,
\begin{align}
\int \vec{Y}^{*}_{JLM}(\uvec{r}) \cdot \vec{Y}_{J'L'M'}(\uvec{r}) \, d\Omega_r = \delta_{JJ'} \delta_{LL'} \delta_{MM'},
\end{align}
then gives
\begin{align}
c_{JLM} = 4 \pi i^L \vec{\ell} \cdot \vec{Y}^{*}_{JLM}(\uvec{q}) . 
\end{align}
Relations (\ref{eq:jeiqrB1}--\ref{eq:jeiqrB3}) finally lead to Eq.~(\ref{eq:jeiqr}).

\section{Sum/average over nuclear spins}
\label{app:sum_nuclear_spins}
We provide here the details leading to
Eq.~(\ref{eq:HfiHfi_groundstate}).  The matrix element of the
effective WIMP--nucleon Hamiltonian between an initial nuclear state
$|J_iM_i\rangle$ and a final nuclear state $\langle J_fM_f|$ can be
expanded into multipoles using
Eq.~\eqref{eq:effective_hamiltonian_multipole_expansion}. Then the
Wigner-Eckart theorem can be applied to the matrix element of each
nuclear current multipole operator $ \op{\curr{X}}^{\tau}_{JM}(q)$ in
the right-hand side of Eqs.~(\ref{eq:HJM}),
\begin{align}
\langle J_fM_f| \op{\curr{X}}^{\tau}_{JM}(q) |J_iM_i\rangle = \frac{1}{\sqrt{2J_f+1}} \,\, C^{J_fM_f}_{J_iM_iJM} \,\, \langle J_f || \op{\curr{X}}^{\tau}_{J}(q) || J_i\rangle ,
\end{align}
where $C^{J_fM_f}_{J_iM_iJM}$ is the Clebsch-Gordan coefficient coupling angular momenta $J_iM_i$ and $JM$ into angular momentum $J_fM_f$, and $ \langle J_f || \op{\curr{X}}^{\tau}_{J}(q) || J_i\rangle $ is the reduced matrix element of the nuclear multipole operator. One then computes the sum/average over nuclear spins
\begin{align}
\overline{H_\rmfi^* H_\rmfi^{}} \equiv \frac{1}{2J_i+1} \sum_{M_iM_f} H_\rmfi^* H_\rmfi
\end{align}
using
\begin{align}
\sum_{M_iM_f} C^{J_fM_f}_{J_iM_iJM} C^{J_fM_f}_{J_iM_iJ'M'} = \frac{2J_f+1}{2J+1} \delta_{JJ'} \delta_{MM'} 
\end{align}
and Eqs.~\eqref{eq:sumoverM2a}--\eqref{eq:sumoverM2b}. 
One obtains
\begin{align}
\overline{H_\rmfi^* H_\rmfi^{}}  = \frac{4\pi}{2J_i+1} \sum_{J} \bigg[ & H_J^* H_J + \frac{1}{2} \big( \delta_{ij} - \uvec{q}_i \uvec{q}_j \big ) \Big ( H_{J,i}^{(\mpTE)*} H_{J,j}^{(\mpTE)} + H_{J,i}^{(\mpTM)*} H_{J,j}^{(\mpTM)} \Big ) \nonumber \\ & -  \uvec{q} \cdot \Re \! \Big ( \vec{H}_{J}^{(\mpTM)*} \times \vec{H}_{J}^{(\mpTE)} \Big ) \bigg ] .
\label{eq:Hfi_Hfi_multipoles}
\end{align}
Here
\begin{align}
\label{eq:HJ_a}
& H_{J} = \sum_{\tau} \bigg( \widetilde{\ell}_M^\tau \, M_{J}^\tau  - i \vec{\widetilde{\ell}}{}_\Sigma^\tau \cdot \uvec{q} \, \Sigma_{J}^{\prime\prime\,\tau}  - \frac{q}{m_N} \vec{\ell}_{\Delta}^\tau \cdot \uvec{q} \, {\widetilde\Delta}_{J}^{\prime\prime\,\tau} - \frac{iq}{m_N} \vec{\ell}_{\Phi}^\tau \cdot \uvec{q} \, \Phi_{J}^{\prime\prime\,\tau} - \frac{iq}{m_N} \ell_{\Omega}^\tau \, \Omega_{J}^\tau \bigg ) , \\
& \vec{H}^{(\mpTE)}_{J} = \sum_{\tau} \bigg( - i \vec{\widetilde{\ell}}{}_\Sigma^\tau\, \Sigma_{J}^{\prime\,\tau} + \frac{q}{m_N} \vec{\ell}_{\Delta}^\tau\, \Delta_{J}^{\prime\,\tau} - \frac{iq}{m_N} \vec{\ell}_\Phi {\widetilde\Phi}_{J}^{\prime\,\tau} \bigg ) , \\
& \vec{H}^{(\mpTM)}_{J} = \sum_{\tau} \bigg( i \vec{\widetilde{\ell}}{}_\Sigma^\tau \, \Sigma_{J}^{\tau} + \frac{q}{m_N} \vec{\ell}_{\Delta}^\tau\, \Delta_{J}^{\tau} + \frac{iq}{m_N} \vec{\ell}_\Phi \Phi_{J}^{\tau} \bigg ) ,
\label{eq:HJ_b}
\end{align}
with
\begin{align}
X_{J}^\tau(q)  = \langle J_f || \op{X}_{J}^\tau || J_i \rangle .
\end{align}

According to Table~\ref{tab:nucleon_current_parities}, the nuclear
matrix elements that do not vanish in the nucleus ground state are
\begin{align}
M^{\tau}_{J}(q) , \Phi^{\prime\,\tau}_{J}(q)  , \Phi^{\prime\prime\,\tau}_{J}(q) ,  \text{for $J$ even;} \qquad
\Delta^{\tau}_{J}(q) , \Sigma^{\prime\,\tau}_{J}(q) , \Sigma^{\prime\prime\,\tau}_{J}(q) ,\text{for $J$ odd.}
\end{align}
This gives
\begin{align}
H_J & = \begin{cases}
\widetilde{\ell}_{M}^\tau \, M_{J}^\tau - \frac{iq}{m_N} \vec{\ell}_\Phi^\tau \cdot \uvec{q} \, \Phi_{J}^{\prime\prime\,\tau} , & \text{for $J$ even,} \\[0.5ex]
- i \vec{\widetilde{\ell}}{}_\Sigma^\tau \cdot \uvec{q} \, \Sigma_{J}^{\prime\prime\,\tau} , & \text{for $J$ odd,} 
\end{cases}
\\
\vec{H}_{JM}^{(\mpTE)} & = \begin{cases}
- \frac{iq}{m_N} \vec{\ell}_\Phi^\tau\, {\widetilde\Phi}_{J}^{\prime\,\tau}  , & \text{for $J$ even,} \\[0.5ex]
- i \vec{\widetilde{\ell}}{}_\Sigma^\tau \, \Sigma_{J}^{\prime\,\tau} , & \text{for $J$ odd,} 
\end{cases}
\\
\vec{H}_{JM}^{(\mpTM)} & = \begin{cases}
0  , & \text{for $J$ even,} \\
\frac{q}{m_N} \vec{\ell}_\Delta^\tau \, \Delta_{J}^{\tau} , & \text{for $J$ odd.} 
\end{cases}
\end{align}

Substituting the latter equations into
Eq.~\eqref{eq:Hfi_Hfi_multipoles} gives
Eq.~(\ref{eq:HfiHfi_groundstate}) in terms of the nuclear response
functions defined in~(\ref{eq:FXYAB}), with $F^{\tau\tau^\prime}_{XY}$
= $F^{\tau\tau^\prime}_{M}$, $F^{\tau\tau^\prime}_{\Sigma'}$,
$F^{\tau\tau^\prime}_{\Sigma''}$, $F^{\tau\tau^\prime}_{\Delta}$,
$F^{\tau\tau^\prime}_{\Phi''}$, $F^{\tau\tau^\prime}_{\tilde{\Psi}'}$,
$F^{\tau\tau^\prime}_{M\Phi''}$ and
$F^{\tau\tau^\prime}_{\Delta\Sigma'}$.

\section{One-nucleon multipole operators}
\label{sec:one_nucleon_multiple_operators}

Here we list the one-nucleon multipole operators defined, for
instance, in~\cite{deforest_walecka, donnelly_walecka, walecka}.  In
the position-space representation, with $\vec{r}$ the position vector
and $\vec{\sigma}$ the Pauli spin matrices, they are given by:
%\begin{subequations}
\begin{align}
%\label{eq:basis_free_nucleon_multipole_operators_a}
\op{M}_{JM}(q,\vec{r}) & = j_J(qr) \, Y_{JM}(\uvec{r}) , \nonumber\\
\op{\Delta}_{JM}(q,\vec{r}) & = \vec{M}_{JJ}^{M}(q,\vec{r}) \cdot \frac{1}{q} \frac{\partial}{\partial\vec{r}} , \nonumber\\
\op{\Delta}'_{JM}(q,\vec{r}) & = - i \bigg( \frac{1}{q} \frac{\partial}{\partial\vec{r}} \times \vec{M}_{JJ}^{M}(q,\vec{r}) \bigg) \cdot \frac{1}{q} \frac{\partial}{\partial\vec{r}} , \nonumber\\
\op{\Delta}''_{JM}(q,\vec{r}) & = \bigg( \frac{1}{q} \frac{\partial M_{JM}(q,\vec{r})}{\partial\vec{r}} \bigg) \cdot \frac{1}{q} \frac{\partial}{\partial\vec{r}} , \nonumber\\
\op{\Sigma}_{JM}(q,\vec{r}) & = \vec{M}_{JJ}^{M}(q,\vec{r}) \cdot \vec{\sigma} , \nonumber\\
\op{\Sigma}'_{JM}(q,\vec{r}) & = - i \bigg( \frac{1}{q} \frac{\partial}{\partial\vec{r}} \times \vec{M}_{JJ}^{M}(q,\vec{r}) \bigg) \cdot \vec{\sigma} , \nonumber \\
\op{\Sigma}''_{JM}(q,\vec{r}) & = \bigg( \frac{1}{q} \frac{\partial M_{JM}(q,\vec{r})}{\partial\vec{r}} \bigg) \cdot \vec{\sigma} , \nonumber\\
\op{\Phi}_{JM}(q,\vec{r}) & = i \vec{M}_{JJ}^{M}(q,\vec{r}) \cdot \bigg( \vec{\sigma} \times  \frac{1}{q} \frac{\partial}{\partial\vec{r}} \bigg) , \nonumber\\
\op{\Phi}'_{JM}(q,\vec{r}) & = \bigg( \frac{1}{q} \frac{\partial}{\partial\vec{r}} \times \vec{M}_{JJ}^{M}(q,\vec{r}) \bigg) \times \bigg( \vec{\sigma} \times  \frac{1}{q} \frac{\partial}{\partial\vec{r}} \bigg) ,\nonumber \\
\op{\Phi}''_{JM}(q,\vec{r}) & = i \bigg( \frac{1}{q} \frac{\partial M_{JM}(q,\vec{r})}{\partial\vec{r}} \bigg) \cdot \bigg( \vec{\sigma} \times  \frac{1}{q} \frac{\partial}{\partial\vec{r}} \bigg), \nonumber\\
\op{\Omega}_{JM}(q,\vec{r}) & = M_{JM}(q,\vec{r}) \vec{\sigma} \cdot  \frac{1}{q} \frac{\partial}{\partial\vec{r}} \, . \end{align}
%\end{subequations}
Moreover, the following definitions are given in~\cite{haxton1} as implementation of Eq.~(\ref{eq:j_sym_tilde}),
\begin{align}
\op{\widetilde{\Delta}}{}^{\prime\prime}_{JM}(q,\vec{r}) & = \op{\Delta}''_{JM}(q,\vec{r}) - \frac{1}{2} \op{M}_{JM}(q,\vec{r}) ,\nonumber \\
\op{\widetilde{\Phi}}{}^{\prime}_{JM}(q,\vec{r}) & = \op{\Phi}'_{JM}(q,\vec{r}) + \frac{1}{2} \op{\Sigma}_{JM}(q,\vec{r}) , \nonumber\\
\op{\widetilde{\Phi}}{}_{JM}(q,\vec{r}) & = \op{\Phi}_{JM}(q,\vec{r}) - \frac{1}{2} \op{\Sigma}'_{JM}(q,\vec{r}) , \nonumber\\
\op{\widetilde{\Omega}}_{JM}(q,\vec{r}) & = \op{\Omega}_{JM}(q,\vec{r}) + \frac{1}{2} \op{\Sigma}''_{JM}(q,\vec{r}) \, .
\label{eq:basis_free_nucleon_multipole_operators_b}
\end{align}
The one-nucleon operators appearing in Eqs.~\eqref{eq:nucleon_current_multipole_expansion_a}--\eqref{eq:nucleon_current_multipole_expansion_c} are then
\begin{align}
\op{X}_{JM}^\tau(q) = \sum_N \op{X}_{JM}(\vec{q},\vec{r}_N) \, t^\tau_N,
\label{eq:one_nucleon_multipoles}
\end{align}
where $X=M,\Delta,\Delta',\Delta'',\Sigma,\Sigma',\Sigma'',\Phi,\Phi',\Phi'',\Omega,\widetilde\Delta,\widetilde\Phi,\widetilde\Omega$.

\section{Some mathematical identities}

\subsection{Sums of products of spherical harmonics over magnetic quantum number}

In the sum/average over nuclear spins, one needs expressions for the sum over $M$ of products of scalar and vector spherical harmonics. The simplest one is for the case of the product of two scalar spherical harmonics. It is
\begin{align}
\sum_{M=-J}^{J} Y_{JM}(\uvec{r}) \, Y^{*}_{JM}(\uvec{r}) = \frac{2J+1}{4\pi} .
\label{eq:sumoverM1}
\end{align}
Here we prove the following equations, written in dyadic notation (namely $\vec{a} \vec{b} \equiv a_i b_j $), with $\uvec{r}$, $\uvec{\uptheta}$, $\uvec{\upphi}$ equal to the unit coordinate vectors in spherical coordinates $(r,\theta,\phi)$.
\begin{align}
\label{eq:sumoverM2a}
& \sum_{M=-J}^{J} \vec{Y}^{(\mpTE)}_{JM}(\uvec{r}) \, Y^{*}_{JM}(\uvec{r}) 
= \sum_{M=-J}^{J} \vec{Y}^{(\mpTM)}_{JM}(\uvec{r}) \, Y^{*}_{JM}(\uvec{r}) 
=  0 , 
\\
& \sum_{M=-J}^{J} \vec{Y}^{(\mpTE)}_{JM}(\uvec{r}) \, \vec{Y}^{(\mpTE)*}_{JM}(\uvec{r}) 
= \sum_{M=-J}^{J} \vec{Y}^{(\mpTM)}_{JM}(\uvec{r}) \, \vec{Y}^{(\mpTM)*}_{JM}(\uvec{r}) 
= \frac{2J+1}{4\pi} \, \frac{1}{2} \left( \uvec{\uptheta} \, \uvec{\uptheta} + \uvec{\upphi} \, \uvec{\upphi} \right)  , 
\\
& \sum_{M=-J}^{J} \vec{Y}^{(\mpTM)}_{JM}(\uvec{r}) \, \vec{Y}^{(\mpTE)*}_{JM}(\uvec{r}) = \frac{2J+1}{4\pi} \, \frac{1}{2} \left(  \uvec{\upphi} \, \uvec{\uptheta}  - \uvec{\uptheta} \, \uvec{\upphi} \right)  .
\label{eq:sumoverM2b}
\end{align}
%\begin{align}
%\label{eq:sumoverM2a}
%\sum_{M=-J}^{J} \vec{Y}^{(\mpL)}_{JM}(\uvec{r}) \, Y^{*}_{JM}(\uvec{r}) & = \frac{2J+1}{4\pi} \, \uvec{r}  , \\
%\sum_{M=-J}^{J} \vec{Y}^{(\mpTE)}_{JM}(\uvec{r}) \, Y^{*}_{JM}(\uvec{r}) & = \sum_{M=-J}^{J} \vec{Y}^{(\mpTM)}_{JM}(\uvec{r}) \, Y^{*}_{JM}(\uvec{r}) =  0 , \\
%\sum_{M=-J}^{J} \vec{Y}^{(\mpL)}_{JM}(\uvec{r}) \, \vec{Y}^{(\mpL)*}_{JM}(\uvec{r}) & = \frac{2J+1}{4\pi} \, \uvec{r} \, \uvec{r} , \\
%\sum_{M=-J}^{J} \vec{Y}^{(\mpTE)}_{JM}(\uvec{r}) \, \vec{Y}^{(\mpTE)*}_{JM}(\uvec{r}) & = \frac{2J+1}{4\pi} \, \frac{1}{2} \left( \uvec{\uptheta} \, \uvec{\uptheta} + \uvec{\upphi} \, \uvec{\upphi} \right)  , \\
%\sum_{M=-J}^{J} \vec{Y}^{(\mpTM)}_{JM}(\uvec{r}) \, \vec{Y}^{(\mpTM)*}_{JM}(\uvec{r}) & = \frac{2J+1}{4\pi} \, \frac{1}{2} \left( \uvec{\uptheta} \, \uvec{\uptheta} + \uvec{\upphi} \, \uvec{\upphi} \right)  , \\
%\sum_{M=-J}^{J} \vec{Y}^{(\mpTM)}_{JM}(\uvec{r}) \, \vec{Y}^{(\mpTE)*}_{JM}(\uvec{r}) & = \frac{2J+1}{4\pi} \, \frac{1}{2} \left( \uvec{\upphi} \, \uvec{\uptheta} - \uvec{\uptheta} \, \uvec{\upphi} \right)  , \\
%\sum_{M=-J}^{J} \vec{Y}^{(\mpL)}_{JM}(\uvec{r}) \, \vec{Y}^{(\mpTE)*}_{JM}(\uvec{r}) & = \sum_{M=-J}^{J} \vec{Y}^{(\mpL)}_{JM}(\uvec{r}) \, \vec{Y}^{(\mpTM)*}_{JM}(\uvec{r}) = 0 
%\label{eq:sumoverM2b}
%\end{align}

To obtain Eqs.~(\ref{eq:sumoverM2a})--(\ref{eq:sumoverM2b}), one first writes
\begin{align}
\label{eq:sumoverM3a}
\vec{Y}^{(\mpL)}_{JM}(\uvec{r}) & = Y_{JM}(\uvec{r}) \, \uvec{r} , \\
\vec{Y}^{(\mpTE)}_{JM}(\uvec{r}) & =  \frac{1}{\sqrt{J(J+1)}} \left[ \frac{\partial Y_{JM}(\uvec{r})}{\partial\theta} \, \uvec{\uptheta} + \frac{1}{\sin\theta} \frac{\partial Y_{JM}(\uvec{r})}{\partial \phi} \, \uvec{\upphi} \right]  , \\
\vec{Y}^{(\mpTM)}_{JM}(\uvec{r}) & = \frac{1}{\sqrt{J(J+1)}} \left[ - \frac{1}{\sin\theta} \frac{\partial Y_{JM}(\uvec{r})}{\partial\phi} \, \uvec{\uptheta} + \frac{\partial Y_{JM}(\uvec{r})}{\partial \theta} \, \uvec{\upphi} \right] .
\label{eq:sumoverM3b}
\end{align}
The sums over $M$ involving derivatives of the $Y_{JM}$ are evaluated by differentiating the addition theorem of spherical harmonics
\begin{align}
\sum_{M=-J}^{J} Y_{JM}(\theta_1,\phi_1) \, Y^{*}_{JM}(\theta_2,\phi_2) = \frac{2J+1}{4\pi}  \, P_J(\mu) ,
\label{eq:sumoverM0}
\end{align}
where $P_J(\mu)$ is the Legendre polynomial of order $J$ with
\begin{align}
\mu = \cos\theta_1 \cos\theta_2 + \sin\theta_1 \sin\theta_2 \cos(\phi_1-\phi_2) .
\end{align}
For example, with the $\to$ indicating the limit  $(\theta_1,\phi_1)\to(\theta_2,\phi_2)$,
\begin{align}
&  \sum_M \frac{\partial Y_{JM}(\theta_1,\phi_1)}{\partial \theta_1}  \frac{\partial Y^{*}_{JM}(\theta_2,\phi_2)}{\partial \theta_2} =   \frac{2J+1}{4\pi} \, \frac{\partial^2 P_J(\mu)}{\partial\theta_1\partial\theta_2} \to  \frac{2J+1}{4\pi} \, P'_J(1) .
\end{align}
In this way one finds, using $P'_J(1)=J(J+1)/2$,
\begin{align}
\label{eq:sumoverM4a}
&  \sum_M \frac{\partial Y_{JM}}{\partial \theta}  \frac{\partial Y^{*}_{JM}}{\partial \theta} =  \sum_M \frac{1}{\sin^2\theta} \frac{\partial Y_{JM}}{\partial \phi}  \frac{\partial Y^{*}_{JM}}{\partial \phi} = \frac{2J+1}{4\pi} \, \frac{J(J+1)}{2} , \\
&  \sum_M \frac{\partial Y_{JM}}{\partial \theta}  \frac{\partial Y^{*}_{JM}}{\partial \phi} =  \sum_M \frac{\partial Y_{JM}}{\partial \theta}  Y^{*}_{JM} =  \sum_M \frac{\partial Y_{JM}}{\partial \phi}  Y^{*}_{JM} = 0 .
\label{eq:sumoverM4b}
\end{align}
Combining Eqs.~(\ref{eq:sumoverM3a})--(\ref{eq:sumoverM3b}) and (\ref{eq:sumoverM4a})--(\ref{eq:sumoverM4b}) one obtains Eqs.~(\ref{eq:sumoverM2a})--(\ref{eq:sumoverM2b}).

\subsection{Some relations between symmetric and symmetric traceless tensors}
\label{sec:symmetric_symmetrictraceless}

By definition, the symmetric traceless part $ \myoverbracket{ A_{i_1
    \cdot i_s} } $ of an rank-$s$ tensor $ A_{i_1 \cdots i_s } $ is
obtained by first symmetrizing $ A_{i_1 \cdots i_s } $ completely with
respect to all of its indices, and then subtracting all the possible
traces, i.e., contractions of pairs of indices, double pairs of
indices, \ldots, $s/2$-tuple pairs of indices. There is a general
formula for the resulting expression (cfr.\ Eq.~(2.2)
in~\cite{Damour:1990gj} and~\cite{Thorne:1980ru}, and~(2.44)
in~\cite{Pirani:1965}, where the connection with Legendre polynomials
is also explained),
\begin{align}
 \myoverbracket{ A_{i_1 \cdots i_s} }  = \frac{1}{N_s} \sum_{p=0}^{\lfloor s/2 \rfloor} C_{s,s-2p} \big \{ \delta_{i_1i_2} \cdots \delta_{i_{2p-1}i_{2p}} S_{i_{2p+1} \cdots i_s k_1 k_1 \cdots k_p k_p} \big \},
\label{eq:Thorne}
\end{align}
where the sum is over the number $p$ of traces (or of Kronecker $\delta$'s) in the right hand side, $\lfloor s/2 \rfloor$ is the largest integer smaller than or equal to $s/2$,
\begin{align}
C_{s,s-2p} = (-1)^p \frac{(2s-2p)!}{2^s (s-p)! (s-2p)!} 
\end{align}
is the coefficient of $x^{s-2p}$ in the Legendre polynomial $P_s(x)$ of order $s$ (in the standard normalization $P_s(1)=1$),
\begin{align}
& N_s=C_{s,s},
\\
& S_{i_1 \cdots i_s} = \big \{ A_{i_1 \cdots i_s} \big \},
\end{align}
and curly brackets indicate complete symmetrization with respect to the free indices inside the brackets, \begin{align}
\big \{ A_{i_1 \cdots i_s} \big \} = \frac{1}{s!} \, \sum_{\pi} A_{i_{\pi(1)} \cdots i_{\pi(s)}} ,
\end{align}
with the sum over the permutations $\pi$ of $12\cdots s$.

For products of spin operators $\vec{S}$ and a vector $\vec{q}$, Eq.~\eqref{eq:Thorne} gives
\begin{align}
\myoverbracket{(\vec{q} \cdot \vec{S})^n} =  \sum_{k=0}^{\lfloor n/2 \rfloor} c_{n,k} \, q^{2k} \vec{S}^{\,2k} \, (\vec{q} \cdot \vec{S})^{n-2k} ,
\label{eq:ST2Sq}
\end{align}
%\begin{align}
%\myoverbracket{(\vec{q} \cdot \vec{S})^n} =  \sum_{l=n,n-2,\ldots} c_{n,l} \, q^{n-l} \vec{S}^{\,n-l} \, (\vec{q} \cdot \vec{S})^{l} ,
%\label{eq:ST2Sq}
%\end{align}
where
\begin{align}
c_{n,k} = (-1)^k \frac{(n!)^2(2n-2k)!}{k!(2n)!(n-2k)!(n-k)!} .
\end{align}
Recall that for a particle of spin $j_\chi$,
\begin{align}
\vec{S}^{\,2}=j_\chi(j_\chi+1), \qquad \vec{S}^{\,2k} = [ j_\chi (j_\chi+1)]^k.
\end{align}
The quantity $c_{n,k}$ is the coefficient of $x^{n-2k}$ in the monic Legendre polynomial $\overline{P}_n(x)$ of degree $n$ (in a monic polynomial, the coefficient of the term of highest degree is equal to 1),
\begin{align}
\overline{P}_n(x) =  \frac{(2n-1)!!}{n!} P_n(x)  = \sum_{k=0}^{\lfloor n/2 \rfloor} c_{n,k} x^{n-2k} .
\end{align}
The first few cases, relevant for WIMPs of spin up to 2, are
\begin{align}
& \myoverbracket{\vec{q}\cdot\vec{S}} = \vec{q} \cdot \vec{S} ,
\\
& \myoverbracket{ (\vec{q}\cdot\vec{S})^2 } =  (\vec{q} \cdot \vec{S})^2 - \frac{1}{3} q^2 \vec{S}^2 ,
\\
& \myoverbracket{ (\vec{q}\cdot\vec{S})^3 } =  (\vec{q} \cdot \vec{S})^3 - \frac{3}{5} q^2 \vec{S}^2 ( \vec{q} \cdot \vec{S}) ,
\\
& \myoverbracket{ (\vec{q}\cdot\vec{S})^4 } =  (\vec{q} \cdot \vec{S})^4 - \frac{6}{7} q^2 \vec{S}^2 ( \vec{q} \cdot \vec{S})^2 + \frac{3}{35} q^4 \vec{S}^4 .
\end{align}
The coefficients can be compared to those appearing in the Legendre polynomials 
\begin{align}
& P_1(x) = x,
\\
& P_2(x) = \frac{3}{2} \bigg( x^2 - \frac{1}{3} \bigg) ,
\\
& P_3(x) = \frac{5}{2} \bigg( x^3 - \frac{3}{5} x \bigg) ,
\\
& P_4(x) = \frac{35}{8} \bigg( x^4 - \frac{6}{7} x^2 + \frac{3}{35} \bigg) .
\end{align}
The reason for the equality of these coefficients is that Legendre polynomials are the expressions in polar angles of the symmetric traceless tensors that define electrostatic multipoles. The identities that connect these quantities are
%\begin{align}
%\partial_{i_1} \cdots \partial_{i_l}  \frac{1}{r} = (-1)^l \frac{(2l-1)!! }{r^{2l+1}} r_{i_1} \cdots r_{i_l}  .
%\end{align}
\begin{align}
\big( \vec{q} \cdot \vec{\del} \big)^l \frac{1}{r} = (-1)^l \frac{(2l-1)!! }{r^{2l+1}} (\vec{q}\cdot\vec{r})^l 
 = \frac{(-1)^l}{l!} \frac{P_l(\cos\theta)}{r^{2l+1}} , \label{eq:electrostatic_multipoles}
\end{align}
\begin{align}
(\vec{q}\cdot\vec{r})^l = \frac{l!}{(2l-1)!!} q^l r^l P_l(\cos\theta) ,
\end{align}
where $\theta$ is the angle between $\vec{q}$ and $\vec{r}$.

A formula for products of spin operators $\vec{S}$ involving two vectors $\vec{q}$ and $\vec{a}$ is
\begin{align}
\myoverbracket{ (\vec{a}\cdot\vec{S}) \, (\vec{q}\cdot\vec{S})^{n-1} } =   \sum_{k=0}^{\lfloor n/2 \rfloor} c_{n,k} \, \bigg\{ & \frac{2k}{n-2k}  \, q^{2k-2} \vec{S}^{\,2k} \, \vec{a}\cdot\vec{q} \, \, \big\lsymm (\vec{q} \cdot \vec{S})^{n-2k} \big\rsymm  \nonumber \\ & + \frac{n-4k}{n-2k} \, q^{2k} \vec{S}^{\, 2k} \,\, \big\lsymm (\vec{a}\cdot\vec{S}) \, (\vec{q}\cdot\vec{S})^{n-2k-1} \big\rsymm \bigg\} .
\label{eq:ST2Saq}
\end{align}
It can be obtained by replacing one of the directional derivatives $\vec{q} \cdot \vec{\del}$ in Eq.~\eqref{eq:electrostatic_multipoles} with $\vec{a} \cdot \vec{\del}$, leading to the polynomials
\begin{align}
P_l(a,x) = P_l(x) + \frac{a-x}{l} \frac{dP_l(x)}{dx} .
\end{align}
The first few cases are
\begin{align}
& \myoverbracket{\vec{a}\cdot\vec{S}} = \vec{a} \cdot \vec{S} ,
\\
& \myoverbracket{ (\vec{a}\cdot\vec{S}) \, (\vec{q}\cdot\vec{S}) } =  \big\lsymm (\vec{a}\cdot\vec{S}) \, (\vec{q} \cdot \vec{S})\big\rsymm - \frac{1}{3} \, \vec{a}\cdot\vec{q}  \, \vec{S}^2 ,
\\
& \myoverbracket{ (\vec{a}\cdot\vec{S}) \, (\vec{q}\cdot\vec{S})^2 } =  \big\lsymm (\vec{a}\cdot\vec{S}) \, (\vec{q} \cdot \vec{S})^2\big\rsymm  - \frac{2}{5} \, (\vec{a}\cdot\vec{q}) \, \vec{S}^2  \, ( \vec{q} \cdot \vec{S}) - \frac{1}{5} q^2 \vec{S}^{\, 2} (\vec{a}\cdot\vec{S}) ,
\\
& \myoverbracket{ (\vec{a}\cdot\vec{S}) \, (\vec{q}\cdot\vec{S})^3  } =  \big\lsymm (\vec{a}\cdot\vec{S}) \, (\vec{q} \cdot \vec{S})^3 \big\rsymm - \frac{3}{7}  \, (\vec{a}\cdot\vec{q}) \, \vec{S}^2  \, ( \vec{q} \cdot \vec{S})^2 \nonumber \\ & \qquad\qquad\qquad\quad\,\, - \frac{3}{7} q^2 \vec{S}^2 \big\lsymm (\vec{a}\cdot\vec{S}) \, (\vec{q} \cdot \vec{S}) \big\rsymm  + \frac{3}{35} \, (\vec{a}\cdot\vec{q})  \, q^2 \vec{S}^4 .
\end{align}
Compare the coefficients to those in $P_l(a,x)$,
\begin{align}
& P_1(a,x) = a,
\\
& P_2(a,x) = \frac{3}{2} \bigg( a x - \frac{1}{3} \bigg) ,
\\
& P_3(a,x) = \frac{5}{2} \bigg( a x^2 - \frac{2}{5} x - \frac{1}{5} a \bigg) ,
\\
& P_4(a,x) = \frac{35}{8} \bigg( a x^3 - \frac{3}{7} x^2 - \frac{3}{7} a x + \frac{3}{35} \bigg) .
\end{align}
For completeness, we recall that
\begin{align}
& \big\lsymm (\vec{a}\cdot\vec{S}) \, (\vec{q} \cdot \vec{S})\big\rsymm = \frac{1}{2} \Big[ (\vec{a}\cdot\vec{S}) \, (\vec{q} \cdot \vec{S}) + (\vec{q}\cdot\vec{S}) \, (\vec{a} \cdot \vec{S})\Big] ,
\\
& \big\lsymm (\vec{a}\cdot\vec{S}) \, (\vec{q} \cdot \vec{S})^2 \big\rsymm = \frac{1}{3} \Big[ (\vec{a}\cdot\vec{S}) \, (\vec{q} \cdot \vec{S})^2 + (\vec{q}\cdot\vec{S}) \, (\vec{a} \cdot \vec{S}) \, (\vec{q} \cdot \vec{S})+ (\vec{q}\cdot\vec{S})^2 \, (\vec{a} \cdot \vec{S}) \Big] ,
\\
& \big\lsymm (\vec{a}\cdot\vec{S}) \, (\vec{q} \cdot \vec{S})^3 \big\rsymm = \frac{1}{4} \Big[ (\vec{a}\cdot\vec{S}) \, (\vec{q} \cdot \vec{S})^3 + (\vec{q}\cdot\vec{S}) \, (\vec{a} \cdot \vec{S}) \, (\vec{q} \cdot \vec{S})^2 \nonumber \\ & \qquad\qquad\qquad\qquad\quad\,\,\, + (\vec{q}\cdot\vec{S})^2 \, (\vec{a} \cdot \vec{S}) \, (\vec{q} \cdot \vec{S}) + (\vec{q}\cdot\vec{S})^3 \, (\vec{a} \cdot \vec{S}) \Big] ,
\end{align}
and so on.

Inverse relations to Eqs.~\eqref{eq:ST2Sq} and~\eqref{eq:ST2Saq}, giving the symmetric products of spin operators in terms of the symmetric traceless products, are
\begin{align}
\big\lsymm (\vec{q} \cdot \vec{S} )^n \big\rsymm =  \sum_{k=0}^{\lfloor n/2 \rfloor}  d_{n,k} \, q^{2k} \vec{S}^{\,2k} \, \myoverbracket{ (\vec{q} \cdot \vec{S} )^{n-2k} } ,
\label{eq:S2STq}
\end{align}
and
\begin{align}
\big\lsymm (\vec{a}\cdot\vec{S}) \, (\vec{q} \cdot \vec{S} )^{n-1} \big\rsymm =  \sum_{k=0}^{\lfloor n/2 \rfloor}  d_{n,k} \, \bigg[ 
& \frac{2k}{n}  \, q^{2k-2} \vec{S}^{\,2k} \, \vec{a}\cdot\vec{q} \, \, \myoverbracket{ (\vec{q} \cdot \vec{S})^{n-2k} }  \nonumber \\ 
& + \frac{n-2k}{n} \, q^{2k} \vec{S}^{\, 2k} \,\, \myoverbracket{ (\vec{a}\cdot\vec{S}) \, (\vec{q}\cdot\vec{S})^{n-2k-1} } 
\bigg] ,
\label{eq:S2STaq}
\end{align}
where
\begin{align}
%d_{n,k} = \frac{(n-l-1)!!(2l+1)!!}{(n+l+1)!!} {n \choose l} . with l=n-2k
d_{n,k} = \frac{(2k-1)!!(2n-4k+1)!!}{(2n-2k+1)!!} {n \choose 2k} . 
%\\
%d_{n,k} = \frac{(2n-4k+2)!(n-k+1)!}{k!(n-2k+1)!(2n-2k+2)!} \frac{n!}{(n-2k)!} .
\end{align}
The quantities $d_{n,k}$ appear in the expansion of powers of $x$ in monic Legendre polynomials,
\begin{align}
x^n  =  \sum_{k=0}^{\lfloor n/2 \rfloor} d_{n,k}  \overline{P}_{n-2k}(x) .
\end{align}
The first few cases are
\begin{align}
& \big\lsymm \vec{q} \cdot \vec{S} \big\rsymm =  \myoverbracket{ \vec{q} \cdot \vec{S} } , 
\\
& \big\lsymm (\vec{q} \cdot \vec{S} )^2 \big\rsymm =  \myoverbracket{ (\vec{q} \cdot \vec{S} )^2 } + \frac{1}{3} q^2 \vec{S}^{\,2} , 
\\
& \big\lsymm (\vec{q} \cdot \vec{S} )^3 \big\rsymm =  \myoverbracket{ (\vec{q} \cdot \vec{S} )^3 }  + \frac{3}{5} q^2 \vec{S}^{\,2} \myoverbracket{ \vec{q}\cdot\vec{S} }  , 
\\
& \big\lsymm (\vec{q} \cdot \vec{S} )^4 \big\rsymm =  \myoverbracket{ (\vec{q} \cdot \vec{S} )^4 }  + \frac{6}{7} q^2 \vec{S}^{\,2} \myoverbracket{ (\vec{q}\cdot\vec{S})^2 } + \frac{1}{5} q^4 \vec{S}^{\, 4} , 
\end{align}
\begin{align}
& \big\lsymm \vec{a} \cdot \vec{S} \big\rsymm =  \myoverbracket{ \vec{a} \cdot \vec{S} } , 
\\
& \big\lsymm (\vec{a} \cdot \vec{S}) \, (\vec{q} \cdot \vec{S} ) \big\rsymm =  \myoverbracket{ (\vec{a} \cdot \vec{S}) \, (\vec{q} \cdot \vec{S} ) } + \frac{1}{3} \, \vec{a}\cdot\vec{q}  \, \vec{S}^{\,2}, 
\\
& \big\lsymm (\vec{a} \cdot \vec{S}) \, (\vec{q} \cdot \vec{S} )^2 \big\rsymm =  \myoverbracket{ (\vec{a} \cdot \vec{S}) \, (\vec{q} \cdot \vec{S} )^2 }   + \frac{2}{5} \, \vec{a}\cdot\vec{q}  \, \vec{S}^{\,2} \myoverbracket{ \vec{q} \cdot \vec{S} } + \frac{1}{5} q^2 \vec{S}^{\,2} \myoverbracket{ \vec{a}\cdot\vec{S} }  , 
\\
& \big\lsymm (\vec{a} \cdot \vec{S}) \, (\vec{q} \cdot \vec{S} )^3 \big\rsymm =  \myoverbracket{ (\vec{a} \cdot \vec{S}) \, (\vec{q} \cdot \vec{S} )^3 }  + \frac{3}{7} \, \vec{a}\cdot\vec{q}  \, \vec{S}^{\,2} \myoverbracket{ (\vec{q} \cdot \vec{S})^2 } \nonumber \\ & \qquad\qquad\qquad\qquad + \frac{3}{7} q^2 \vec{S}^{\,2} \myoverbracket{ (\vec{a}\cdot\vec{S}) \, (\vec{q} \cdot \vec{S} )^2 } + \frac{1}{5}  \, \vec{a}\cdot\vec{q}  \, q^2 \vec{S}^{\,2} .
\end{align}

\subsection{Formulas for WIMP spin averages}
\label{sec:WIMP_spin_averages}
To prove
Eqs.~\eqref{eq:WIMP_spin_average_master_equation_a}--\eqref{eq:WIMP_spin_average_master_equation_b},
we make use of the formula~\cite{hess},
\begin{align}
\frac{1}{2j_\chi+1} \tr 
\Big ( \myoverbracket{S_{i_1} S_{i_2} \cdots S_{i_{s}}} 
\myoverbracket{S_{j_1} S_{j_2} \cdots S_{j_{s'}}} 
\Big )
= \delta_{ss'} \, \frac{s!}{(2s+1)!!} K_0 K_1 \cdots K_{s-1} \,  
\mathit\Delta^{(s)}_{i_1i_2 \cdots i_{s},j_1j_2 \cdots j_{s}} ,
\label{eq:Hess13.27}
\end{align}
Here the tensor $\mathit\Delta^{(s)}_{i_1i_2\cdots i_s,j_1j_2\cdots
  j_s} $ projects the symmetric traceless part of a
rank-$s$ tensor~\cite{hess}. In other words, it is defined by
\begin{align}
  \myoverbracket{S_{i_s} \cdots S_{i_s}}  = \mathit\Delta^{(s)}_{i_1i_2\cdots i_s,i'_1i'_2\cdots i'_s} S_{i_s'} \cdots S_{i_s'}.
  \label{eq:delta_tensor}
\end{align}

\noindent Saturating all the free indices of Eq~(\ref{eq:Hess13.27}) with the
product of momenta $\uvec{q}_{j_1} \cdots \uvec{q}_{j_{s}}$,
  $\uvec{q}_{j_1} \cdots \uvec{q}_{j_{s'}}$ one gets

\begin{align}
\frac{1}{2j_\chi+1} \tr 
& \Big (\myoverbracket{S_{i_1} \cdots S_{i_{s}}} 
\uvec{q}_{i_1} \cdots \uvec{q}_{i_s}
\myoverbracket{S_{j_1} \cdots S_{j_{s'}}} 
\uvec{q}_{j_1} \cdots \uvec{q}_{j_{s'}} 
\Big )
= \nonumber \\ & 
\delta_{ss'} \, \frac{s!}{(2s+1)!!} K_0 K_1 \cdots K_{s-1}  \, 
\myoverbracket{ \uvec{q}_{i_1} \cdots \uvec{q}_{i_s} } \,
\myoverbracket{ \uvec{q}_{j_1} \cdots \uvec{q}_{j_s} } .
\end{align}
\noindent Eq.~\eqref{eq:WIMP_spin_average_master_equation_a} follows
from the identity
\begin{align}
\myoverbracket{ \uvec{q}_{i_1} \cdots \uvec{q}_{i_s} } \,
\myoverbracket{ \uvec{q}_{j_1} \cdots \uvec{q}_{j_s} }  = \frac{1}{N_s} ,
\label{eq:WIMP_spin_average_master_equation_proof_3}
\end{align}
where
\begin{align}
N_s = \frac{(2s-1)!!}{s!}
\end{align}
is the coefficient of $x^s$ in the Legendre polynomial $P_s(x)$ of order $s$ (in the standard normalization $P_s(1)=1$).

To prove Eq.~\eqref{eq:WIMP_spin_average_master_equation_b}, write 
\begin{align}
\frac{1}{2j_\chi+1} \tr 
& \Big (\myoverbracket{S_{i_1} \cdots S_{i_{s}}} 
\uvec{q}_{i_1} \cdots \uvec{q}_{i_{s-1}} a_{i_s}
\myoverbracket{S_{j_1} \cdots S_{j_{s'}}}
\uvec{q}_{j_1} \cdots \uvec{q}_{j_{s'-1}}  b_{j_{s'}} 
\Big )
= \nonumber \\ & 
\delta_{ss'} \, \frac{s!}{(2s+1)!!} K_0 K_1 \cdots K_{s-1}  \, 
\myoverbracket{ \uvec{q}_{i_1} \cdots \uvec{q}_{i_{s-1}}   a_{i_s} } \,
\myoverbracket{ \uvec{q}_{j_1} \cdots \uvec{q}_{j_{s-1}} b_{i_s} } .
\label{eq:WIMP_spin_average_master_equation_proof_1}
\end{align}
In section~\ref{sec:qqa_qqb} we show that 
\begin{align}
\myoverbracket{ \uvec{q}_{i_1} \cdots \uvec{q}_{i_{s-1}}   a_{i_s} } \,
\myoverbracket{ \uvec{q}_{i_1} \cdots \uvec{q}_{i_{s-1}} b_{i_s} } = \frac{s+1}{2sN_s} \, (\vec{a} \cdot \vec{b})
\qquad \text{for $\vec{a} \cdot \vec{q} = 0$.}
\label{eq:WIMP_spin_average_master_equation_proof_2}
\end{align}
Thus write
\begin{align}
a_{i_s} = \uvec{q}_{i_s} \, a^{||} + a{}^{\perp}_{i_s} ,
\end{align}
where
\begin{align}
a^{||} = \uvec{q}_i a_i, 
\qquad
a{}^{\perp}_i =  ( \delta_{ij} - \uvec{q}_i \uvec{q}_j ) \, a{}^{\perp}_j ,
\end{align}
and similarly for $b_{i_s}$. Then using Eqs.~\eqref{eq:WIMP_spin_average_master_equation_proof_3}
 and \eqref{eq:WIMP_spin_average_master_equation_proof_2},
\begin{align}
\myoverbracket{ \uvec{q}_{i_1} \cdots \uvec{q}_{i_{s-1}} a_{i_s} } \,
\myoverbracket{ \uvec{q}_{j_1} \cdots \uvec{q}_{j_{s-1}} b_{i_s} }
= & 
\myoverbracket{ \uvec{q}_{i_1} \cdots \uvec{q}_{i_{s-1}} \uvec{q}_{i_s} } \,
\myoverbracket{ \uvec{q}_{j_1} \cdots \uvec{q}_{j_{s-1}} \uvec{q}_{i_s} } \, a^{||} \, b^{||}
\nonumber \\ 
& +
\myoverbracket{ \uvec{q}_{i_1} \cdots \uvec{q}_{i_{s-1}} \uvec{q}^{\phantom{\perp}}_{i_s} } \,
\myoverbracket{ \uvec{q}_{j_1} \cdots \uvec{q}_{j_{s-1}} b{}^{\perp}_{i_s} } \, a^{||} 
\nonumber \\ 
& +
\myoverbracket{ \uvec{q}_{i_1} \cdots \uvec{q}_{i_{s-1}} a{}^{\plus}_{i_s} } \,
\myoverbracket{ \uvec{q}_{j_1} \cdots \uvec{q}_{j_{s-1}} \uvec{q}^{\phantom{\perp}}_{i_s} } \, b^{||}
\nonumber \\ 
& +
\myoverbracket{ \uvec{q}_{i_1} \cdots \uvec{q}_{i_{s-1}} a{}^{\perp}_{i_s} } \,
\myoverbracket{ \uvec{q}_{j_1} \cdots \uvec{q}_{j_{s-1}} b{}^{\perp}_{i_s} } 
\\
= & 
\frac{1}{N_s} \, a^{||} \, b^{||}
+ \frac{s+1}{2sN_s} \, ( \vec{a}{}^{\perp} \cdot \vec{b}{}^{\perp} )
\\
= & 
\frac{1}{N_s} \Big [ \, \uvec{q}_i \uvec{q}_j 
+ \frac{s+1}{2s} ( \delta_{ij} - \uvec{q}_i \uvec{q}_j ) \Big ] a_i b_j 
\label{eq:WIMP_spin_average_master_equation_proof_4}
\end{align}
Then Eq.~\eqref{eq:WIMP_spin_average_master_equation_b} follows from combining Eqs.~\eqref{eq:WIMP_spin_average_master_equation_proof_1} and \eqref{eq:WIMP_spin_average_master_equation_proof_4}.

\subsection{Proof of Eq.~\protect{\eqref{eq:WIMP_spin_average_master_equation_proof_2}}}
\label{sec:qqa_qqb}

We apply formula~\eqref{eq:Thorne} to the product $ \uvec{q}_{i_1} \cdots \uvec{q}_{i_{s-1}} a_{i_s} $. The symmetrization gives
\begin{align}
S^{(a)}_{i_1 \cdots i_s} \equiv \big \{  \uvec{q}_{i_1} \cdots \uvec{q}_{i_{s-1}} a_{i_s}  \big \}  = \frac{1}{s!} \, \sum_{\pi} \uvec{q}_{i_{\pi(1)}} \cdots \uvec{q}_{i_{\pi(s-1)}} a_{i_{\pi(s)}} .
\end{align}
Separating the permutations involving $a_{i_1}$, $a_{i_2}$, \ldots, $a_{i_s}$ leads to
\begin{align}
S^{(a)}_{i_1 \cdots i_s} = \frac{1}{s!} \Big [ & a_{i_1} \!\!\!\! \sum_{\substack{\pi \\ \text{without } i_1}} \uvec{q}_{i_{\pi(1)}} \cdots \uvec{q}_{i_{\pi(s-1)}} + a_{i_2} \!\!\!\! \sum_{\substack{\pi \\ \text{without } i_2}} \uvec{q}_{i_{\pi(1)}} \cdots \uvec{q}_{i_{\pi(s-1)}} + \cdots \nonumber \\ & + a_{i_s} \!\!\!\! \sum_{\substack{\pi \\ \text{without } i_s}} \uvec{q}_{i_{\pi(1)}} \cdots \uvec{q}_{i_{\pi(s-1)}} \big ] 
\\
= \frac{1}{s!} \Big [ & a_{i_1} \uvec{q}_{i_2} \cdots \uvec{q}_{i_{s-1}} (s-1)! + a_{i_2} \uvec{q}_{i_{1}}  \uvec{q}_{i_3} \cdots \uvec{q}_{i_{s-1}} (s-1)! + \cdots \nonumber \\ & + a_{i_s} \uvec{q}_{i_{1}} \cdots \uvec{q}_{i_{s-1}} (s-1)! \big ] 
\\
= \frac{1}{s} \Big [ & a_{i_1} \uvec{q}_{i_2} \cdots \uvec{q}_{i_{s-1}} + \uvec{q}_{i_{1}}  a_{i_2}  \uvec{q}_{i_3} \cdots \uvec{q}_{i_{s-1}} + \cdots + \uvec{q}_{i_{1}} \cdots \uvec{q}_{i_{s-1}} a_{i_s}  \big ] .
\end{align}
In subtracting the traces, the assumption $\vec{a} \cdot \vec{q} = 0$ simplifies the expressions considerably, because all contractions involving one index from $a$ and the other from $\uvec{q}$ vanish. Contraction of one pair of indices gives 
\begin{align}
S^{(a)}_{i_1 \cdots i_{s-2} k k } = & \frac{1}{s} \Big [
a_{i_1} \uvec{q}_{i_2} \cdots \uvec{q}_{i_{s-2}} \uvec{q}_k \uvec{q}_k +  
\uvec{q}_{i_1} a_{i_2} \uvec{q}_{i_3} \cdots \uvec{q}_{i_{s-2}} \uvec{q}_k \uvec{q}_k +  
\cdots +
\uvec{q}_{i_1} \cdots \uvec{q}_{i_{s-3}} a_{i_{s-2}} \uvec{q}_k \uvec{q}_k +  
\nonumber \\ & \qquad
\uvec{q}_{i_1} \cdots \uvec{q}_{i_{s-3}} a_{i_{s-1}} a_k \uvec{q}_k +  
\uvec{q}_{i_1} \cdots \uvec{q}_{i_{s-3}} a_{i_{s-1}} \uvec{q}_k a_k 
 \Big ]
\nonumber\\
= & \frac{1}{s} \Big [
a_{i_1} \uvec{q}_{i_2} \cdots \uvec{q}_{i_{s-2}}  +  
\uvec{q}_{i_1} a_{i_2} \uvec{q}_{i_3} \cdots \uvec{q}_{i_{s-2}}  +  
\cdots +
\uvec{q}_{i_1} \cdots \uvec{q}_{i_{s-3}} a_{i_{s-2}} 
 \Big ]
\nonumber\\
= & \frac{s-2}{s} \, S^{(a)}_{i_1 \cdots i_{s-2}} . 
\end{align}
The contractions of double pairs, triple pairs, etc., follow by recursion as
\begin{align}
S^{(a)}_{i_1 \cdots i_{s-4} k_1k_1 k_2k_2} & = \frac{s-2}{s} \, \frac{s-4}{s-2} \, S^{(a)}_{i_1 \cdots i_{s-4}} = \frac{s-4}{s} \, S^{(a)}_{i_1 \cdots i_{s-2}} ,
\end{align}
and in general
\begin{align}
S^{(a)}_{i_1 \cdots i_{s-2p} k_1 k_1 \cdots k_p k_p}  = \frac{s-2p}{s} \, S^{(a)}_{i_1 \cdots i_{s-2p}} .
\end{align}
Inserting the latter expression into formula~\eqref{eq:Thorne} leads to
\begin{align}
\myoverbracket{ \uvec{q}_{i_1} \cdots \uvec{q}_{i_{s-1}} a_{i_s}  } = \frac{1}{N_s} \sum_{p=0}^{\lfloor s/2 \rfloor} C_{s,s-2p} \big \{ \delta_{i_1i_2} \cdots \delta_{i_{2p-1}i_{2p}} S^{(a)}_{i_{2p+1} \cdots i_s} \big \} \frac{s-2p}{s} .
\label{eq:prove_Thorne_1}
\end{align}

We now consider the product $ \myoverbracket{ \uvec{q}_{i_1} \cdots \uvec{q}_{i_{s-1}}   a_{i_s} } \,
\myoverbracket{ \uvec{q}_{j_1} \cdots \uvec{q}_{j_{s-1}} b_{i_s} } $, with $\vec{a} \cdot \vec{q} = 0$. The symmetric traceless operation on the left forces a symmetric traceless operation on the right, so we can write
\begin{align}
\myoverbracket{ \uvec{q}_{i_1} \cdots \uvec{q}_{i_{s-1}}   a_{i_s} } \,
\myoverbracket{ \uvec{q}_{i_1} \cdots \uvec{q}_{i_{s-1}} b_{i_s} }  = \myoverbracket{ \uvec{q}_{i_1} \cdots \uvec{q}_{i_{s-1}}   a_{i_s} } \, \uvec{q}_{i_1} \cdots \uvec{q}_{i_{s-1}} b_{i_s} .
\end{align}
Inserting Eq.~\eqref{eq:prove_Thorne_1}, and using 
\begin{align}
\delta_{i_1i_2} \cdots \delta_{i_{2p-1}i_{2p}} S_{i_1 \cdots i_s} = S_{i_{2p+1} \cdots i_s k_1k_1 \cdots k_pk_p} = \frac{s-2p}{s} S_{i_{2p+1} \cdots i_s} ,
\end{align}
we obtain
\begin{align}
&
\myoverbracket{ \uvec{q}_{i_1} \cdots \uvec{q}_{i_{s-1}}   a_{i_s} } \,
\myoverbracket{ \uvec{q}_{j_1} \cdots \uvec{q}_{j_{s-1}} b_{i_s} }  
\nonumber 
\nonumber\\
& \qquad = 
\sum_{p=0}^{\lfloor s/2 \rfloor} \frac{C_{s,s-2p}}{N_s} \, \frac{s-2p}{s} \big \{ \delta_{i_1i_2} \cdots \delta_{i_{2p-1}i_{2p}} S^{(a)}_{i_{2p+1} \cdots i_s} \big \} \uvec{q}_{i_1} \cdots \uvec{q}_{i_{s-1}} b_{i_s} 
\nonumber\\
& \qquad = 
\sum_{p=0}^{\lfloor s/2 \rfloor} \frac{C_{s,s-2p}}{N_s} \, \frac{s-2p}{s} \big \{ \delta_{i_1i_2} \cdots \delta_{i_{2p-1}i_{2p}} S^{(a)}_{i_{2p+1} \cdots i_s} \big \} \, \big \{ \uvec{q}_{i_1} \cdots \uvec{q}_{i_{s-1}} b_{i_s} \big \}
\nonumber\\
& \qquad = 
\sum_{p=0}^{\lfloor s/2 \rfloor} \frac{C_{s,s-2p}}{N_s} \, \frac{s-2p}{s} \big \{ \delta_{i_1i_2} \cdots \delta_{i_{2p-1}i_{2p}} S^{(a)}_{i_{2p+1} \cdots i_s} \big \} \, S^{(b)}_{i_1 \cdots i_s} 
\nonumber\\
& \qquad = 
\sum_{p=0}^{\lfloor s/2 \rfloor} \frac{C_{s,s-2p}}{N_s} \, \frac{s-2p}{s} \delta_{i_1i_2} \cdots \delta_{i_{2p-1}i_{2p}} S^{(a)}_{i_{2p+1} \cdots i_s} \, S^{(b)}_{i_1 \cdots i_s} 
\nonumber\\
& \qquad = 
\sum_{p=0}^{\lfloor s/2 \rfloor} \frac{C_{s,s-2p}}{N_s} \, \Big ( \frac{s-2p}{s} \Big )^2 S^{(a)}_{i_{2p+1} \cdots i_s} \, S^{(b)}_{i_{2p+1} \cdots i_s} 
\end{align}
To find the product $S^{(a)}_{i_1 \cdots i_n} S^{(b)}_{i_1 \cdots i_n}$, we write
\begin{align}
S_{i_1 \cdots i_n} S_{i_1 \cdots i_n} = \frac{1}{n^2} 
& 
\Big [ 
a_{i_1} \uvec{q}_{i_2} \cdots \uvec{q}_{i_n}  +  
\uvec{q}_{i_1} a_{i_2} \uvec{q}_{i_3} \cdots \uvec{q}_{i_n}  +  
\cdots +
\uvec{q}_{i_1} \cdots \uvec{q}_{i_{n-1}} a_{i_n} 
\Big ] 
\nonumber \\
& 
\Big [ 
b_{i_1} \uvec{q}_{i_2} \cdots \uvec{q}_{i_n}  +  
\uvec{q}_{i_1} b_{i_2} \uvec{q}_{i_3} \cdots \uvec{q}_{i_n}  +  
\cdots +
\uvec{q}_{i_1} \cdots \uvec{q}_{i_{n-1}} b_{i_n} 
\Big ] .
\end{align}
Now all cross terms in the product of the two square brackets have $a_i q_i =0$ and so are zero. Only the square terms remain, and there are $n$ of them. Thus,
\begin{align}
S_{i_1 \cdots i_n} S_{i_1 \cdots i_n} = \frac{1}{n} (\vec{a} \cdot \vec{b})  .
\end{align}
Hence,
\begin{align}
\myoverbracket{ \uvec{q}_{i_1} \cdots \uvec{q}_{i_{s-1}}   a_{i_s} } \,
\myoverbracket{ \uvec{q}_{j_1} \cdots \uvec{q}_{j_{s-1}} b_{i_s} }  
& =
\sum_{p=0}^{\lfloor s/2 \rfloor} \frac{C_{s,s-2p}}{N_s} \, \Big ( \frac{s-2p}{s} \Big )^2  \frac{1}{s-2p} (\vec{a} \cdot \vec{b})  
\nonumber\\
& = 
(\vec{a} \cdot \vec{b})  \frac{1}{s^2 N_s} \sum_{p=0}^{\lfloor s/2 \rfloor} (s-2p) \, C_{s,s-2p} .
\end{align}
To evaluate the last sum, we recall that by definition of $C_{s,s-2p}$,
\begin{align}
P_s(x) =  \sum_{p=0}^{\lfloor s/2 \rfloor} C_{s,s-2p} \, x^{s-2p} .
\end{align}
Taking one derivative,
\begin{align}
P'_s(x) =  \sum_{p=0}^{\lfloor s/2 \rfloor} (s-2p) C_{s,s-2p} \, x^{s-2p} .
\end{align}
Thus
\begin{align}
\sum_{p=0}^{\lfloor s/2 \rfloor} (s-2p) \, C_{s,s-2p}  = P'_s(1) = \frac{s(s+1)}{2} .
\end{align}
We conclude that
\begin{align}
&
\myoverbracket{ \uvec{q}_{i_1} \cdots \uvec{q}_{i_{s-1}} a_{i_s} } \,
\myoverbracket{ \uvec{q}_{j_1} \cdots \uvec{q}_{j_{s-1}} b_{i_s} }  
=
\frac{s+1}{2sN_s} (\vec{a} \cdot \vec{b}) ,
\end{align}
which is Eq.~\eqref{eq:WIMP_spin_average_master_equation_proof_2}.


\begin{thebibliography}{99}
%\bibliographystyle{JHEP} \bibliography{bibliography}

\bibitem{chang_momentum_dependence_2010}
S.~Chang, A.~Pierce and N.~Weiner, \emph{{Momentum Dependent Dark Matter
  Scattering}},
  \href{https://doi.org/10.1088/1475-7516/2010/01/006}{\emph{JCAP} {\bfseries
  1001} (2010) 006}, [\href{https://arxiv.org/abs/0908.3192}{{\ttfamily
  0908.3192}}].

\bibitem{dobrescu_nreft}
B.~A. Dobrescu and I.~Mocioiu, \emph{{Spin-dependent macroscopic forces from
  new particle exchange}},
  \href{https://doi.org/10.1088/1126-6708/2006/11/005}{\emph{JHEP} {\bfseries
  11} (2006) 005}, [\href{https://arxiv.org/abs/hep-ph/0605342}{{\ttfamily
  hep-ph/0605342}}].

\bibitem{fan_2010}
J.~Fan, M.~Reece and L.-T. Wang, \emph{{Non-relativistic effective theory of
  dark matter direct detection}},
  \href{https://doi.org/10.1088/1475-7516/2010/11/042}{\emph{JCAP} {\bfseries
  1011} (2010) 042}, [\href{https://arxiv.org/abs/1008.1591}{{\ttfamily
  1008.1591}}].

\bibitem{Hisano_vector_effective}
J.~Hisano, K.~Ishiwata, N.~Nagata and M.~Yamanaka, \emph{{Direct Detection of
  Vector Dark Matter}}, \href{https://doi.org/10.1143/PTP.126.435}{\emph{Prog.
  Theor. Phys.} {\bfseries 126} (2011) 435--456},
  [\href{https://arxiv.org/abs/1012.5455}{{\ttfamily 1012.5455}}].

\bibitem{Hisano_EW_DM}
J.~Hisano, K.~Ishiwata, N.~Nagata and T.~Takesako, \emph{{Direct Detection of
  Electroweak-Interacting Dark Matter}},
  \href{https://doi.org/10.1007/JHEP07(2011)005}{\emph{JHEP} {\bfseries 07}
  (2011) 005}, [\href{https://arxiv.org/abs/1104.0228}{{\ttfamily 1104.0228}}].

\bibitem{hill_solon_nreft}
R.~J. Hill and M.~P. Solon, \emph{{WIMP-nucleon scattering with heavy WIMP
  effective theory}},
  \href{https://doi.org/10.1103/PhysRevLett.112.211602}{\emph{Phys. Rev. Lett.}
  {\bfseries 112} (2014) 211602},
  [\href{https://arxiv.org/abs/1309.4092}{{\ttfamily 1309.4092}}].

\bibitem{peter_nreft}
V.~Gluscevic and A.~H.~G. Peter, \emph{{Understanding WIMP-baryon interactions
  with direct detection: A Roadmap}},
  \href{https://doi.org/10.1088/1475-7516/2014/09/040}{\emph{JCAP} {\bfseries
  1409} (2014) 040}, [\href{https://arxiv.org/abs/1406.7008}{{\ttfamily
  1406.7008}}].

\bibitem{cirelli_tools_2013}
M.~Cirelli, E.~Del~Nobile and P.~Panci, \emph{{Tools for model-independent
  bounds in direct dark matter searches}},
  \href{https://doi.org/10.1088/1475-7516/2013/10/019}{\emph{JCAP} {\bfseries
  1310} (2013) 019}, [\href{https://arxiv.org/abs/1307.5955}{{\ttfamily
  1307.5955}}].

\bibitem{effective_wimps_2014}
S.~Chang, R.~Edezhath, J.~Hutchinson and M.~Luty, \emph{{Effective WIMPs}},
  \href{https://doi.org/10.1103/PhysRevD.89.015011}{\emph{Phys. Rev.}
  {\bfseries D89} (2014) 015011},
  [\href{https://arxiv.org/abs/1307.8120}{{\ttfamily 1307.8120}}].

\bibitem{catena_nreft}
R.~Catena, \emph{{Prospects for direct detection of dark matter in an effective
  theory approach}},
  \href{https://doi.org/10.1088/1475-7516/2014/07/055}{\emph{JCAP} {\bfseries
  1407} (2014) 055}, [\href{https://arxiv.org/abs/1406.0524}{{\ttfamily
  1406.0524}}].

\bibitem{catena_directionality_nreft_2015}
R.~Catena, \emph{{Dark matter directional detection in non-relativistic
  effective theories}},
  \href{https://doi.org/10.1088/1475-7516/2015/07/026}{\emph{JCAP} {\bfseries
  1507} (2015) 026}, [\href{https://arxiv.org/abs/1505.06441}{{\ttfamily
  1505.06441}}].

\bibitem{Hisano_Wino_DM}
J.~Hisano, K.~Ishiwata and N.~Nagata, \emph{{QCD Effects on Direct Detection of
  Wino Dark Matter}},
  \href{https://doi.org/10.1007/JHEP06(2015)097}{\emph{JHEP} {\bfseries 06}
  (2015) 097}, [\href{https://arxiv.org/abs/1504.00915}{{\ttfamily
  1504.00915}}].

\bibitem{Catena_Gondolo_global_fits}
R.~Catena and P.~Gondolo, \emph{{Global fits of the dark matter-nucleon
  effective interactions}},
  \href{https://doi.org/10.1088/1475-7516/2014/09/045}{\emph{JCAP} {\bfseries
  1409} (2014) 045}, [\href{https://arxiv.org/abs/1405.2637}{{\ttfamily
  1405.2637}}].

\bibitem{cerdeno_nreft}
{\scshape SuperCDMS} collaboration, K.~Schneck et~al., \emph{{Dark matter
  effective field theory scattering in direct detection experiments}},
  \href{https://doi.org/10.1103/PhysRevD.91.092004}{\emph{Phys. Rev.}
  {\bfseries D91} (2015) 092004},
  [\href{https://arxiv.org/abs/1503.03379}{{\ttfamily 1503.03379}}].

\bibitem{Catena_Gondolo_global_limits}
R.~Catena and P.~Gondolo, \emph{{Global limits and interference patterns in
  dark matter direct detection}},
  \href{https://doi.org/10.1088/1475-7516/2015/08/022}{\emph{JCAP} {\bfseries
  1508} (2015) 022}, [\href{https://arxiv.org/abs/1504.06554}{{\ttfamily
  1504.06554}}].

\bibitem{nreft_bayesian}
H.~Rogers, D.~G. Cerdeno, P.~Cushman, F.~Livet and V.~Mandic,
  \emph{{Multidimensional effective field theory analysis for direct detection
  of dark matter}},
  \href{https://doi.org/10.1103/PhysRevD.95.082003}{\emph{Phys. Rev.}
  {\bfseries D95} (2017) 082003},
  [\href{https://arxiv.org/abs/1612.09038}{{\ttfamily 1612.09038}}].

\bibitem{xenon100_nreft}
{\scshape XENON} collaboration, E.~Aprile et~al., \emph{{Effective field theory
  search for high-energy nuclear recoils using the XENON100 dark matter
  detector}}, \href{https://doi.org/10.1103/PhysRevD.96.042004}{\emph{Phys.
  Rev.} {\bfseries D96} (2017) 042004},
  [\href{https://arxiv.org/abs/1705.02614}{{\ttfamily 1705.02614}}].

\bibitem{cresst_nreft}
{\scshape CRESST} collaboration, G.~Angloher et~al., \emph{{Limits on Dark
  Matter Effective Field Theory Parameters with CRESST-II}},
  \href{https://doi.org/10.1140/epjc/s10052-018-6523-4}{\emph{Eur. Phys. J.}
  {\bfseries C79} (2019) 43},
  [\href{https://arxiv.org/abs/1809.03753}{{\ttfamily 1809.03753}}].

\bibitem{matching_solon1}
R.~J. Hill and M.~P. Solon, \emph{{Universal behavior in the scattering of
  heavy, weakly interacting dark matter on nuclear targets}},
  \href{https://doi.org/10.1016/j.physletb.2012.01.013}{\emph{Phys. Lett.}
  {\bfseries B707} (2012) 539--545},
  [\href{https://arxiv.org/abs/1111.0016}{{\ttfamily 1111.0016}}].

\bibitem{matching_solon2}
R.~J. Hill and M.~P. Solon, \emph{{Universal behavior in the scattering of
  heavy, weakly interacting dark matter on nuclear targets}},
  \href{https://doi.org/10.1016/j.physletb.2012.01.013}{\emph{Phys. Lett.}
  {\bfseries B707} (2012) 539--545},
  [\href{https://arxiv.org/abs/1111.0016}{{\ttfamily 1111.0016}}].

\bibitem{chiral_eft}
M.~Hoferichter, P.~Klos and A.~Schwenk, \emph{{Chiral power counting of one-
  and two-body currents in direct detection of dark matter}},
  \href{https://doi.org/10.1016/j.physletb.2015.05.041}{\emph{Phys. Lett.}
  {\bfseries B746} (2015) 410--416},
  [\href{https://arxiv.org/abs/1503.04811}{{\ttfamily 1503.04811}}].

\bibitem{hoferichter_si}
M.~Hoferichter, P.~Klos, J.~Men{\'e}ndez and A.~Schwenk, \emph{{Analysis
  strategies for general spin-independent WIMP-nucleus scattering}},
  \href{https://doi.org/10.1103/PhysRevD.94.063505}{\emph{Phys. Rev.}
  {\bfseries D94} (2016) 063505},
  [\href{https://arxiv.org/abs/1605.08043}{{\ttfamily 1605.08043}}].

\bibitem{bishara_2017}
F.~Bishara, J.~Brod, B.~Grinstein and J.~Zupan, \emph{{From quarks to nucleons
  in dark matter direct detection}},
  \href{https://doi.org/10.1007/JHEP11(2017)059}{\emph{JHEP} {\bfseries 11}
  (2017) 059}, [\href{https://arxiv.org/abs/1707.06998}{{\ttfamily
  1707.06998}}].

\bibitem{GoodmanWitten}
M.~W. Goodman and E.~Witten, \emph{{Detectability of Certain Dark Matter
  Candidates}}, \href{https://doi.org/10.1103/PhysRevD.31.3059}{\emph{Phys.
  Rev. D} {\bfseries 31} (1985) 3059}.

\bibitem{haxton1}
A.~L. Fitzpatrick, W.~Haxton, E.~Katz, N.~Lubbers and Y.~Xu, \emph{{The
  Effective Field Theory of Dark Matter Direct Detection}},
  \href{https://doi.org/10.1088/1475-7516/2013/02/004}{\emph{JCAP} {\bfseries
  1302} (2013) 004}, [\href{https://arxiv.org/abs/1203.3542}{{\ttfamily
  1203.3542}}].

\bibitem{haxton2}
N.~Anand, A.~L. Fitzpatrick and W.~C. Haxton, \emph{{Weakly interacting massive
  particle-nucleus elastic scattering response}},
  \href{https://doi.org/10.1103/PhysRevC.89.065501}{\emph{Phys. Rev.}
  {\bfseries C89} (2014) 065501},
  [\href{https://arxiv.org/abs/1308.6288}{{\ttfamily 1308.6288}}].

\bibitem{krauss_spin_1}
J.~B. Dent, L.~M. Krauss, J.~L. Newstead and S.~Sabharwal, \emph{{General
  analysis of direct dark matter detection: From microphysics to observational
  signatures}}, \href{https://doi.org/10.1103/PhysRevD.92.063515}{\emph{Phys.
  Rev.} {\bfseries D92} (2015) 063515},
  [\href{https://arxiv.org/abs/1505.03117}{{\ttfamily 1505.03117}}].

\bibitem{catena_krauss_spin_1}
R.~Catena, K.~Fridell and M.~B. Krauss, \emph{{Non-relativistic Effective
  Interactions of Spin 1 Dark Matter}},
  \href{https://doi.org/10.1007/JHEP08(2019)030}{\emph{JHEP} {\bfseries 08}
  (2019) 030}, [\href{https://arxiv.org/abs/1907.02910}{{\ttfamily
  1907.02910}}].

\bibitem{barger_2008}
V.~{Barger}, W.-Y. {Keung} and G.~{Shaughnessy}, \emph{{Spin dependence of dark
  matter scattering}},
  \href{https://doi.org/10.1103/PhysRevD.78.056007}{\emph{Phys. Rev.}
  {\bfseries 78} (Sep, 2008) 056007},
  [\href{https://arxiv.org/abs/0806.1962}{{\ttfamily 0806.1962}}].

\bibitem{Barello:2014uda}
G.~Barello, S.~Chang and C.~A. Newby, \emph{{A Model Independent Approach to
  Inelastic Dark Matter Scattering}},
  \href{https://doi.org/10.1103/PhysRevD.90.094027}{\emph{Phys. Rev. D}
  {\bfseries 90} (2014) 094027},
  [\href{https://arxiv.org/abs/1409.0536}{{\ttfamily 1409.0536}}].

\bibitem{two_body_1}
\emph{{Large-scale nuclear structure calculations for spin-dependent WIMP
  scattering with chiral effective field theory currents}},
  \href{https://doi.org/10.1103/PhysRevD.89.029901,
  10.1103/PhysRevD.88.083516}{\emph{Phys. Rev.} {\bfseries D88} (2013) 083516},
  [\href{https://arxiv.org/abs/1304.7684}{{\ttfamily 1304.7684}}].

\bibitem{two_body_2}
\emph{{Nuclear structure aspects of spin-independent WIMP scattering off
  xenon}}, \href{https://doi.org/10.1103/PhysRevD.91.043520}{\emph{Phys. Rev.}
  {\bfseries D91} (2015) 043520},
  [\href{https://arxiv.org/abs/1412.6091}{{\ttfamily 1412.6091}}].

\bibitem{catena}
R.~Catena and B.~Schwabe, \emph{{Form factors for dark matter capture by the
  Sun in effective theories}},
  \href{https://doi.org/10.1088/1475-7516/2015/04/042}{\emph{JCAP} {\bfseries
  1504} (2015) 042}, [\href{https://arxiv.org/abs/1501.03729}{{\ttfamily
  1501.03729}}].

\bibitem{Yennie:1957}
D.~R. Yennie, M.~M. L\'evy and D.~G. Ravenhall, \emph{Electromagnetic structure
  of nucleons}, \href{https://doi.org/10.1103/RevModPhys.29.144}{\emph{Rev.
  Mod. Phys.} {\bfseries 29} (Jan, 1957) 144--157}.

\bibitem{Friar:1973}
J.~L. Friar, \emph{{Relativistic Corrections to Electron Scattering by ${}^2$H,
  ${}^3$He, and ${}^4$He$^*$}}, {\emph{Annals Phys.} {\bfseries 81} (1973)
  332--363}.

\bibitem{Kelly:2002if}
J.~J. Kelly, \emph{{Nucleon charge and magnetization densities from Sachs
  form-factors}}, \href{https://doi.org/10.1103/PhysRevC.66.065203}{\emph{Phys.
  Rev. C} {\bfseries 66} (2002) 065203},
  [\href{https://arxiv.org/abs/hep-ph/0204239}{{\ttfamily hep-ph/0204239}}].

\bibitem{Perdrisat:2006hj}
C.~Perdrisat, V.~Punjabi and M.~Vanderhaeghen, \emph{{Nucleon Electromagnetic
  Form Factors}}, \href{https://doi.org/10.1016/j.ppnp.2007.05.001}{\emph{Prog.
  Part. Nucl. Phys.} {\bfseries 59} (2007) 694--764},
  [\href{https://arxiv.org/abs/hep-ph/0612014}{{\ttfamily hep-ph/0612014}}].

\bibitem{hess}
S.~Hess, \emph{Tensors for Physics}.
\newblock Springer, 2015.

\bibitem{deforest_walecka}
J.~De~Forest, T. and J.~Walecka, \emph{{Electron scattering and nuclear
  structure}}, \href{https://doi.org/10.1080/00018736600101254}{\emph{Adv.
  Phys.} {\bfseries 15} (1966) 1--109}.

\bibitem{donnelly_walecka}
T.~W. Donnelly and J.~D. Walecka, \emph{{Electron Scattering and Nuclear
  Structure}},
  \href{https://doi.org/10.1146/annurev.ns.25.120175.001553}{\emph{Ann. Rev.
  Nucl. Part. Sci.} {\bfseries 25} (1975) 329--405}.

\bibitem{walecka}
J.~Walecka, \emph{Semileptonic weak interactions in nuclei},  in \emph{Muon
  Physics} (V.~W. Hughes and C.~Wu, eds.), pp.~113 -- 218.
\newblock Academic Press, 1975.
\newblock
  \href{https://doi.org/https://doi.org/10.1016/B978-0-12-360602-0.50010-5}{DOI}.

\bibitem{Damour:1990gj}
T.~Damour and B.~R. Iyer, \emph{{Multipole analysis for electromagnetism and
  linearized gravity with irreducible cartesian tensors}},
  \href{https://doi.org/10.1103/PhysRevD.43.3259}{\emph{Phys. Rev.} {\bfseries
  D43} (1991) 3259--3272}.

\bibitem{Thorne:1980ru}
K.~S. Thorne, \emph{{Multipole Expansions of Gravitational Radiation}},
  \href{https://doi.org/10.1103/RevModPhys.52.299}{\emph{Rev. Mod. Phys.}
  {\bfseries 52} (1980) 299--339}.

\bibitem{Pirani:1965}
F.~A.~E. Pirani, \emph{{Introduction to Gravitational Radiation Theory}},  in
  \emph{{Lectures on general relativity}} (A.~{Trautmann}, F.~A.~E. {Pirani}
  and H.~{Bondi}, eds.), pp.~249--374.
\newblock Prentice Hall, Inc., Englewood Cliffs, NJ, 1965.

\end{thebibliography}
\end{document}